\documentclass[twocolumn]{aastex63}

\usepackage{amsmath,bm}	
\usepackage{enumitem}
\usepackage[space]{grffile}  

\def\Kepler{\textit{Kepler}} 
\def\Gaia{\textit{Gaia}} 

\defcitealias{HFR2019}{Paper I}

\received{14 August 2020}
\revised{28 October 2020}
\accepted{7 December 2020}
\shorttitle{Clustered Planetary Systems with Host Star Color}
\shortauthors{He, Ford, and Ragozzine}
\graphicspath{{./}{Figures/}}

\begin{document}

\title{Architectures of Exoplanetary Systems. II: An Increase in Inner Planetary System Occurrence Toward Later Spectral Types for Kepler's FGK Dwarfs}

\correspondingauthor{Matthias Yang He}
\email{myh7@psu.edu}

\author[0000-0002-5223-7945]{Matthias Y. He}
\affiliation{Department of Astronomy \& Astrophysics, 525 Davey Laboratory, The Pennsylvania State University, University Park, PA 16802, USA}
\affiliation{Center for Exoplanets \& Habitable Worlds, 525 Davey Laboratory, The Pennsylvania State University, University Park, PA 16802, USA}
\affiliation{Center for Astrostatistics, 525 Davey Laboratory, The Pennsylvania State University, University Park, PA 16802, USA}
\affiliation{Institute for Computational \& Data Sciences, 525 Davey Laboratory, The Pennsylvania State University, University Park, PA 16802, USA}

\author[0000-0001-6545-639X]{Eric B. Ford}
\affiliation{Department of Astronomy \& Astrophysics, 525 Davey Laboratory, The Pennsylvania State University, University Park, PA 16802, USA}
\affiliation{Center for Exoplanets \& Habitable Worlds, 525 Davey Laboratory, The Pennsylvania State University, University Park, PA 16802, USA}
\affiliation{Center for Astrostatistics, 525 Davey Laboratory, The Pennsylvania State University, University Park, PA 16802, USA}
\affiliation{Institute for Computational \& Data Sciences, 525 Davey Laboratory, The Pennsylvania State University, University Park, PA 16802, USA}

\author[0000-0003-1080-9770]{Darin Ragozzine}
\affiliation{Department of Physics \& Astronomy, N283 ESC, Brigham Young University, Provo, UT 84602, USA}



\begin{abstract}

The \Kepler{} mission observed thousands of transiting exoplanet candidates around hundreds of thousands of FGK dwarf stars.  
He, Ford, \& Ragozzine (2019) applied forward modeling to infer the distribution of intrinsic architectures of planetary systems, 
developed a clustered Poisson point process model for exoplanetary systems (\texttt{SysSim}) to reproduce the marginal distributions of the observed \Kepler{} population, and they showed that orbital periods and planet radii are clustered within a given planetary system.
Here, we extend the clustered model to explore correlations between planetary systems and their host star properties.
We split the sample of \Kepler{} FGK dwarfs into two halves and model the fraction of stars with planets (between 0.5--10 $R_\oplus$ and 3--300 days), $f_{\rm swpa}$, as a linear function of the Gaia DR2 color.
We confirm previous findings that the occurrence of these planetary systems rises significantly toward later type (redder) stars.
The fraction of stars with planets increases from $f_{\rm swpa} = 0.32_{-0.11}^{+0.12}$ for F2V dwarfs to $f_{\rm swpa} = 0.96_{-0.19}^{+0.04}$ for mid K-dwarfs.
About half ($f_{\rm swpa} = 0.57_{-0.10}^{+0.14}$) of all solar-type (G2V) dwarfs harbor a planetary system between 3 and 300 days.
This simple model can closely match the observed multiplicity distributions of both the bluer and redder halves in our sample, suggesting that the architectures of planetary systems around stars of different spectral types may be similar aside from a shift in the overall fraction of planet hosting stars.

\end{abstract}



\section{Introduction} \label{sec:intro}

NASA's \Kepler{} mission \citep{B2010, B2011a, B2011b, B2013} boosted the number of strong exoplanet candidates by surveying $\sim 200,000$ stars for nearly four years. 
It revealed a large number of transiting super-Earth to sub-Neptune size planets ($R_p \lesssim 4 R_\oplus$) at short orbital periods ($P \lesssim 1$ yr) \citep{La2011, Li2011a, Li2011b, H2012, Li2014, R2014} and an abundance of tightly-spaced multitransiting planetary systems.
These offer key clues about their architectures and formation histories \citep{RH2010, F2014, WF2015, HFR2019}. 
In addition to enabling robust calculations of the planet population statistics themselves, the \Kepler{} catalog also allows for the detailed study of the correlations between planetary systems and their host stars.

Many previous studies have used the census of exoplanet candidates from \Kepler{} to infer the occurrence rates of planets around primarily main sequence stars of F, G, and K spectral types \citep{CS2011, H2012, F2013, PMH2013b, H2018, Mu2018, H2019, ZH2019}. Of these studies, \citet{H2012} was the first to report a dependence of the planet occurrence rate on host star spectral type. They used 1235 planet candidates (with orbital periods $< 50$ days) from the first three quarters of \Kepler{} data \citep{B2011b}, around dwarf stars spanning $T_{\rm eff} = 3600-7100$ K, to explore how the occurrence of planets varies as a function of stellar effective temperature. By splitting the stellar sample into 500 K bins, they found a strong inverse relationship between the occurrence of small ($R_p = 2-4 R_\oplus$) planets and $T_{\rm eff}$, for which they fit a linear model, $f(T_{\rm eff}) = f_0 + k_T(T_{\rm eff} - 5100 {\rm K})/1000 {\rm K}$ where $f_0 = 0.165 \pm 0.011$ and $k_T = -0.081 \pm 0.011$. Interestingly, they did not find any such correlation for larger planets ($R_p = 4-32 R_\oplus$).

In contrast, \citet{F2013} found no dependence between planet occurrence and spectral type for the same planet sizes, using the first 16 quarters of the \Kepler{} data containing $\sim 2300$ planet candidates \citep{B2013} combined with a new model for the detection efficiency and accounting for false positives. They argue that the increase in planet occurrence toward later type stars is a result of observational bias, manifesting for three reasons: (1) the \Kepler{} planet candidate list is incomplete for sub-Neptunes, such that many of these planets transiting the larger (earlier type) stars have not been recovered; (2) the distribution of planet radii rises towards smaller sizes, which are easier to detect around smaller (later type) stars; and (3) the false positive rate is slightly higher for later-type stars, artificially boosting the occurrence rate if not corrected for.
In this work, we are able to address these concerns by taking advantage of more recent improvements in modeling \Kepler{}'s detection efficiency \citep{BC2017a, BC2017b, BC2017c, C2020, Co2017}, including accounting for the rate of false alarms due to instrument systematics and astrophysical false positives due to background blends and eclipsing binaries that can be recognized by the transit shape and/or pixel offset from the \textit{Robovetter} and using the \Kepler{} DR25 catalog that was vetted using this fully automated pipeline \citep{T2018}. We show that the limit of transit detectability for a given planet is more complicated than what would result from only considering the stellar radius.

\citet{Mu2015} extended the above studies to include a large sample of F, G, K, and M dwarfs, dividing these four spectral types using $T_{\rm eff}$ and computing the planet occurrence rates in each bin. With an eye toward exploring how the planetary system architectures, not just the overall rate of planets, may differ across stellar types, they calculated the occurrence rate as a function of semi-major axis for each spectral type. They find that the occurrence rate of planets between $1-4 R_\oplus$ increases toward later spectral type at all separations out to $\sim 150$ days, in agreement with the findings of \citet{H2012}. \citet{Mu2015} also suggest that the cut-off semi-major axis (potentially indicative of the inner disk edge, where planets become less common interior of) shifts toward smaller separations for planets around later type stars.

Numerous studies have also constrained the occurrence of planets around M-dwarf stars, including those using surveys other than \Kepler{} (e.g., \citealt{E2006, C2008, G2013, Bon2013, L2016}). The earlier works relied on radial velocity (RV) data and more massive planets, attempting to correct for survey completeness; for example, \citet{C2008} used a sample of 585 FGKM stars with RV measurements and found that the occurrence of giant planets ($>0.3 M_{\rm Jup}$) within 2000 days around M-dwarfs is about an order of magnitude lower than that of FGK dwarfs. Similarly, \citet{Bon2013} also found a lower occurrence rate of giant planets around M-dwarfs compared to earlier type dwarfs, although at shorter periods $<100$ days. Results from direct imaging are qualitatively consistent; \citet{L2016} observed 58 M-dwarfs and inferred that giant planet companions (with masses $>1 M_{\rm Jup}$ and low mass ratios $<1\%$) tend to be less common around low mass stars.

Returning to studies with the \Kepler{} data, which mainly consists of smaller, super-Earth to sub-Neptune sized planets, \citet{DC2013} used a sample of $\sim 3900$ stars then estimated to have $T_{\rm eff} < 4000$ K hosting 95 planet candidates to measure the occurrence rate of small planets ($R_p = 0.5-4 R_\oplus$). Their results for the occurrence rates of small planets are generally larger than the values for FGK dwarfs found by other studies. Intriguingly, however, they find that the occurrence of planets ($R_p = 1.4-4 R_\oplus$) may actually increase from the cooler (mid) to hotter (early) M-dwarfs, although the number of planet candidates driving this result is relatively small (and they find no such trend for smaller planets with $R_p = 0.5-1.4 R_\oplus$). More recently, \citet{Hu2019} also find an increased occurrence rate of planets around M-dwarfs, although with much larger uncertainties, as well as evidence for an increasing occurrence toward later M-dwarfs. The higher occurrence of small planets around M-dwarfs has also been suggested by \citet{G2016}, who estimated an average of $2.2 \pm 0.3$ planets ($R_p = 1-4 R_\oplus$) per star between 1.5--180 days. Finally, \citet{HFT2020} and \citet{Bryson2020} leveraged \Kepler{} DR25, \Gaia{} DR2, and 2MASS data to compute the planet occurrence rates around M-dwarfs. While their findings corroborate these previous results, they also show that the increased occurrence rates compared to that of FGK stars largely disappear when normalizing by stellar irradiance.

At the time of writing this paper, \citet{YXZ2020} also used the \Kepler{} DR25 catalog of exoplanet candidates to study the occurrence of planetary systems, namely the fraction of stars with planets, as a function of stellar type. They split a sample of stars between 3000--7500 K into ten quantiles and modeled the fraction of stars with planets, the mean planet multiplicity, and the mutual inclination dispersion power-law index $\alpha$ (assuming the same mutual inclination $\sigma_i$--planet multiplicity $k$ relation from \citealt{Z2018}, $\sigma_i \propto k^\alpha$) in each quantile. They also find that the fraction of stars with planets, and to a lesser significance the mean number of planets per system, increases with decreasing stellar effective temperature.
In this paper, we take an approach similar to \citet{YXZ2020} to focus on the fraction of stars with planets as opposed to just the mean number of planets per star, as both of these quantities can be computed given knowledge of the intrinsic planet multiplicity distribution, which we constrain using our forward model. We extend the methodology described in \citet{HFR2019} (hereafter \citetalias{HFR2019}) to explore a clustered model describing the relation between planetary architectures and host star properties, using Gaia $b_p-r_p-E^*$ colors as a proxy for stellar effective temperature (and equivalently, spectral type), where $E^*$ is a reddening correction. We summarize our forward modeling procedure in \S\ref{Methods}, focusing on the key features and updates while leaving the full details in \citetalias{HFR2019} (\S2 therein).
We model the fraction of stars with planets ($f_{\rm swpa}$) as a linear function of $b_p-r_p-E^*$ for our FGK sample and show that the occurrence of planetary systems increases significantly toward later type stars. We also consider an alternative model in which the period power--law index ($\alpha_P$) is a linear function of $b_p-r_p-E^*$. In \S\ref{Results}, we present our results for our new clustered models. 
We discuss the implications of our results in \S\ref{Discussion}. 
Finally, we summarize our conclusions in \S\ref{Conclusions}.

\section{Methods} \label{Methods}

As in \citetalias{HFR2019} (and described therein), our models are built in the context of the Exoplanets Systems Simulator (``SysSim'') codebase, which can be installed as the ExoplanetsSysSim.jl package \citep{F2018b}. Step-by-step instructions on how to install, as well as our forward models, can be accessed at \url{https://github.com/ExoJulia/SysSimExClusters}. The SysSim project is also described in \citet{H2018, H2019}.

Our previous models for planetary systems (non-clustered, clustered periods, and clustered periods and sizes models) and our multi-stage approach to performing an approximate Bayesian computing (ABC) analysis are fully described in \citetalias{HFR2019}. Here, we modify our best model, the clustered periods and sizes model, to explore the dependence on host star properties. In this section, we first summarize our full procedure and then describe the updates:

\begin{itemize}[leftmargin=*, label={}]
 \item \textbf{Step 0: Define a statistical description for the intrinsic distribution of exoplanetary systems.}
 \item \textbf{Step 1: Generate an underlying population of exoplanetary systems (\textit{physical catalog}).}
 \item \textbf{Step 2: Generate an observed population (\textit{observed catalog}) from the \textit{physical catalog}.}
 \item \textbf{Step 3: Compare the simulated \textit{observed catalog} with the \Kepler{} data.}
 \item \textbf{Step 4: Optimize a distance function to find the best-fit model parameters.}
 \item \textbf{Step 5: Explore the posterior distribution of model parameters using a Gaussian process (GP) emulator.}
 \item \textbf{Step 6: Compute credible intervals for model parameters and simulated catalogs using ABC.}
\end{itemize}

Our updates to each step are described in the following subsections.

\subsection{Clustered Model Updates}

In \citetalias{HFR2019}, we explored three models simulating planetary systems as a clustered Poisson point process and used forward modeling to fit to the key properties (e.g., marginal distributions of the principal observables for detected planets) of the \Kepler{} planet catalog around a clean sample of FGK stars \citep{H2019}. We found that the occurrence of multi-transiting systems, and their distributions of period ratios and radius ratios are highly clustered to the extent that a simple non-clustered model cannot reproduce. We showed that, instead, a model involving clustered periods and planet sizes provides the best fit to the observed data.

In this paper, we adopt this fully clustered model from \citetalias{HFR2019} and modify it slightly with a re-parametrization, before adding a dependence on the host star color. To review, our clustered periods and sizes model as described in \citetalias{HFR2019} consists of the following features:

\begin{itemize}[leftmargin=*, label={}]
 \item \textbf{Planet clusters:} each planetary system is composed of ``clusters'' of planets.  We attempt to assign a number of clusters drawn from a Poisson distribution (with mean parameter $\lambda_c$), but some may be rejected due to stability concerns (see \S2.2 of \citetalias{HFR2019} for the exact procedure).  The number of planets for each cluster is drawn from a zero-truncated Poisson (ZTP) distribution (with mean parameter $\lambda_p$).
 \item \textbf{Orbital periods:} a power-law (with slope index $\alpha_P$) describes the distribution of cluster period scales $P_c$, and the period of each planet in the cluster is drawn from a log-normal distribution centred on $P_c$ with cluster width $N_p \sigma_P$ (where $N_p$ is the number of planets in the cluster and $\sigma_P$ is a width scale parameter), between 3 and 300 days.
 \item \textbf{Planet radii:} a broken power-law (with slope indices $\alpha_{R1}$, $\alpha_{R2}$, and break radius $R_{p,\rm break} = 3 R_\oplus$) describes the distribution of cluster radius scales $R_{p,c}$, and the radius of each planet in the cluster is drawn from a log-normal distribution centered on $R_{p,c}$ with cluster width $\sigma_R$, between 0.5 and $10 R_\oplus$. We note that this parametrization is not flexible enough to produce any radius valley.
 \item \textbf{Planet masses:} a non-parametric, probabilistic mass--radius relation from \citet{NWG2018} is used to draw the masses of the planets conditioned on their radii.
 \item \textbf{Eccentricities:} the orbital eccentricities are drawn from a Rayleigh distribution (with scale $\sigma_e$).
 \item \textbf{Mutual inclinations:} two Rayleigh distributions for the mutual inclinations are used, corresponding to a high and a low mutual inclination population (with scales $\sigma_{i,\rm high}$ and $\sigma_{i,\rm low}$, respectively, such that $\sigma_{i,\rm high} \geq \sigma_{i,\rm low}$), where the fraction of systems belonging to the high inclination population is $f_{\sigma_{i,\rm high}}$.
 \item \textbf{Planets near resonance:} peaks near the first-order mean motion resonances (MMRs) in the observed period ratio distribution are produced by drawing low mutual inclinations for the planets ``near an MMR'' with another planet (which we define as cases where the period ratio is in the range $[\mathcal{P}_{\rm mmr}, 1.05 \mathcal{P}_{\rm mmr}]$ for any $\mathcal{P}_{\rm mmr}$ in \{2:1, 3:2, 4:3, 5:4\}), such that these planets have mutual inclinations drawn from the Rayleigh distribution with $\sigma_{i,\rm low}$ regardless of which mutual inclination population the system belongs to.
 \item \textbf{Stability criteria:} adjacent planets are separated by at least $\Delta_c = 8$ mutual Hill radii, and orbital periods are resampled until this criteria is met. For clusters where a maximum number of resampling attempts has been met, the entire cluster is discarded. We note that for nominal (i.e. best-fitting) model parameters, $\sim 20\%$ of all attempted clusters are rejected, although this rate is greater for larger values of $\lambda_c$ and $\lambda_p$. Thus in both \citetalias{HFR2019} and this paper, we also report the final (i.e. true) mean numbers of clusters and planets per cluster, which are more accurate than $\lambda_c$ and $\lambda_p$ respectively.
\end{itemize}

\subsubsection{Modified clustered model: constant $f_{\rm swpa} + \alpha_P$} \label{Baseline_model}

The model described above induces a link between the way planets are distributed between clusters and the number of zero planet systems, since the number of clusters per system is drawn from a Poisson distribution which also controls the number of zero-cluster (and thus zero-planet) draws.
In other words, the fraction of stars with planets in the above model (which we referred to as ``FSWP'' in \citetalias{HFR2019}) is the fraction of draws with at least one cluster that is successfully attempted (and thus is implicitly a function of $\lambda_c$).
To decouple these two, we modify the model by introducing an additional parameter, the fraction of stars with planets attempted $f_{\rm swpa}$, and replace the Poisson distribution for the number of clusters per system with also a ZTP. This way, $f_{\rm swpa}$ controls the fraction of stars we draw planetary systems for, while $\lambda_c$ and $\lambda_p$ both parametrize the (ZTP-distributed) numbers of clusters and planets per cluster, respectively, for such systems. We call $f_{\rm swpa}$ the fraction of stars with planets \textit{attempted} because systems for which we draw planets can still end up with zero planets in the case where all planets are discarded after the maximum number of attempts due to the stability criteria, although this is very rare ($< 0.5$\% of attempted systems).

We make this modification to our clustered periods and sizes model from \citetalias{HFR2019} not only to decouple the number of zero--planet systems from the underlying multiplicity distribution, but also to serve as a baseline model for a more natural comparison to the new models we introduce below. For the remainder of this paper, we refer to this baseline model as the ``constant $f_{\rm swpa}+\alpha_P$'' model. While all of the model parameters are ``constant'' (i.e., not functions of stellar type) in this model, we use this name to differentiate from the models below in which $f_{\rm swpa}$ or $\alpha_P$ is dependent on the host stars.

\begin{figure}
 \includegraphics[scale=0.425,trim={0.4cm 0 0 0.2cm},clip]{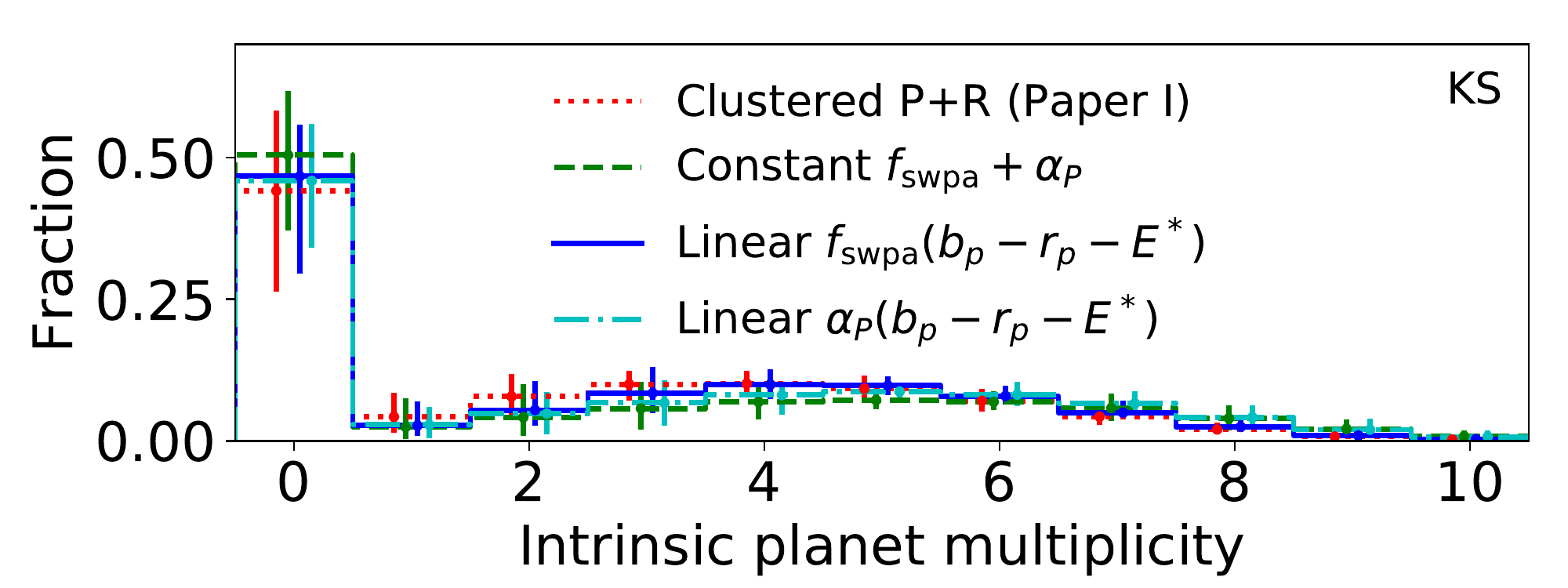}
 \includegraphics[scale=0.425,trim={0.4cm 0 0 0.2cm},clip]{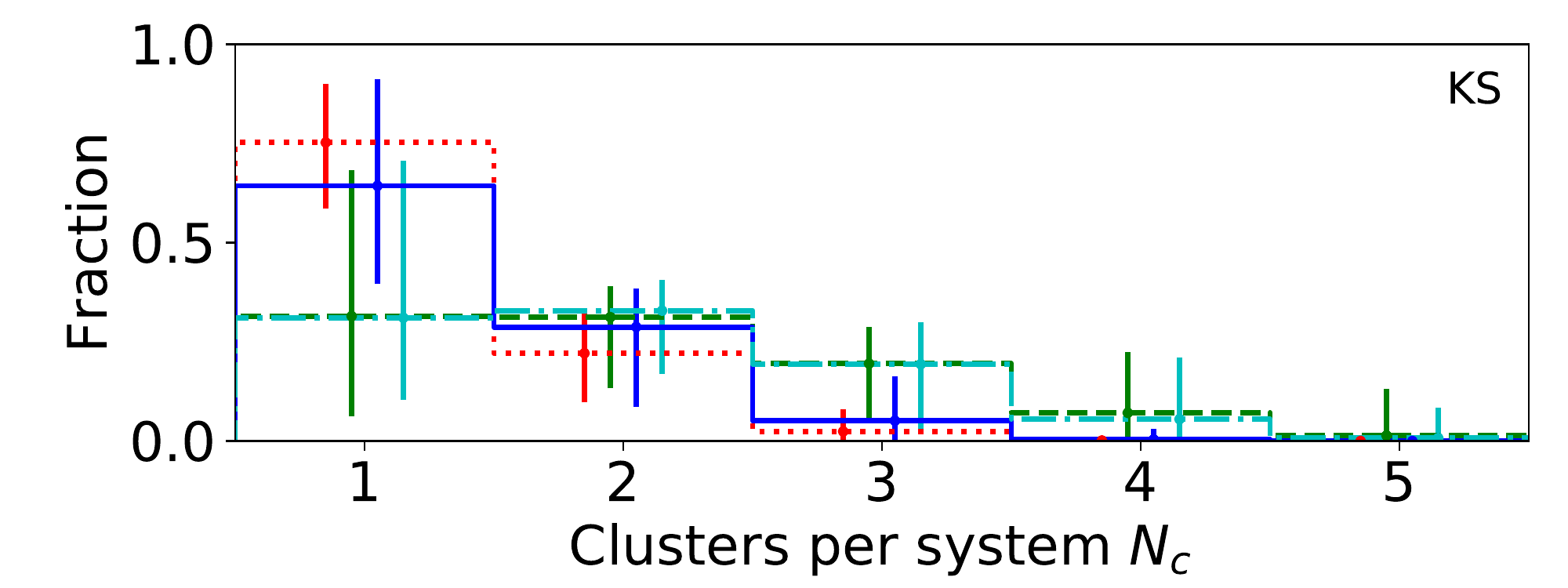}
 \includegraphics[scale=0.425,trim={0.4cm 0 0 0.2cm},clip]{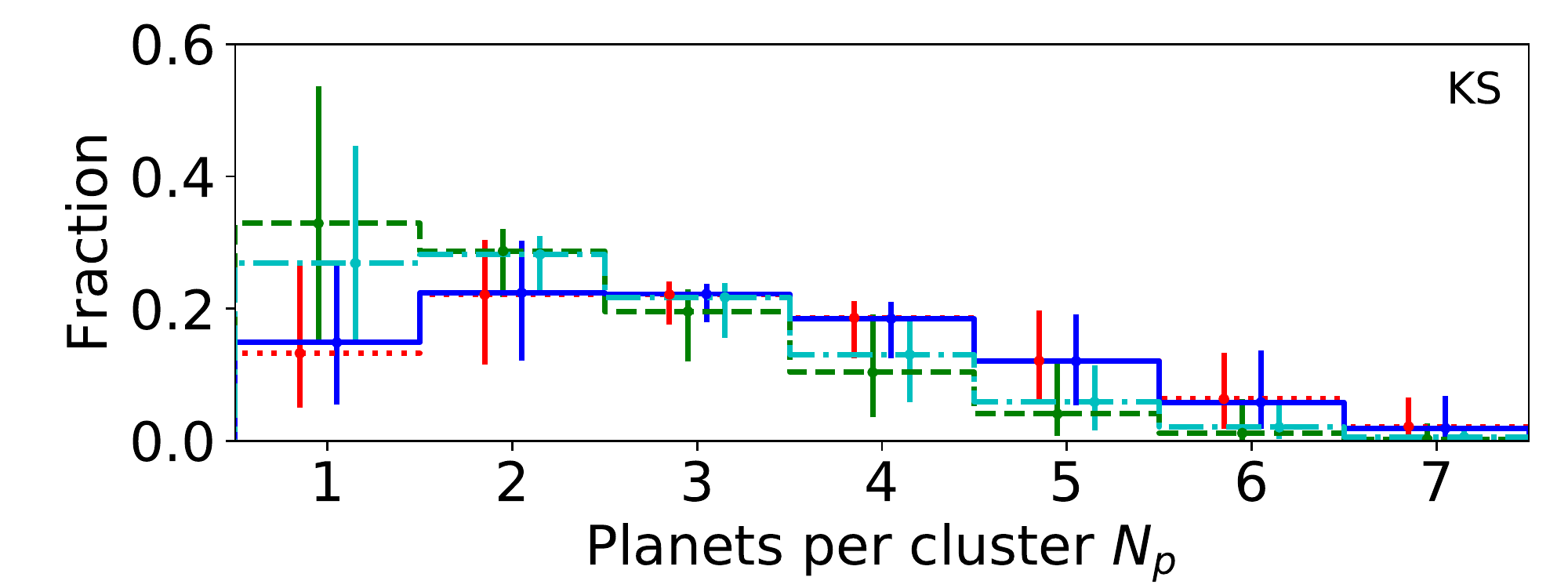}
\caption{Intrinsic distributions of total planet multiplicity (\textbf{top panel}), cluster multiplicity $N_c$ (\textbf{middle panel}), and planets per cluster $N_p$ (\textbf{bottom panel}) drawn from our models.
In \citetalias{HFR2019}, our clustered models (red dotted lines) are parametrized by Poisson($\lambda_c$) and zero-truncated Poisson (ZTP($\lambda_p$)) distributions for $N_c$ and $N_p$, respectively. Thus, zero-planet systems result from draws of $N_c = 0$. In this paper, we parametrize our constant $f_{\rm swpa}+\alpha_P$ model (green dashed lines), linear $f_{\rm swpa}(b_p-r_p-E^*)$ model (blue solid lines), and linear $\alpha_P(b_p-r_p-E^*)$ model (cyan dash-dotted lines) with ZTP distributions for both $N_c$ and $N_p$, with a separate parameter for the overall fraction of stars with planets $f_{\rm swpa}$ (so zero-planet systems make up $1 - f_{\rm swpa}$ of all systems). Error bars denote the 68\% credible regions computed from 100 catalogs passing our distance thresholds for each model. While our new parametrization results in somewhat more clusters per system and fewer planets per cluster, the total planet multiplicity distribution is very similar between the models. In particular, the (overall) fraction of stars with planets is well constrained in all four models.}
\label{fig:intrinsic_mults}
\end{figure}

In Figure \ref{fig:intrinsic_mults}, we plot the intrinsic distributions of total planet multiplicity, clusters per system $N_c$, and planets per cluster $N_p$, for our old clustered model from \citetalias{HFR2019} (red dotted line) and our constant $f_{\rm swpa}+\alpha_P$ model (green dashed line). While our parametrization is different between these models, we find that the resulting intrinsic planet multiplicity distribution is very similar (top panel). We discuss these results in more detail in \S\ref{Baseline_results}.

\subsubsection{Model dependence on host star color} \label{Gaia_colors}

The occurrence of planetary systems is likely not independent of the host star properties \citep{H2012, DC2013, Mu2015, Hu2019, YXZ2020}. As we motivate in \S\ref{Results}, we find that while our clustered periods and sizes model performs well for our sample of \Kepler{} planetary systems around FGK stars as a whole \citepalias{HFR2019}, there are differences in the way these planets are distributed for different stars. These differences are complicated by the complex detection biases present in the \Kepler{} survey. Since we are forward modeling planetary systems in detail using SysSim, our approach allows us to test various models in order to distinguish between these observational effects and real trends in the data.

In order to account for the potential differences in planet occurrence as a function of stellar properties, we introduce a host-star dependence in our clustered model. We adopt the cross--matched \Gaia{} DR2 $b_p-r_p$ colors \citep{Gaia2018} as a proxy for spectral type, and we explore several model parameters as simple functions of color. We use the \Gaia{} $b_p-r_p$ color instead of the more obvious choice of stellar effective temperature, $T_{\rm eff}$, because of known limitations and caveats in the estimates of $T_{\rm eff}$ from Gaia DR2 photometry \citep{A2018}.  First, the characteristic scatter in $T_{\rm eff}$ relative to literature values is over 300K.  In contrast, $b_p-r_p$ is measured precisely, as demonstrated by the narrowness of the main sequence in the $b_p-r_p$ versus $L_\star$ color-luminosity diagram.  Additionally, the small number of stellar models used for training by the \Gaia{} team results in substantial ``banding'' of the inferred $T_{\rm eff}$ values.  Indeed, we find three significant modes in the distribution of \Gaia{} DR2 $T_{\rm eff}$ values for our sample.  This would be particularly problematic for making precise comparisons of stellar temperatures.
Therefore, we conclude that $b_p-r_p$ provides a superior proxy for making comparisons of the temperatures of stars within our particular stellar sample.

Stars in the \Kepler{} field are moderately affected by reddening due to interstellar dust \citep{A2018}; thus, we apply a correction for reddening.  We find that there is a significant scatter in the reddening values of $E(b_p-r_p)$ from \Gaia{} DR2 for our \Kepler{} target stars, and that the distribution of $b_p-r_p-E(b_p-r_p)$ is trimodal, analogous to the $T_{\rm eff}$ values from \Gaia{} DR2, and similarly due to the limited number of stellar models used. While the median $E(b_p-r_p)$ is 0.15 mag, some targets have values as large as $\sim 0.8$ mag. Further, not all stars have valid $E(b_p-r_p)$ values. 
Indeed, \citet{A2018} caution that $E(b_p-r_p)$ should not be used for individual stars, but advise that it be used for for statistical studies, such as this study.  
Therefore, instead of directly using the \Gaia{} DR2 $E(b_p-r_p)$ values for each star, we construct a simple model for $E(b_p-r_p)$ as a function of $b_p-r_p$ by interpolation in order to account for differential reddening.
While such an approach would not be wise for all science cases, it works well for the purposes of this study.
First, for our analysis, we are only interested in the effects of differential reddening across a subset of the FGK main sequence stars observed by \Kepler{}.  The uncertainties due to differential reddening are significantly smaller for our sample than for \Gaia{} targets in general.
As explained below, we have applied multiple cuts to avoid stars that have evolved off the main sequence (e.g., subgiants, giants) as well as pre-main sequence or binary stars, so as to obtain a clean sample of FGK main sequence stars.
The careful filtering of stars results in a smooth relation between \Gaia{} color and reddening, after normalizing by distance.

\begin{figure}
\centering
\includegraphics[scale=0.42,trim={0.2cm 0.5cm 0.2cm 0.2cm},clip]{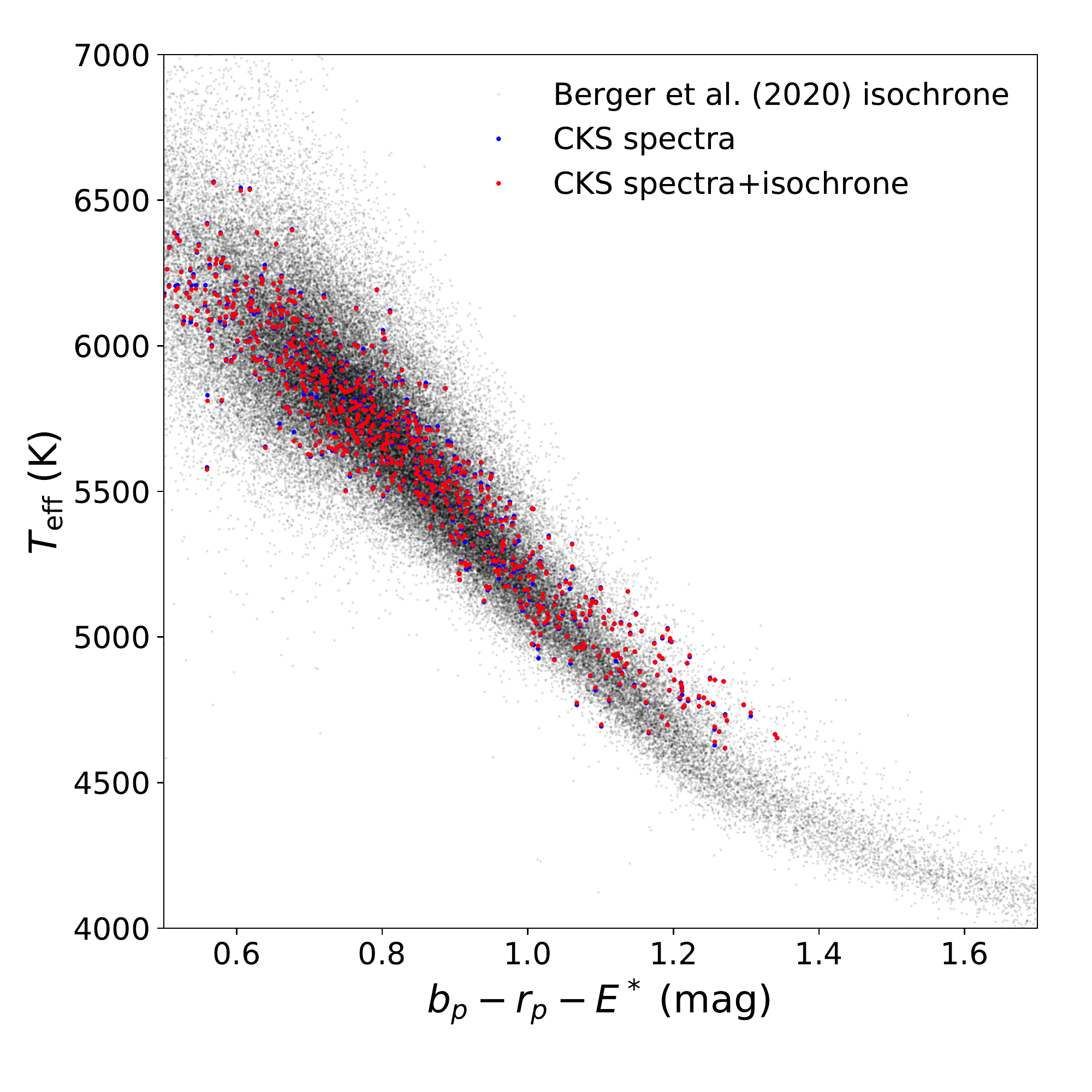}
\caption{Scatter plot of stellar effective temperature ($T_{\rm eff}$) vs. our reddening--corrected color ($b_p-r_p-E^*$). The gray points denote stellar temperatures from the \citet{B2020} catalog (which are available for almost all of the stars in our catalog), while the blue and red points denote stellar temperatures derived by the California--Kepler Survey \citep{J2017}, using just spectra and spectra with isochrones, respectively (available only for 779 stars in our catalog).}
\label{fig:color_teff}
\end{figure}

Our procedure for applying a differential reddening correction is as follows. First, we perform all of the cuts described in \citet{H2019} (see \S3.1 therein) on the \Kepler{} DR25 stellar catalog, which includes requiring targets to have:
\begin{itemize}[leftmargin=*]
 \item a \Gaia{} fractional parallax error less than 10\%,
 \item $0.5 \leq b_p-r_p \leq 1.7$ (for pre-selecting FGK stars), and
 \item a luminosity $L \leq 1.75 L_{\rm MS}(b_p-r_p)$, where $L_{\rm MS}(b_p-r_p)$ is determined by iteratively fitting to the main sequence for the remaining stars, six times.
\end{itemize}
This results in 70,477 targets with \Gaia{}--derived $E(b_p-r_p)$ values.
Then, we bin the stars into 20 quantiles by $b_p-r_p$ (about $\sim 3500$ stars, or 5\%, in each bin). While there is large scale structure to the distribution of interstellar dust as a function of galactic latitude, the primary \Kepler{} field is highly localized, and we do not find that including galactic latitude makes a significant difference.
Thus, main sequence stars of a similar spectral type or color at a similar distance should experience similar reddening. We compute the median distance--normalized reddening, $E(b_p-r_p)/d$ where $d = 1/\pi$ is the distance and $\pi$ is the parallax, for each bin. Although more reliable distances can be obtained using a probabilistic model (e.g., \citealt{BJ2018}), 
our inversion of the parallax yields good distance estimates for our sample of target stars with small parallax uncertainties and thus strongly peaked likelihoods.
The estimated distance--normalized reddening for each target is then computed by interpolating $E(b_p-r_p)/d$ as a function of $b_p-r_p$.
Finally, we multiply by the distances again to get the interpolated reddening for each target, which we denote as $E^*(b_p-r_p)$. For the remainder of this paper, we will simply refer to the interpolated reddening values as $E^*$, and likewise the colors corrected using the interpolated reddening values as $b_p-r_p-E^*$ (i.e., a measure of the intrinsic stellar colors).
We note that the main analyses in this paper were also performed on the measured colors without a reddening correction (i.e. $b_p-r_p$) and the results are very similar, indicating that the results are insensitive to our choices for modeling colors and reddening.

With our model for estimating $E^*$ as a smooth function of $b_p-r_p$, we then apply the reddening correction to all target stars before the color and luminosity cuts described earlier. We re--cut and re--fit the FGK main sequence using the corrected colors, with $0.5 \leq b_p-r_p-E^* \leq 1.7$. This results in a final stellar catalog of 88,912 usable targets (this is greater than the 70,477 targets before applying our interpolated reddening values because we also apply our reddening model to targets without \Gaia{} $E(b_p-r_p)$ values). The median (corrected) color is $b_p-r_p-E^* \simeq 0.81$ mag, which is close to the solar value. In Figure \ref{fig:color_teff}, we plot our reddening--corrected color ($b_p-r_p-E^*$) versus stellar effective temperature ($T_{\rm eff}$) from both the \citet{B2020} catalog (photometrically derived, isochrone fitted using \Gaia{} DR2 data) and the California--Kepler Survey (spectroscopically derived; \citealt{J2017}). There is a strong correlation between our corrected colors and the stellar temperatures.

\subsubsection{Linear $f_{\rm swpa}(b_p - r_p - E^*)$ model} \label{Linear_fswp_model}

A simple way to allow for planet occurrence to vary with host star color in our clustered model is to allow the fraction of stars with planets, $f_{\rm swpa}$, to be a function of $b_p-r_p-E^*$. This is simpler than allowing for the multiplicity to vary with stellar color, since in our clustered models we have two parameters controlling the number of planets per planetary system, $\lambda_c$ and $\lambda_p$. In this model, we assume the form of a linear relation between $f_{\rm swpa}$ and $b_p-r_p-E^*$:
\begin{align}
\begin{split}
 & f_{\rm swpa}(b_p-r_p-E^*) = \\
 &\quad \max \Big\{ 0,\min \Big[ m \Big( ({b_p-r_p-E^*}) - ({b_p-r_p-E^*})_{\rm med} \Big) \\
 &\quad\quad\quad + f_{\rm swpa,med}, 1 \Big] \Big\} \label{eq_fswp_bprp}
\end{split}
\end{align}
where $m = d{f_{\rm swpa}}/d(b_p-r_p-E^*)$ is the slope of the line, and $f_{\rm swpa,med} = f_{\rm swpa}((b_p-r_p-E^*)_{\rm med})$ is the $y$-intercept (which we have chosen to parametrize at the median color, $(b_p-r_p-E^*)_{\rm med} \simeq 0.81$ mag). We enforce $f_{\rm swpa}(b_p-r_p-E^*)$ to be bounded between 0 and 1, since the fraction of systems with planets cannot be negative or greater than 1. We refer to this model as the ``linear $f_{\rm swpa}(b_p-r_p-E^*)$'' model. Our previous (constant) model is essentially a special case of this more general model, where the slope is set to $m = d{f_{\rm swpa}}/d(b_p-r_p-E^*) = 0$.

The blue solid lines in Figure \ref{fig:intrinsic_mults} show the intrinsic planet and cluster multiplicity distributions for this linear $f_{\rm swpa}(b_p-r_p-E^*)$ model. While the fraction of stars with planets (and thus the fraction of zero-planet systems) is a strong function of host star color as we will show in \S\ref{Results}, the intrinsic planet multiplicity distribution marginalized over all of the FGK stars in our sample is very similar to that of our baseline model (constant $f_{\rm swpa}+\alpha_P$; green dashed line). There is a slight tradeoff between the distributions of the number of clusters and planets per cluster compared to the constant $f_{\rm swpa}+\alpha_P$ model. We emphasize that the numbers of clusters and planets per cluster are \textit{not} host star dependent since we have not made $\lambda_c$ or $\lambda_p$ functions of $b_p-r_p-E^*$.

To serve as a check on this model, we also explore a ``step $f_{\rm swpa}$'' model in which two parameters, $f_{\rm swpa,bluer}$ and $f_{\rm swpa,redder}$, describe the fraction of stars with planets below and above the median $b_p-r_p-E^*$ respectively. We briefly discuss the results for this model in \S\ref{secFSWP}.

\subsubsection{Linear $\alpha_P(b_p - r_p - E^*)$ model} \label{Linear_alphaP_model}

Another way in which planetary systems can differ as a function of host stellar type is in their distribution of orbital periods. For example, \citet{Mu2015} found that the occurrence rate of small planets increases with decreasing stellar effective temperature at all semi-major axes less than 1 AU. Another motivation for exploring the ``linear $\alpha_P(b_p-r_p-E^*)$'' model is because it can have a similar effect as the fraction of stars with planets on the overall rate of detected planets, as a function of stellar color. We explore the period distribution in a similar way as the fraction of stars with planets, by allowing the period power-law index $\alpha_P$ to vary as a linear function of $b_p-r_p-E^*$:
\begin{align}
\begin{split}
 & \alpha_P(b_p-r_p-E^*) = \\
 &\quad m \Big( ({b_p-r_p-E^*}) - ({b_p-r_p-E^*})_{\rm med} \Big) + \alpha_{P,\rm med} \label{eq_alphaP_bprp}
\end{split}
\end{align}
where $m = d\alpha_P/d(b_p-r_p-E^*)$ is the slope and $\alpha_{P,\rm med} = \alpha_P((b_p-r_p-E^*)_{\rm med})$ is the value at the median color.

\subsection{Observational Comparisons} \label{Obs}

We adopt the same procedure for constraining our model parameters as described in \citetalias{HFR2019}, by defining a similar set of summary statistics and a distance function that accounts for these summary statistics.

\subsubsection{Summary statistics} \label{SummaryStats}

Since we are interested in how planet occurrence varies as a function of stellar color, we divide the stellar sample into two halves, split at the median $b_p-r_p-E^*$ color (a ``bluer'' half, with smaller $b_p-r_p-E^*$ values, and a ``redder'' half, with larger $b_p-r_p-E^*$ values). We compute the same summary statistics for each half, in addition to the full sample (hereafter labeled as ``All''), for each observed catalog:
\begin{enumerate}[leftmargin=*]
 \item the total number of observed planets $N_{p,\rm tot}$ relative to the number of target stars $N_{\rm stars}$, $f = N_{p,\rm tot}/N_{\rm stars}$,
 \item the observed multiplicity distribution, $\{N_m\}$, where $N_m$ is the number of systems with $m$ observed planets and $m = 1,2,3,...$,
 \item the observed orbital period distribution, $\{P\}$,
 \item the observed period ratio distribution, $\{\mathcal{P}\}$,
 \item the observed transit depth distribution, $\{\delta\}$,
 \item the observed transit depth ratio distribution, $\{\delta_{i+1}/\delta_i\}$,
 \item the observed transit duration distribution, $\{t_{\rm dur}\}$,
 \item the observed period-normalized transit duration ratio distribution of adjacent planets apparently near an MMR, $\{\xi_{\rm res}\}$, and not near an MMR, $\{\xi_{\rm non-res}\}$. The normalized transit duration ratio is given by $\xi = (t_{\rm dur,in}/t_{\rm dur,out})(P_{\rm out}/P_{\rm in})^{1/3}$ \citep{S2010, F2014}.
\end{enumerate}

The list above contains nine summary statistics, which we compute for each full observed catalog as well as for each of the bluer and redder halves, totaling 27 summary statistics.

\subsubsection{Distance function}

\begin{deluxetable}{lcccccc}
\centering
\tablecaption{Weights for the individual distance terms as computed from a reference clustered periods and sizes model (from \citetalias{HFR2019}, with a chosen set of parameters as follows: $f_{\sigma_{i,\rm high}} = 0.4$, $\lambda_c = 0.8$, $\lambda_p = 3.7$, $\alpha_P = 0.4$, $\alpha_{R1} = -1$, $\alpha_{R2} = -4.4$, $\sigma_e = 0.02$, $\sigma_{i,\rm high} = 50^\circ$, $\sigma_{i,\rm low} = 1.4^\circ$, $\sigma_R = 0.3$, and $\sigma_P = 0.2$).}
\tablewidth{0pt}
\tablehead{
 \colhead{Distance term} & \multicolumn2c{All} & \multicolumn2c{Bluer} & \multicolumn2c{Redder} \\
 & \colhead{$\hat{\sigma}(\mathcal{D})$} & \colhead{$w$} & \colhead{$\hat{\sigma}(\mathcal{D})$} & \colhead{$w$} & \colhead{$\hat{\sigma}(\mathcal{D})$} & \colhead{$w$}
}
\decimalcolnumbers
\startdata
 $D_f$ & 0.00103 & 971 & 0.00146 & 683 & 0.00154 & 649 \\
 $D_{\rm mult}$ & 0.00593 & 169 & 0.01150 & 87 & 0.01373 & 73 \\
 \hline
 $\mathcal{D}_{\rm KS}$: & & & & & & \\
 $\{P\}$ & 0.02616 & 38 & 0.03544 & 28 & 0.03805 & 26 \\
 $\{\mathcal{P}\}$ & 0.04836 & 21 & 0.06441 & 16 & 0.07167 & 14 \\
 $\{\delta\}$ & 0.02907 & 34 & 0.03988 & 25 & 0.04121 & 24 \\
 $\{\delta_{i+1}/\delta_i\}$ & 0.05106 & 20 & 0.06821 & 15 & 0.07437 & 13 \\
 $\{t_{\rm dur}\}$ & 0.02831 & 35 & 0.03928 & 25 & 0.03995 & 25  \\
 $\{\xi_{\rm res}\}$ & 0.11572 & 9 & 0.16131 & 7 & 0.17897 & 6 \\
 $\{\xi_{\rm non-res}\}$ & 0.05607 & 18 & 0.07361 & 14 & 0.08078 & 12 \\
 \hline
 $\mathcal{D}_{\rm AD'}$: & & & & & & \\
 $\{P\}$ & 0.00113 & 882 & 0.00218 & 459 & 0.00233 & 429 \\
 $\{\mathcal{P}\}$ & 0.00329 & 304 & 0.00602 & 166 & 0.00736 & 136 \\
 $\{\delta\}$ & 0.00138 & 723 & 0.00263 & 380 & 0.00276 & 362 \\
 $\{\delta_{i+1}/\delta_i\}$ & 0.00392 & 255 & 0.00698 & 143 & 0.00862 & 116 \\
 $\{t_{\rm dur}\}$ & 0.00145 & 691 & 0.00291 & 344 & 0.00302 & 331 \\
 $\{\xi_{\rm res}\}$ & 0.02098 & 48 & 0.04515 & 22 & 0.05154 & 19 \\
 $\{\xi_{\rm non-res}\}$ & 0.00479 & 209 & 0.00808 & 124 & 0.00982 & 102 \\
\enddata
\tablecomments{Each weight $w$ is computed as the inverse of the root mean square of the distances $\hat{\sigma}(\mathcal{D})$ between repeated realizations of the same (i.e. ``perfect'') model, $w = 1/\hat{\sigma}(\mathcal{D})$, using the same number of target stars as our \Kepler{} sample. The weights are shown here as rounded whole numbers for guidance purposes only.}
\label{tab:weights}
\end{deluxetable}

In \citetalias{HFR2019}, we used a linear weighted sum of individual distance terms to combine the fits to each summary statistic into a single distance function. Two separate distance functions were used for the analysis, with one adopting the two-sample Kolmogorov--Smirnov (KS; \citealt{K1933, S1948}) distance for each marginal distribution and the other adopting a modified version of the two-sample Anderson--Darling (AD; \citealt{AD1952, P1976}; see equations 23--24 in \citetalias{HFR2019} for our modification) statistic. Both distance functions included a term for the overall rate of planets ($D_f = | f_{\rm sim} - f_{\rm Kepler}|$, where $f_{\rm sim} = N_{p,\rm tot}/N_{\rm stars}$ and likewise for \Kepler{}) and the observed multiplicity distribution ($D_{\rm mult} = \rho_{\rm CRPD} = (9/5)\sum_j O_j \big[{(O_j/E_j)}^{2/3} - 1\big]$, where $O_j$ are the number of ``observed'' systems in our models and $E_j$ are the number of expected systems from the \Kepler{} data, for multiplicity bins $j = 1,2,3,4,5+$; see \citealt{CR1984} and the discussion surrounding equation 19 in \citetalias{HFR2019}).

For this paper, we extend the distance function in the same way as the summary statistics, by computing the individual distance term corresponding to each summary statistic, for each of the bluer, redder, and full samples, and summing them using a set of weights $w_{i'}$.
\begin{align}
 \mathcal{D}_W &= \sum_{\rm samples} \sum_{i'} w_{i'} \mathcal{D}_{i'} \label{eq_dist_general} \\
 &= \sum_{\rm samples} \bigg[ \frac{D_f}{\hat{\sigma}(D_f)} + \frac{D_{\rm mult}}{\hat{\sigma}(D_{\rm mult})} + \sum_{i=1}^{7} \frac{\mathcal{D}_i}{\hat{\sigma}(\mathcal{D}_i)} \bigg], \label{eq_dist}
\end{align}
where $w_{i'} = 1/\hat{\sigma}(\mathcal{D}_{i'})$ and everything within the outer summation refers to the distances computed using the summary statistics in a given sample only. The distances $\mathcal{D}_i$ within the inner summation are either KS or AD distances, where the summation is over the indices labeling the summary statistics (iii)--(viii).
As in \citetalias{HFR2019}, we repeat the analyses of this paper using both KS and AD distances, to serve as an additional check for the consistency of our results. The KS (two-sample) distance is simply the maximum difference between two cumulative distributions and thus primarily compares the modes of the distributions, while the AD (two-sample) distance is an integral that more heavily weights the tails of the distributions. As shown in \S\ref{Results}, the results are very similar between both analyses, and thus we focus our discussion on the results from the (easier to interpret) KS distance function for the remainder of the paper.

\emph{Old weights:}
In order to compute the weights in \citetalias{HFR2019}, a single reference catalog was generated (using a nominal set of model parameters for the clustered periods and sizes model, and the same number of targets as our \Kepler{} sample) and 1000 repeated catalogs were simulated (with five times as many targets, to reduce stochastic noise) from the same (i.e. ``perfect'') model. The summary statistics and distances were then computed for each of the 1000 catalogs compared to the reference catalog, and the weight for each distance term was taken as the reciprocal of the rms of that distance term, $w_i = 1/\hat{\sigma}(\mathcal{D}_i)$, where $\mathcal{D}_i$ is the $i^{\rm th}$ distance term, and $\hat{\sigma}$ is the rms. While the weights computed in this way are reasonable, we find that they are prone to stochastic noise, even with larger numbers of repeated catalogs, since this method essentially treats one realization of the ``perfect'' model as the true data (i.e., the one used as the reference catalog). Also, the use of five times the number of targets for the repeated catalogs inflates the weights, due to reduced Monte Carlo noise and thus smoother distributions of the summary statistics.

\emph{New weights:}
In this study we compute the weights in a revised manner to resolve both of the above points: we simulate 100 catalogs assuming a single model, each with the same number of targets as the \Kepler{} sample, and then compute the individual distances for each unique pair of catalogs before computing the rms for each distance term. This way, each of the 100 realizations of the same model are treated equally, with no single catalog serving as \textit{the} reference catalog (or equivalently, \textit{all} of the catalogs serve as the reference catalog), and there are effectively more evaluations of each distance term.\footnote{With just 100 simulated catalogs, each distance term is computed ${100 \choose 2} = 4950$ times instead of 1000 as before.} We find that the weights generated in this way are significantly more reliable, even with the reduced number of model evaluations. The rms distances and weights are listed in Table \ref{tab:weights}. We also use a set of best-fitting model parameters from \citetalias{HFR2019} for the reference catalog (the parameters are listed in the Table \ref{tab:weights} caption).

\subsubsection{The Kepler catalog}

Our stellar catalog is described in \S\ref{Gaia_colors}, where we detailed our procedure for accounting for differential reddening. To summarize, it is derived from a series of cuts on the \Kepler{} DR25 target list based on updated stellar parameters from \Gaia{} DR2 \citep{Gaia2018} as explained in \citet{H2019} (\S3.1 therein), with some modification to account for differential reddening using interpolated reddening values ($E^*$). The list of cuts also includes \Gaia{} GOF\_AL $\leq 20$ and astrometric excess noise $\leq 5$ to filter out targets with a poor astrometric fit. In combination with the main sequence luminosity fitting described in \S\ref{Gaia_colors}, this filters out likely close--in binary stars. This results in a clean sample of 88,912 FGK main sequence stars, with corrected colors ranging from $b_p-r_p-E^* \simeq 0.5$ (~F2V) at the bluest end to $b_p-r_p-E^* \simeq 1.7$ (~K7V) at the reddest end. The planet catalog is derived from the \Kepler{} DR25 KOI table (only keeping planet candidates around stars in our stellar catalog), where we also:
\begin{enumerate}[leftmargin=*]
 \item replace the transit depths and durations with the median values from the posterior samples in \citet{R2015},
 \item replace the planet radii based on the transit depths and the updated \Gaia{} DR2 stellar radii, and
 \item only keep planets in the period range $[3, 300]$ days and planet radii range $[0.5, 10] R_\oplus$.
\end{enumerate}
Our final \Kepler{} planet catalog consists of 2216 planet candidates. Of these, 982 are around stars in our bluer sample and 1234 are around stars in our redder sample.

\subsection{Model Optimization}

We adopt the same multi-stage approach in \citetalias{HFR2019} for performing approximate Bayesian inference on our model parameters. Our forward model is complex and relatively expensive, taking $\sim 10$s to generate a physical and observed catalog with the same number of targets as our \Kepler{} catalog. The model is also stochastic, due to Monte Carlo noise and the finite catalog size, resulting in a noisy distance function even for repeated evaluations using the exact same model parameters. Finally, the parameter space we are optimizing over is large, even larger than that of our clustered periods and sizes model due to the introduction of the $f_{\rm swpa}$ parameter for our constant $f_{\rm swpa}+\alpha_P$ model, and $f_{\rm swpa,med}$ and $d{f_{\rm swpa}}/d(b_p-r_p-E^*)$ for our linear $f_{\rm swpa}(b_p-r_p-E^*)$ model (and likewise for the linear $\alpha_P(b_p-r_p-E^*)$ model). Thus, we begin with an optimization stage before training and using a fast emulator for inference with ABC.

\begin{deluxetable*}{lccccccccc}
\centering
\tablecaption{Optimizer bounds, GP length scale hyperparameters $\lambda_i$, and emulator bounds for each free parameter of the models.}
\tablehead{
 \colhead{Parameter} & \multicolumn3c{Constant $f_{\rm swpa}+\alpha_P$} & \multicolumn3c{Linear $f_{\rm swpa}(b_p - r_p - E^*)$} & \multicolumn3c{Linear $\alpha_P(b_p - r_p - E^*)$} \\
 & \colhead{Optimizer} & \colhead{$\lambda_i$} & \colhead{Emulator} & \colhead{Optimizer} & \colhead{$\lambda_i$} & \colhead{Emulator} & \colhead{Optimizer} & \colhead{$\lambda_i$} & \colhead{Emulator}
}
\decimalcolnumbers
\startdata
 $f_{\sigma_{i,\rm high}}$ & $(0, 1)$ & 0.2 & $(0.1, 0.7)$ & $(0, 1)$ & 0.2 & $(0.1, 0.7)$ & $(0, 1)$ & 0.2 & $(0.1, 0.7)$ \\
 $f_{\rm swpa}$\tablenotemark{a} & $(0, 1)$ & 0.2 & $(0.3, 0.9)$ & $(0, 1)$ & 0.2 & $(0.3, 0.9)$ & $(0, 1)$ & 0.2 & $(0.3, 0.9)$ \\
 $\frac{df_{\rm swpa}}{d(b_p-r_p-E^*)}$ & - & - & - & $(-1, 1)$ & 1 & $(0, 2)$ & - & - & - \\
 $\ln{(\lambda_c)}$ & $(\ln(0.2), \ln(10))$ & - & - & $(\ln(0.2), \ln(10))$ & - & - & $(\ln(0.2), \ln(10))$ & - & - \\
 $\ln{(\lambda_p)}$ & $(\ln(0.2), \ln(10))$ & - & - & $(\ln(0.2), \ln(10))$ & - & - & $(\ln(0.2), \ln(10))$ & - & - \\
 $\ln{(\lambda_c \lambda_p)}$ & - & 1 & $(0, 3)$ & - & 1 & $(0, 3)$ & - & 1 & $(0, 3)$ \\
 $\ln{(\frac{\lambda_p}{\lambda_c})}$ & - & 2 & $(-2, 3)$ & - & 2 & $(-2, 3)$ & - & 2 & $(-2, 3)$ \\
 $\alpha_P$\tablenotemark{b} & $(-2, 2)$ & 1 & $(-0.8, 1.6)$ & $(-2, 2)$ & 1 & $(-0.8, 1.6)$ & $(-2, 2)$ & 1 & $(-0.8, 2)$ \\
 $\frac{d\alpha_P}{d(b_p-r_p-E^*)}$ & - & - & - & - & - & - & $(-2, 2)$ & 1 & $(-2, 0)$ \\
 $\alpha_{R1}$ & $(-4, 2)$ & 1 & $(-2.5, -0.5)$ & $(-4, 2)$ & 1 & $(-2.5, -0.5)$ & $(-4, 2)$ & 1 & $(-2.5, -0.5)$ \\
 $\alpha_{R2}$ & $(-6, 0)$ & 1.5 & $(-6, -3)$ & $(-6, 0)$ & 1.5 & $(-6, -3)$ & $(-6, 0)$ & 1.5 & $(-6, -3)$ \\
 $\sigma_e$ & $(0, 0.1)$ & 0.02 & $(0, 0.04)$ & $(0, 0.1)$ & 0.02 & $(0, 0.04)$ & $(0, 0.1)$ & 0.02 & $(0, 0.04)$ \\
 $\sigma_{i,\rm high}$ ($^\circ$) & $(0, 90)$ & 30 & $(0, 90)$ & $(0, 90)$ & 30 & $(0, 90)$ & $(0, 90)$ & 30 & $(0, 90)$ \\
 $\sigma_{i,\rm low}$ ($^\circ$) & $(0, \sigma_{i,\rm high})$ & 1 & $(0, 2.4)$ & $(0, \sigma_{i,\rm high})$ & 1 & $(0, 2.4)$ & $(0, \sigma_{i,\rm high})$ & 1 & $(0, 2.4)$ \\
 $\sigma_R$ & $(0, 0.5)$ & 0.2 & $(0.1, 0.5)$ & $(0, 0.5)$ & 0.2 & $(0.1, 0.5)$ & $(0, 0.5)$ & 0.2 & $(0.1, 0.5)$ \\
 $\sigma_P$ & $(0, 0.3)$ & 0.1 & $(0.1, 0.3)$ & $(0, 0.3)$ & 0.1 & $(0.1, 0.3)$ & $(0, 0.3)$ & 0.1 & $(0.1, 0.3)$ \\
\enddata
\tablecomments{We varied the parameters $\ln(\lambda_c)$ and $\ln(\lambda_p)$ separately in the optimization stage, while we trained and predicted on $\ln(\lambda_c \lambda_p)$ and $\ln(\lambda_p/\lambda_c)$ during the emulator stage (since these transformed parameters, the sum and difference of the log-rates of clusters and planets per cluster, appear more Gaussian). The same values are used for both of the analyses involving the KS and AD distance terms.}
\tablenotetext{a}{This is $f_{\rm swpa,med}$ for the clustered model with linear $f_{\rm swpa}(b_p-r_p-E^*)$.}
\tablenotetext{b}{This is $\alpha_{P,\rm med}$ for the clustered model with linear $\alpha_P(b_p-r_p-E^*)$.}
\label{tab:optim_GP}
\end{deluxetable*}

\subsubsection{Optimization stage}

The first step in our procedure for model optimization involves passing the distance function given by equation \ref{eq_dist} into a Ddifferential evolution optimizer from the ``BlackBoxOptim'' package.\footnote{\url{https://github.com/robertfeldt/BlackBoxOptim.jl}} This package provides several algorithms for general optimization problems. We choose the ``adaptive\_de\_rand\_1\_bin\_radiuslimited'' optimizer, which implements a population-based genetic algorithm to minimize the target fitness function (i.e. our distance function), and set the population size to four times the number of free model parameters. We run the optimizer for 5000 model evaluations, with $N_{\rm stars,sim} = 88,912$ targets per evaluation, saving the results (model parameters and distances) at each iteration. Finally, we repeat the optimization process 50 times for each model and distance function (KS or AD) combination, each with a different starting point in the parameter space. The search bounds for the model parameters are listed in Table \ref{tab:optim_GP}.

For each model and distance function, the optimization stage results in a pool of $5000 \times 50 = 2.5 \times 10^5$ model evaluations. We first rank-order these sets of model parameters by their evaluated distance, and keep every tenth point in the top $10^5$ points so that we have a wide range of parameters (i.e. points both close to and far from the minima found by each run) for training the GP emulator. After ranking, we re-evaluate the distances at each of these points by regenerating a new simulated catalog, in order to avoid the bias that would result in keeping smaller-than-average distances at these points due to the combination of the stochastic nature of our simulations and the mere process of ranking.

\subsubsection{GP emulator stage}

The evaluations of the full forward model during the optimization stage are then used to train an emulator, which can ``predict'' the outputs of the model (i.e. the distance function) given similar inputs (i.e. model parameters). For our emulator, we use a Gaussian process (GP) model \citep{RW2006} that is described by a prior mean function $m(\bm{x})$ and a covariance (i.e. kernel) function $k(\bm{x},\bm{x'};\bm{\phi})$:
\begin{align}
 f(\bm{x}) &\sim \mathcal{GP}\big(m(\bm{x}), k(\bm{x},\bm{x'};\bm{\phi})\big), \label{eq_GP} \\
 k(\bm{x},\bm{x'};\bm{\phi}) &= \sigma_f^2 {\rm exp} \Bigg[-\frac{1}{2} \sum_i \frac{(x_i - {x_i}')^2}{\lambda_i^2} \Bigg], \label{eq_kernel}
\end{align}
where $f(\bm{x}) = \mathcal{D}_W$ is the distance function we wish to model, $\bm{x}$ (and $\bm{x^\prime}$) are the model parameters, and $\bm{\phi} = (\sigma_f, \lambda_1, \lambda_2,..., \lambda_d)$ are the hyperparameters of the kernel. The values of the hyperparameters are also listed in Table \ref{tab:optim_GP}. In particular, $\sigma_f$ determines the strength of correlation between points and also acts as the standard deviation of the Gaussian prior (i.e., for points far away from any training data, the emulator effectively returns draws from a Gaussian distribution with mean $m(\bm{x})$ and standard deviation $\sigma_f$), while $\lambda_i$ are the length scales in each dimension over which points are correlated.

We choose a constant prior mean function, with a value that is set toward the higher end of the distances of the training points. In this way, emulated distances at points far away from any training points will be significantly worse than the best distances achievable by our model, while emulated distances will only be lower than the mean function if they are near training points and these points suggest that the model is good in the vicinity. In practice, we also find that the AD distance is significantly more sensitive to deviations from a perfect model than the KS distance; we thus set $m(\bm{x}) = 75$ for the distance function involving KS distances, and $m(\bm{x}) = 150$ for the distance function involving AD distances. We verify that our results do not change much with differing choices for the mean function, as long as it is well above the minimum distances found by the optimizer and the distance threshold for constructing the ABC posterior.

\subsubsection{ABC for model inference}

In order to compute the credible regions for the model parameters, we construct an ABC posterior distribution by using the emulator to predict the model at a large number of points and accept those that pass a distance threshold. These points are drawn from the prior, for which we assume a uniform distribution in the $d$-dimensional box (with bounds based on inspection of the training points, as listed in Table \ref{tab:optim_GP}). In this paper, our distance function given by equation \ref{eq_dist} (KS or AD) includes 27 individual distance terms, weighted and summed such that even a perfect model results in a distance of $\sim 27 \pm 2.7$. While the lowest distances found during the optimization stage set the best distance threshold possible for a given model, the emulator performs a weighted average of points, and it becomes exceedingly computationally expensive to accept points passing thresholds approaching such distances. Thus, we choose somewhat larger distance thresholds that result in an efficiency of roughly $10^{-5}$ or better for the fraction of drawn points accepted (drawn uniformly in our box). This results in distance thresholds of $\mathcal{D}_{W,\rm KS} = 47$ and $\mathcal{D}_{W,\rm AD^\prime} = 90$ for our linear models (and slightly larger thresholds, $\mathcal{D}_{W,\rm KS} = 50$ and $\mathcal{D}_{W,\rm AD^\prime} = 100$, for our constant $f_{\rm swpa}+\alpha_P$ model, since this is a worse model, as we will show in \S\ref{Results}). We collect $5 \times 10^4$ points with emulated distances passing the distance threshold to serve as the ABC posterior for the model parameters. For more detailed calculations, we also simulate the full forward model and require that both the emulated and true distances pass the distance threshold.

\section{Results} \label{Results}

\begin{figure*}
\includegraphics[scale=0.25,trim={0 0 0 0},clip]{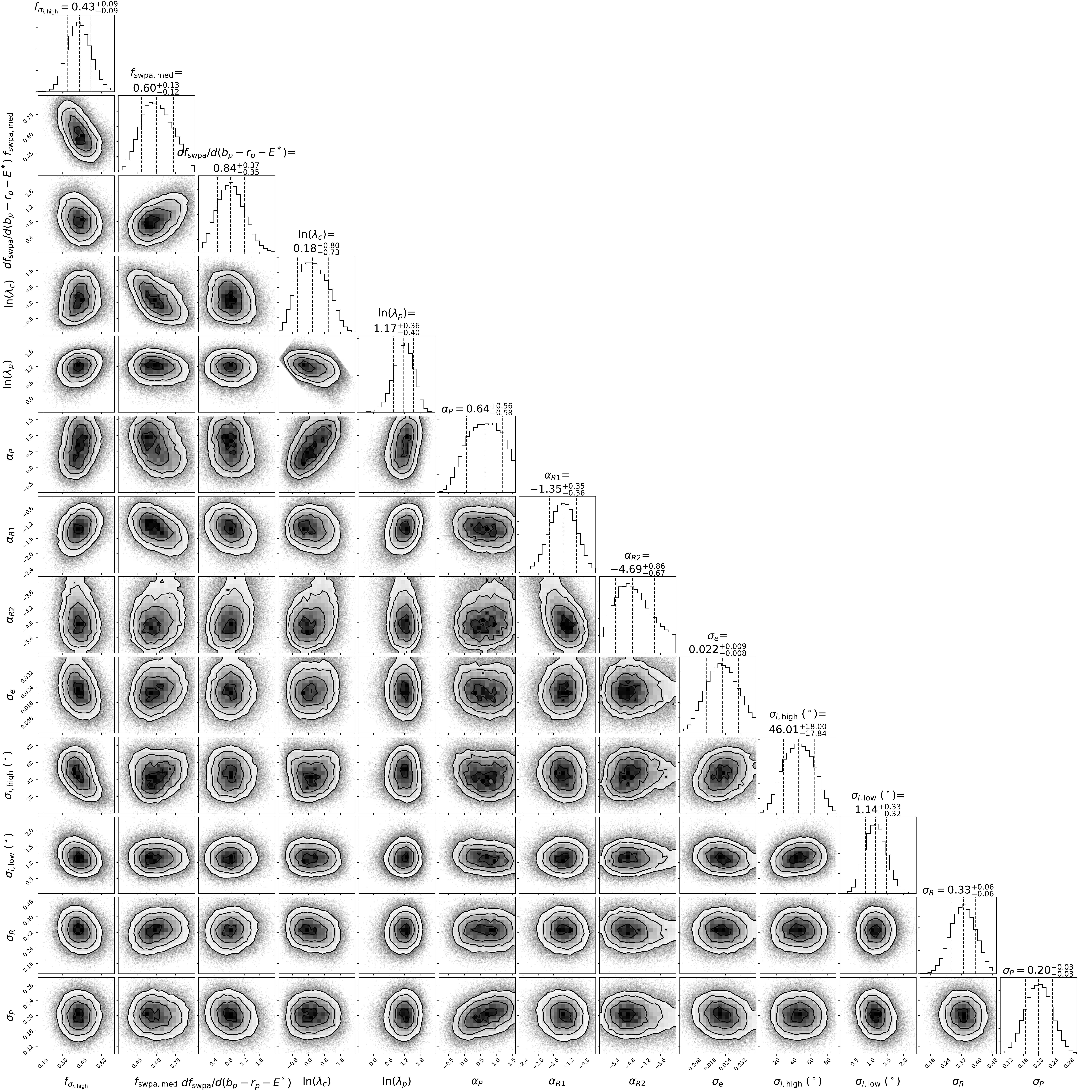}
\caption{ABC posterior distributions of the free model parameters of the linear $f_{\rm swpa}(b_p - r_p - E^*)$ model. A total of $5\times10^4$ points passing a distance threshold of $\mathcal{D}_{W,\rm KS} = 47$ as drawn from the GP emulator are shown. The prior mean function was set to a constant value of 75.}
\label{fig:linear_fswp_corner_KS}
\end{figure*}

In this section, we first report and discuss the main results, beginning with a comparison to the clustered periods and sizes model in \citetalias{HFR2019}, before describing our findings for the model stellar dependence. In Table \ref{tab:param_fits}, we list the best-fitting values and 68.3\% credible regions for the free parameters of the constant $f_{\rm swpa}+\alpha_P$, linear $f_{\rm swpa}(b_p-r_p-E^*)$, and linear $\alpha_P(b_p-r_p-E^*)$ models. We show the same credible regions (i.e. ABC posterior distributions) from our KS analysis in Figure \ref{fig:linear_fswp_corner_KS} (linear $f_{\rm swpa}(b_p-r_p-E^*)$ model), supplemental Figure \ref{fig:linear_alphaP_corner_KS} (linear $\alpha_P(b_p-r_p-E^*)$ model), and supplemental Figure \ref{fig:const_fswp_corner_KS} (constant $f_{\rm swpa}+\alpha_P$ model).
Our discussions will focus on the results from the KS distance function, as those from the AD distance function are very similar, except where noted (and both are listed in Table \ref{tab:param_fits} for completeness).

\begin{deluxetable*}{lcccccccc}
\centering
\tablecaption{Best-fitting values for the free parameters of each model.}
\tablehead{
 \colhead{Parameter} & \multicolumn2c{Clustered P+R (\citetalias{HFR2019})} & \multicolumn2c{Constant $f_{\rm swpa}+\alpha_P$} & \multicolumn2c{Linear $f_{\rm swpa}(b_p-r_p-E^*)$} & \multicolumn2c{Linear $\alpha_P(b_p-r_p-E^*)$} \\
 & \colhead{Best-fit KS} & \colhead{Best-fit AD} & \colhead{Best-fit KS} & \colhead{Best-fit AD} & \colhead{Best-fit KS} & \colhead{Best-fit AD} & \colhead{Best-fit KS} & \colhead{Best-fit AD}
}
\decimalcolnumbers
\startdata
 $f_{\sigma_{i,\rm high}}$ & $0.42_{-0.07}^{+0.08}$ & $0.40_{-0.12}^{+0.11}$ & $0.45_{-0.09}^{+0.09}$ & $0.45_{-0.13}^{+0.10}$ & $0.43_{-0.09}^{+0.09}$ & $0.44_{-0.09}^{+0.10}$ & $0.45_{-0.08}^{+0.09}$ & $0.47_{-0.10}^{+0.10}$ \\[5pt]
 $f_{\rm swpa}$\tablenotemark{a} & - & - & $0.52_{-0.11}^{+0.17}$ & $0.58_{-0.15}^{+0.13}$ & $0.60_{-0.12}^{+0.13}$ & $0.57_{-0.11}^{+0.12}$ & $0.55_{-0.11}^{+0.14}$ & $0.49_{-0.09}^{+0.15}$ \\[5pt]
 $\frac{df_{\rm swpa}}{d(b_p-r_p-E^*)}$ & - & - & - & - & $0.84_{-0.35}^{+0.37}$ & $1.15_{-0.36}^{+0.35}$ & - & - \\[5pt]
 $\ln{(\lambda_c)}$ & $-0.10_{-0.35}^{+0.44} \dagger$ & $0.24_{-0.40}^{+0.33} \dagger$ & $0.73_{-0.98}^{+0.72}$ & $0.85_{-0.67}^{+0.68}$ & $0.18_{-0.73}^{+0.80}$ & $0.99_{-0.84}^{+0.60}$ & $0.90_{-1.02}^{+0.65}$ & $1.43_{-0.80}^{+0.45}$ \\[5pt]
 $\lambda_c$ & $0.90_{-0.26}^{+0.50} \dagger$ & $1.27_{-0.42}^{+0.50} \dagger$ & $2.08_{-1.29}^{+2.18}$ & $2.34_{-1.15}^{+2.29}$ & $1.20_{-0.62}^{+1.46}$ & $2.68_{-1.52}^{+2.23}$ & $2.47_{-1.58}^{+2.28}$ & $4.19_{-2.30}^{+2.38}$ \\[5pt]
 $\ln{(\lambda_p)}$ & $1.35_{-0.44}^{+0.36}$ & $0.73_{-0.56}^{+0.60}$ & $0.94_{-0.55}^{+0.47}$ & $0.52_{-0.51}^{+0.53}$ & $1.17_{-0.40}^{+0.36}$ & $0.77_{-0.55}^{+0.54}$ & $0.80_{-0.50}^{+0.43}$ & $0.62_{-0.39}^{+0.40}$ \\[5pt]
 $\lambda_p$ & $3.86_{-1.38}^{+1.67}$ & $2.08_{-0.89}^{+1.70}$ & $2.55_{-1.07}^{+1.52}$ & $1.68_{-0.67}^{+1.17}$ & $3.22_{-1.05}^{+1.41}$ & $2.15_{-0.91}^{+1.55}$ & $2.23_{-0.87}^{+1.21}$ & $1.85_{-0.60}^{+0.92}$ \\[5pt]
 $\alpha_P$\tablenotemark{b} & $0.40_{-0.56}^{+0.64}$ & $0.07_{-0.45}^{+0.66}$ & $0.56_{-0.52}^{+0.56}$ & $0.56_{-0.49}^{+0.48}$ & $0.64_{-0.58}^{+0.56}$ & $0.81_{-0.44}^{+0.43}$ & $0.71_{-0.45}^{+0.47}$ & $0.92_{-0.48}^{+0.43}$ \\[5pt]
 $\frac{d\alpha_P}{d(b_p-r_p-E^*)}$ & - & - & - & - & - & - & $-1.29_{-0.37}^{+0.45}$ & $-1.27_{-0.35}^{+0.38}$ \\[5pt]
 $\alpha_{R1}$ & $-1.02_{-0.70}^{+0.64}$ & $-1.27_{-0.25}^{+0.26}$ & $-1.23_{-0.36}^{+0.35}$ & $-1.28_{-0.26}^{+0.25}$ & $-1.35_{-0.36}^{+0.35}$ & $-1.48_{-0.29}^{+0.29}$ & $-1.37_{-0.36}^{+0.34}$ & $-1.37_{-0.27}^{+0.27}$ \\[5pt]
 $\alpha_{R2}$ & $-4.41_{-0.79}^{+1.36}$ & $-5.08_{-0.54}^{+0.71}$ & $-4.75_{-0.60}^{+0.69}$ & $-4.91_{-0.55}^{+0.58}$ & $-4.69_{-0.67}^{+0.86}$ & $-4.92_{-0.56}^{+0.62}$ & $-4.55_{-0.67}^{+0.77}$ & $-4.86_{-0.55}^{+0.59}$ \\[5pt]
 $\sigma_e$ & $0.020_{-0.010}^{+0.014}$ & $0.014_{-0.008}^{+0.010}$ & $0.020_{-0.009}^{+0.009}$ & $0.013_{-0.007}^{+0.008}$ & $0.022_{-0.008}^{+0.009}$ & $0.016_{-0.008}^{+0.008}$ & $0.021_{-0.008}^{+0.009}$ & $0.019_{-0.008}^{+0.008}$ \\[5pt]
 $\sigma_{i,\rm high}$ ($^\circ$) & $48_{-17}^{+17}$ & $49_{-25}^{+23}$ & $48_{-19}^{+19}$ & $43_{-18}^{+22}$ & $46_{-18}^{+18}$ & $48_{-18}^{+17}$ & $43_{-18}^{+17}$ & $47_{-18}^{+18}$ \\[5pt]
 $\sigma_{i,\rm low}$ ($^\circ$) & $1.40_{-0.39}^{+0.54}$ & $1.29_{-0.32}^{+0.35}$ & $1.17_{-0.31}^{+0.34}$ & $1.21_{-0.27}^{+0.30}$ & $1.14_{-0.32}^{+0.33}$ & $1.24_{-0.33}^{+0.37}$ & $1.13_{-0.31}^{+0.32}$ & $1.22_{-0.30}^{+0.32}$ \\[5pt]
 $\sigma_R$ & $0.31_{-0.07}^{+0.07}$ & $0.32_{-0.07}^{+0.07}$ & $0.31_{-0.07}^{+0.07}$ & $0.32_{-0.08}^{+0.07}$ & $0.33_{-0.06}^{+0.06}$ & $0.32_{-0.08}^{+0.07}$ & $0.31_{-0.07}^{+0.07}$ & $0.32_{-0.07}^{+0.07}$ \\[5pt]
 $\sigma_P$ & $0.21_{-0.04}^{+0.04}$ & $0.20_{-0.04}^{+0.04}$ & $0.20_{-0.03}^{+0.03}$ & $0.19_{-0.03}^{+0.03}$ & $0.20_{-0.03}^{+0.03}$ & $0.18_{-0.04}^{+0.04}$ & $0.20_{-0.03}^{+0.03}$ & $0.18_{-0.03}^{+0.03}$ \\[5pt]
\enddata
\tablecomments{While we trained the emulator on the transformed parameters $\ln(\lambda_c \lambda_p)$ and $\ln(\lambda_p/\lambda_c)$, we transform back to $\ln(\lambda_c)$ and $\ln(\lambda_p)$ to report the credible intervals. Unlogged rates $\lambda_c$ and $\lambda_p$ are shown for interpretability and are equivalent to the rows with log-values. The 68.3\% credible regions are computed from the ABC posterior using distance thresholds of $\mathcal{D}_{W,\rm KS} = 50$, 47, and 47 for the constant $f_{\rm swpa}$, linear $f_{\rm swpa}(b_p-r_p-E^*)$, and linear $\alpha_P(b_p-r_p-E^*)$ models, respectively, in the KS analyses, while the distance thresholds are $\mathcal{D}_{W,\rm AD^\prime} = 100$, 90, and 90, respectively, in the AD analyses.}
\tablenotetext{a}{This is $f_{\rm swpa,med}$ for the clustered model with linear $f_{\rm swpa}(b_p-r_p-E^*)$.}
\tablenotetext{b}{This is $\alpha_{P,\rm med}$ for the clustered model with linear $\alpha_P(b_p-r_p-E^*)$.}
\tablenotetext{\dagger}{Although the symbol for this parameter is the same, the parameter is not: in \citetalias{HFR2019}, a Poisson distribution for the number of clusters is used, while in this paper, a zero-truncated Poisson distribution is used. Thus, $\lambda_c$ can be interpreted as the mean number of attempted clusters per system in \citetalias{HFR2019}, and as the mean number of attempted clusters per system with planets in this paper. In both cases, some clusters are rejected due to stability.}
\label{tab:param_fits}
\end{deluxetable*}

\subsection{The Overall Fraction of Stars with Planets} \label{Baseline_results}

The only difference between our constant $f_{\rm swpa}+\alpha_P$ model and the clustered periods and sizes model in \citetalias{HFR2019} is a re-parametrization from a Poisson to a ZTP distribution for the number of clusters per system, with the addition of an explicit parameter $f_{\rm swpa}$ for the fraction of stars with planets. As explained in \S\ref{Baseline_model}, this change was made to (1) decouple the number of intrinsic zero-planet systems from the planet--hosting stars and their underlying multiplicity distribution and (2) produce a baseline model for comparison to our more general, linear $f_{\rm swpa}(b_p-r_p-E^*)$ (and linear $\alpha_P(b_p-r_p-E^*)$) model. Methodologically, we also used a distance function with three times the number of terms (fitting to the bluer half, redder half, and full samples) and recomputed weights, in order to facilitate a direct comparison with our new models. We first discuss how these results compare with the results from \citetalias{HFR2019}.

\emph{Fraction of stars with planets:}
For our baseline model, we find that the fraction of stars with planets (in our entire range explored, with periods between 3 and 300 days and planet radii between 0.5 and $10 R_\oplus$) is well constrained. The KS and AD analyses result in similar values of $f_{\rm swpa} = 0.52_{-0.11}^{+0.17}$ and $f_{\rm swpa} = 0.58_{-0.15}^{+0.13}$, respectively. These credible regions are also fully consistent with the results from the clustered periods and sizes model in \citetalias{HFR2019}, where the fraction of stars with planets in the same range is $0.56_{-0.15}^{+0.18}$ (via the KS analysis). This is noteworthy, given that we find a meaningful and comparable constraint on the $f_{\rm swpa}$ despite a re-parametrization of the intrinsic multiplicity distribution and the extra dimensionality of the model optimization problem (and thus a larger parameter space).

\emph{Distribution of planets between and within clusters:}
While our baseline model and the clustered models from \citetalias{HFR2019} all have a parameter $\lambda_c$ (for the \textit{mean} rate of attempted clusters per planetary system), this parameter is not the same in these models. In \citetalias{HFR2019}, $\lambda_c$ is the mean number of clusters per system before any rejection sampling and, due to the draws from a Poisson distribution, is also tied to the number of true zero-planet systems. Loosely, it could be interpreted as the mean number of clusters per star, including those that harbor no planets (between 3 and 300 days), although we found that the true mean is somewhat lower than what the parameter value suggests due to the rejected clusters. For our clustered periods and sizes model, about $\sim79\%$ and $\sim19\%$ of planet--hosting stars have just one and two clusters, respectively (Figure \ref{fig:intrinsic_mults}). Since the fraction of stars with planets in that model is $\sim56\%$, this means that the true mean number of clusters per system is about $\sum_{n_c \geq 0} (f_c n_c) \sim 0.7$, where $f_c$ is the fraction of all stars with $n_c$ clusters, while the mean number of clusters per system with planets is $\sum_{n_c \geq 1} (f_c/0.56) n_c \sim 1.2$.

In this paper, we have decoupled the $\lambda_c$ parameter from the number of zero-planet systems and thus the fraction of stars with planets. However, we find that the mean numbers of clusters (and planets per cluster) are rather poorly constrained. For our constant $f_{\rm swpa}+\alpha_P$ model, we find that the mean rate of clusters per system is $2.08_{-1.29}^{+2.18}$. Part of the reason we find such a large range of $\lambda_c$ values is due to our algorithm for drawing planetary systems: as in \citetalias{HFR2019}, we draw clusters one by one, drawing their period scales after the unscaled periods for each planet in a cluster have been drawn, and only keeping clusters that can fit. In this manner, each individual cluster is deemed ``stable'' (given our minimum spacing in mutual Hill radii for adjacent planet pairs) before their period scale has been drawn, but the period scale drawn must also allow all of the planets in the cluster to be stable with the previously drawn clusters. As such, the actual number of accepted clusters can be significantly less than what would be suggested by the value of the $\lambda_c$ parameter. Thus, we warn that our parameter values of $\lambda_c$ (and $\lambda_p$) should not be misinterpreted as the average number of clusters per system (or average number of planets per cluster). 
Instead, we count the true numbers of clusters and planets per cluster, as shown in Figure \ref{fig:intrinsic_mults} for the KS analysis. The AD analysis (not shown) leads to somewhat more clusters per system and fewer planets per cluster, but a similar intrinsic planet multiplicity distribution (i.e. the top panel).
We conclude that our decoupling of the $\lambda_c$ parameter suggests that the true mean number of clusters per system is greater than what we found in \citetalias{HFR2019}, but the extent is unclear and is degenerate with the mean number of planets per cluster.

As noted earlier in \S\ref{Baseline_model}, it is interesting that our re-parametrization of the intrinsic multiplicity distribution did not change the inferred distribution of total planet multiplicity in any noticeable way (i.e. compared to the red dotted line in the top panel of Figure \ref{fig:intrinsic_mults}). We also note that our intrinsic planet multiplicity distribution is very different than of that inferred by \citet{SKC2019}, who found that a single Zipfian distribution (a discrete power-law distribution) that peaks at unity and falls off rapidly toward larger multiplicities best fits the observed \Kepler{} counts. In contrast, our distribution peaks at a count of about four planets (for planet hosting stars), with similar fractions of seven-planet systems and one-planet systems. We attribute this difference to the constraints from the total rate of observed planets to stars and the fraction of stars with planets, which cannot exceed unity. In contrast, \citet{SKC2019} did not fit the rate of zero-planet systems (i.e. the fraction of stars with planets). As discussed in \citetalias{HFR2019}, we argue that the apparent excess of observed single--planet systems is unlikely to be due to a large fraction of true single--planet systems, since there are not enough stars available to host single planets to replace our high mutual--inclination population. These results highlight the importance of modeling the true fraction of stars with planets and simultaneously fitting the additional observables (period ratio distribution, and so on, as we have done here) when inferring the underlying multiplicity distribution, in order to distinguish between competing models for the \Kepler{} dichotomy.

\subsection{Planetary System Architectures}

Broadly, we find consistent results for the model parameters describing the planetary system architectures between our new models and our old clustered periods and sizes model. We summarize the results here, quoting the KS results for our constant $f_{\rm swpa}+\alpha_P$ model. The results for our linear models and the AD results are listed in Table \ref{tab:param_fits} and are very similar. For a more detailed discussion of what these parameters mean and comparisons to other values in the literature, see \S3 of \citetalias{HFR2019}.

\emph{Period distribution ($\alpha_P$):} the overall period distribution is described by a single power-law between 3 and 300 days, with a shallowly increasing occurrence in log--period given by $\alpha_P = 0.56_{-0.52}^{+0.56}$.
 
\emph{Radius distribution ($\alpha_{R1}$, $\alpha_{R2}$):} the overall radius distribution is described by a broken power-law between 0.5 and $10 R_\oplus$, where we have set the break at $R_{p,\rm break} = 3 R_\oplus$. For planets below the break, the distribution is consistent with flat, $\alpha_{R1} = -1.23_{-0.36}^{+0.35}$; above, there is a sharp fall-off with $\alpha_{R2} = -4.75_{-0.60}^{+0.69}$. As before, our models do not have the flexibility of producing a radius valley.
 
\emph{Eccentricity distribution ($\sigma_e$):} we find small orbital eccentricities, described by a Rayleigh scale $\sigma_e = 0.020_{-0.009}^{+0.009}$. While our models reproduce the transit duration and period-normalized transit duration ratio ($\xi$) distributions reasonably well, there may be evidence for a higher eccentricity component based on comparisons to the circular-normalized transit duration distribution ($t_{\rm dur}/t_{\rm circ}$, where $t_{\rm circ} = R_\star P/(\pi{a})$ is the duration assuming a circular orbit with impact parameter $b = 0$), although we have not included this distribution in our distance function for reasons described in \S\ref{secEcc}.

\emph{Mutual inclination distribution ($f_{\sigma_{i,\rm high}}$, $\sigma_{i,\rm high}$, $\sigma_{i,\rm low}$):} we still find clear evidence for a dichotomous population of planetary systems in terms of their mutual inclinations, with $f_{\sigma_{i,\rm high}} = 0.45 \pm 0.09$ of systems assigned to the high mutual inclination population. We note that other solutions to the \Kepler{} dichotomy exist \citep{Z2018, ZCH2019}, but are beyond the scope of this paper.
As in our models from \citetalias{HFR2019} (except where noted), in our new models we still draw mutual inclinations from $\sigma_{i,\rm low}$ for planets near an MMR with another planet, which affects about $\simeq 30\%$ of all planets. The high mutual inclination scale $\sigma_{i,\rm high}$, while clearly and significantly greater than $\sigma_{i,\rm low}$, is still largely unconstrained at larger values since these systems primarily affect only the observed number of single-transiting systems, and our simple stability criteria do not directly account for the mutual inclinations. The low mutual inclination scale, $\sigma_{i,\rm low} = 1.17_{-0.31}^{+0.34}$ degrees, suggests that most multi-planet systems are nearly but not exactly coplanar. This is consistent with our findings in \citetalias{HFR2019} and those of many previous studies, including \citet{Li2011b} (who found an excellent fit to the multis with $\sigma_i = 2^\circ$), \citet{FM2012} ($\sigma_i = 1^\circ$), and \citet{F2014} ($1-2.2^\circ$); see Section 3.6 in \citetalias{HFR2019} for a more detailed discussion.

\emph{Period and radius clustering ($\sigma_P$, $\sigma_R$):} the periods and planet radii of planets in the same cluster are each highly correlated. We find that $\sigma_P = 0.20 \pm 0.03$, where this parameter quantifies the width per planet in the cluster for each cluster, in log-period. On the other hand, $\sigma_R = 0.31 \pm 0.07$, which is the cluster scale in log-radius (regardless of the number of planets in the cluster). These two parameters are most directly constrained by the period ratio and transit depth (i.e. radius) ratio distributions, respectively.

\subsection{Occurrence of Planetary Systems with Spectral Type} \label{secFSWP}

\begin{figure*}
\centering
\includegraphics[scale=0.5,trim={0.5cm 0.6cm 0 0},clip]{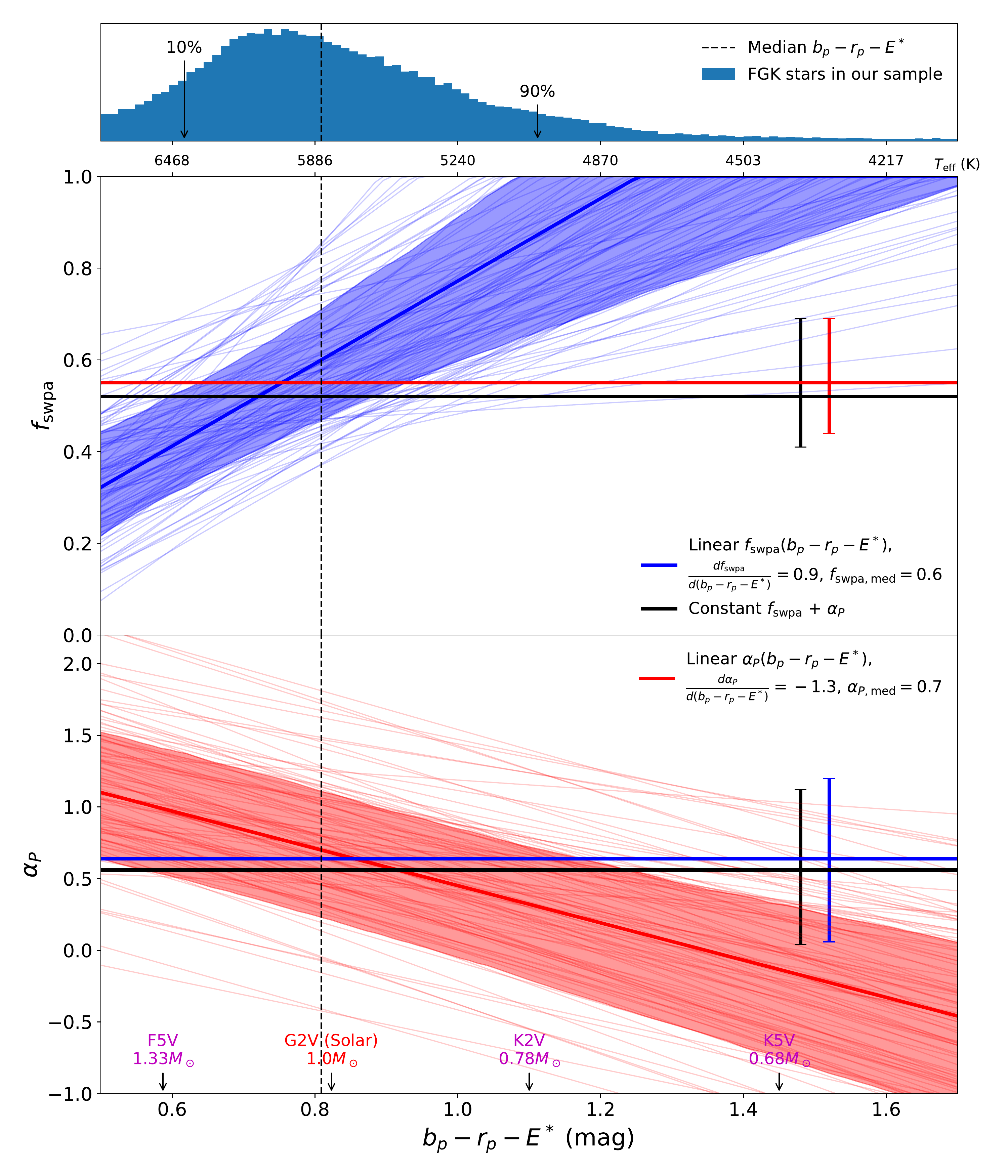}
\caption{Best-fitting relations for the fraction of stars with planets, $f_{\rm swpa}$ (\textbf{middle panel}), and the power-law index of the period distribution, $\alpha_P$ (\textbf{bottom panel}), as functions of \Gaia{} $b_p-r_p-E^*$ color. The constant $f_{\rm swpa}+\alpha_P$, linear $f_{\rm swpa}(b_p-r_p-E^*)$, and linear $\alpha_P(b_p-r_p-E^*)$ models are denoted by black, blue, and red lines, respectively.
Each thick line denotes a single model close to the best-fit median values (as labeled), while the thin lines show 100 models each passing our KS distance thresholds (the shaded regions or error bars denote the 68.3\% credible regions).
\newline For the linear $f_{\rm swpa}(b_p-r_p-E^*)$ model (\textbf{middle panel}), the fraction of Sun-like stars with planets (between $3-300$ days and $0.5-10 R_\oplus$) is around 60\%. A few other example values are labeled with arrows (magenta), where we have used a table relating $T_{\rm eff}$ and $b_p-r_p$ from \citet{PM2013}. The fraction increases by over a factor of two from the bluest (early F) to the reddest (late K) dwarfs in our sample. Alternatively, the linear $\alpha_P(b_p-r_p-E^*)$ model (\textbf{bottom panel}) exhibits a decrease in $\alpha_P$ with increasing $b_p-r_p-E^*$, corresponding to a shallower rise in occurrence toward longer periods for planets around redder stars (for reference, a value of $\alpha_P = -1$ corresponds to a flat distribution in log--period).}
\label{fig:linear_fswp_alphaP}
\end{figure*}

We find a significant positive slope for the linear relation between $f_{\rm swpa}$ and $b_p-r_p-E^*$ color, suggesting that the occurrence rate of planetary systems between 3 and 300 days increases toward later type (redder, higher $b_p-r_p-E^*$) stars. The slope is $df_{\rm swpa}/d(b_p-r_p-E^*) = 0.84_{-0.35}^{+0.37}$ ($1.15_{-0.36}^{+0.35}$) using KS (AD) analyses. Both analyses result in a similar slope and strongly disfavor a flat or negative slope. As expected, the fraction of stars with planets at the median color ($f_{\rm swpa,med}$), is comparable to the overall fraction of stars with planets (i.e. $f_{\rm swpa}$) for our baseline model; we find that $f_{\rm swpa,med} = 0.60_{-0.12}^{+0.13}$.

In Figure \ref{fig:linear_fswp_alphaP} (middle panel), we plot our best--fitting (KS) relations for the linear $f_{\rm swpa}(b_p-r_p-E^*)$ model. The solid blue line shows an example line that is close to the median relation, with $df_{\rm swpa}/d(b_p-r_p-E^*) = 0.9$ and $f_{\rm swpa,med} = 0.6$.
We also plot 100 individual models, each passing our KS distance threshold, as light blue lines, and we denote the 68.3\% credible region by the shaded blue region. The values of $f_{\rm swpa}$ for the constant $f_{\rm swpa}+\alpha_P$ and linear $\alpha_P(b_p-r_p-E^*)$ models are also plotted as horizontal black and red lines with 68.3\% error bars, respectively, for comparison.
The distribution of $b_p-r_p-E^*$ for our stellar sample is plotted as a histogram in the top panel, with a vertical dashed line denoting the median value. We use the table from \citet{PM2013} to adopt a relation between \Gaia{} $b_p-r_p$ color and stellar effective temperature $T_{\rm eff}$, which we label as a secondary x-axis in Figure \ref{fig:linear_fswp_alphaP}.

Given that our stellar sample ranges from $b_p-r_p-E^* \simeq 0.5$ to $\simeq 1.7$, the large slope for our linear relation suggests that the fraction of stars with planets changes by over a factor of two going from the bluest stars (early F dwarfs) to the reddest stars (late K dwarfs) in our sample. For F2V dwarfs, $f_{\rm swpa}(0.5) = 0.34_{-0.11}^{+0.12}$, while this value increases to $f_{\rm swpa}(1.3) = 0.96_{-0.19}^{+0.04}$ for mid K dwarfs. The $f_{\rm swpa}$ rises to unity beyond $b_p-r_p-E^* \simeq 1.3$ ($T_{\rm eff} \simeq 4600$ K), implying that inner planetary systems are extremely common around cooler stars. A more complex model than a simple linear relation may be necessary to model differences at this range of stellar types. We also note that while we assume a linear relation, we fit it using two samples split at the median stellar color, around which most of our stars are concentrated. Finally, the fraction of solar-type (G2V) dwarfs harboring at least one planet between 3 and 300 days is $f_{\rm swpa}(0.823) = 0.57_{-0.10}^{+0.14}$, or roughly half.

\begin{figure}
 \includegraphics[scale=0.42,trim={0 0.2cm 0.5cm 0},clip]{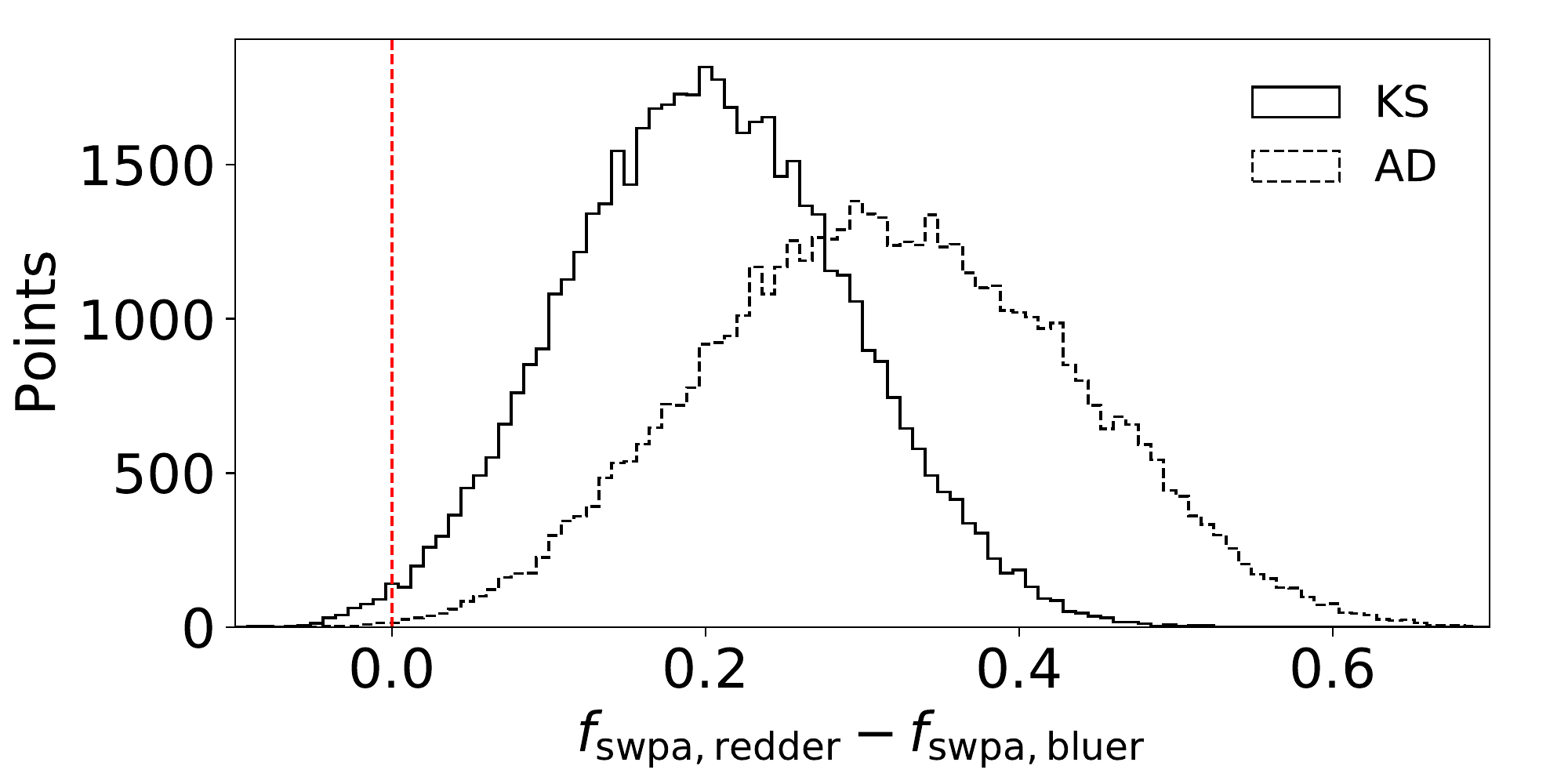}
\caption{Histograms of the difference in $f_{\rm swpa}$ between the bluer and redder halves from our step $f_{\rm swpa}$ model. The solid and dashed lines each show $5\times10^4$ points passing our KS and AD distance thresholds, respectively. For reference, the vertical dashed line denotes zero difference. We find a significantly higher $f_{\rm swpa}$ for the redder sample compared to the bluer sample, with a difference of $0.20\pm0.09$ ($0.31\pm0.12$) using KS (AD). Less than 0.8\% (0.1\%) of the points passing our KS (AD) distance thresholds have $f_{\rm swpa,redder}-f_{\rm swpa,bluer} < 0$.}
\label{fig:step_fswp_diff}
\end{figure}

\medskip
\emph{Step $f_{\rm swpa}$ model:} As a check on the results of our linear $f_{\rm swpa}(b_p-r_p-E^*)$ model, we also explore a model in which the fraction of stars with planets is set to one constant ($f_{\rm swpa,bluer}$) below and another constant ($f_{\rm swpa,redder}$) above the median color. This model has the same number of parameters as the linear $f_{\rm swpa}(b_p-r_p-E^*)$ model. We find that $f_{\rm swpa,bluer} = 0.47_{-0.11}^{+0.12}$ and $f_{\rm swpa,redder} = 0.68_{-0.15}^{+0.14}$ using KS distances. In Figure \ref{fig:step_fswp_diff}, we plot the distribution of $f_{\rm swpa,redder}-f_{\rm swpa,bluer}$ for $5\times10^4$ points passing our KS and AD distance thresholds. Thus, in both analyses, $f_{\rm swpa,bluer} < f_{\rm swpa,redder}$ consistently, supporting our strong positive slope for the linear $f_{\rm swpa}(b_p-r_p-E^*)$ model. While $f_{\rm swpa,bluer}$ and $f_{\rm swpa,redder}$ are correlated, $f_{\rm swpa,redder}$ is $0.20 \pm 0.09$ ($0.31 \pm 0.12$) higher than $f_{\rm swpa,bluer}$ using KS (AD) analyses. We strongly rule out models with $f_{\rm swpa,bluer} > f_{\rm swpa,redder}$ to more than 99\%, as shown by the vertical dashed line. The other parameters of this model are also fully in agreement with those of our other models.

To further verify the consistency between the linear and step $f_{\rm swpa}$ models, we compute a color--weighted average using our linear $f_{\rm swpa}(b_p-r_p-E^*)$ relation for each of the bluer and redder halves: $f_{\rm swpa,avg} = \sum_{\rm stars}[c f_{\rm swpa}(c)]/\sum_{\rm stars} c $, where $c = b_p-r_p-E^*$ is the color. To illustrate using the KS results, we find that $f_{\rm swpa,avg} = 0.47_{-0.09}^{+0.17}$ and $0.71_{-0.12}^{+0.15}$ for the bluer and redder halves, respectively, values that are very close to the $f_{\rm swpa,bluer}$ and $f_{\rm swpa,redder}$ of our step model. While the results of this model serve to corroborate our linear $f_{\rm swpa}(b_p-r_p-E^*)$ model, the linear model is still preferred over the step model for two main reasons: (1) it is a more physically plausible model (the rise in $f_{\rm swpa}$ toward later types is continuous and does not depend on the median color), and (2) the best distances for the linear model are slightly better than those of the step model.

\subsection{The Period Distribution as a Function of Spectral Type: Which is the Preferred Model?} \label{secFSWP_alphaP}

\begin{figure}
 \includegraphics[scale=0.42,trim={0 1cm 0 1cm},clip]{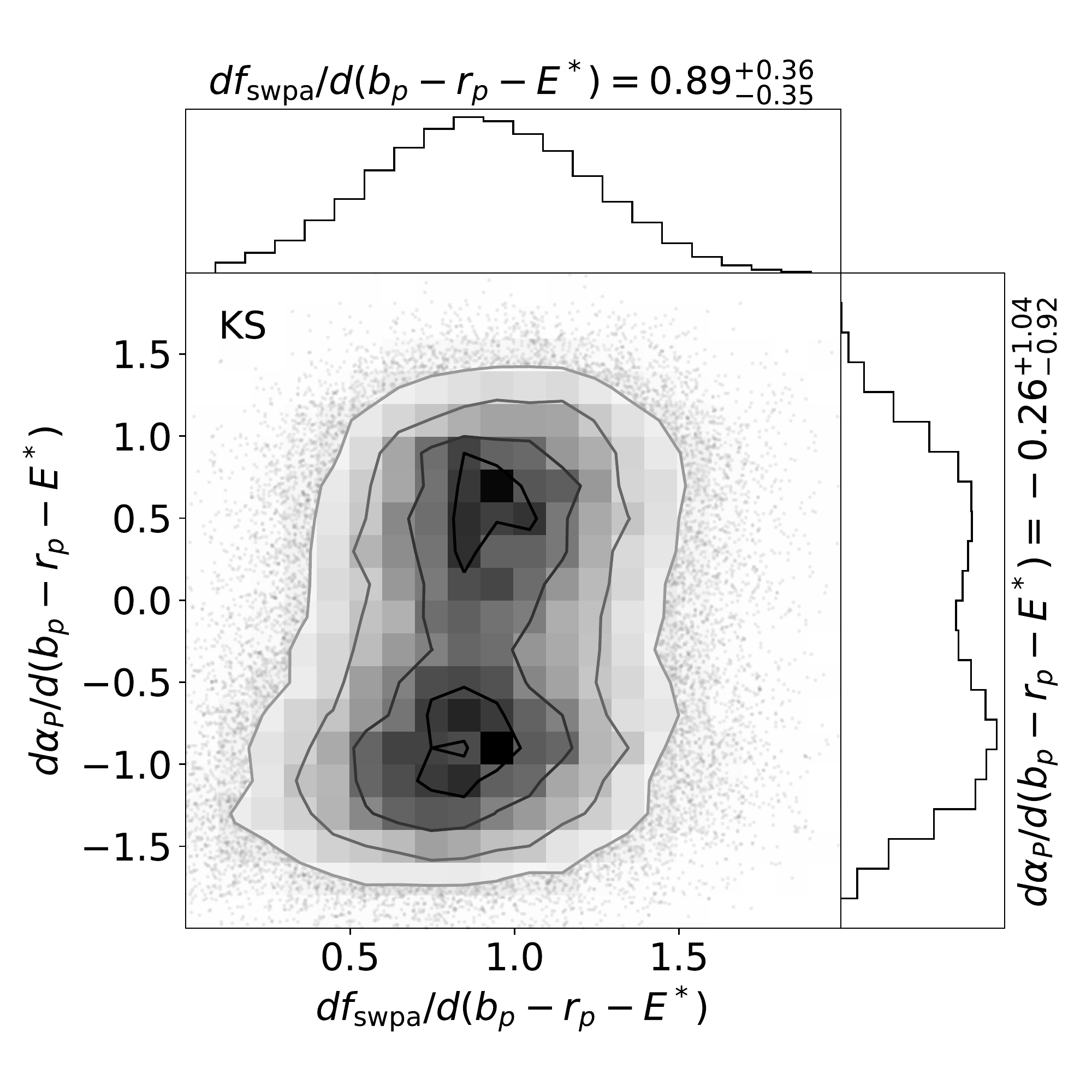}
 \includegraphics[scale=0.42,trim={0 1cm 0 0},clip]{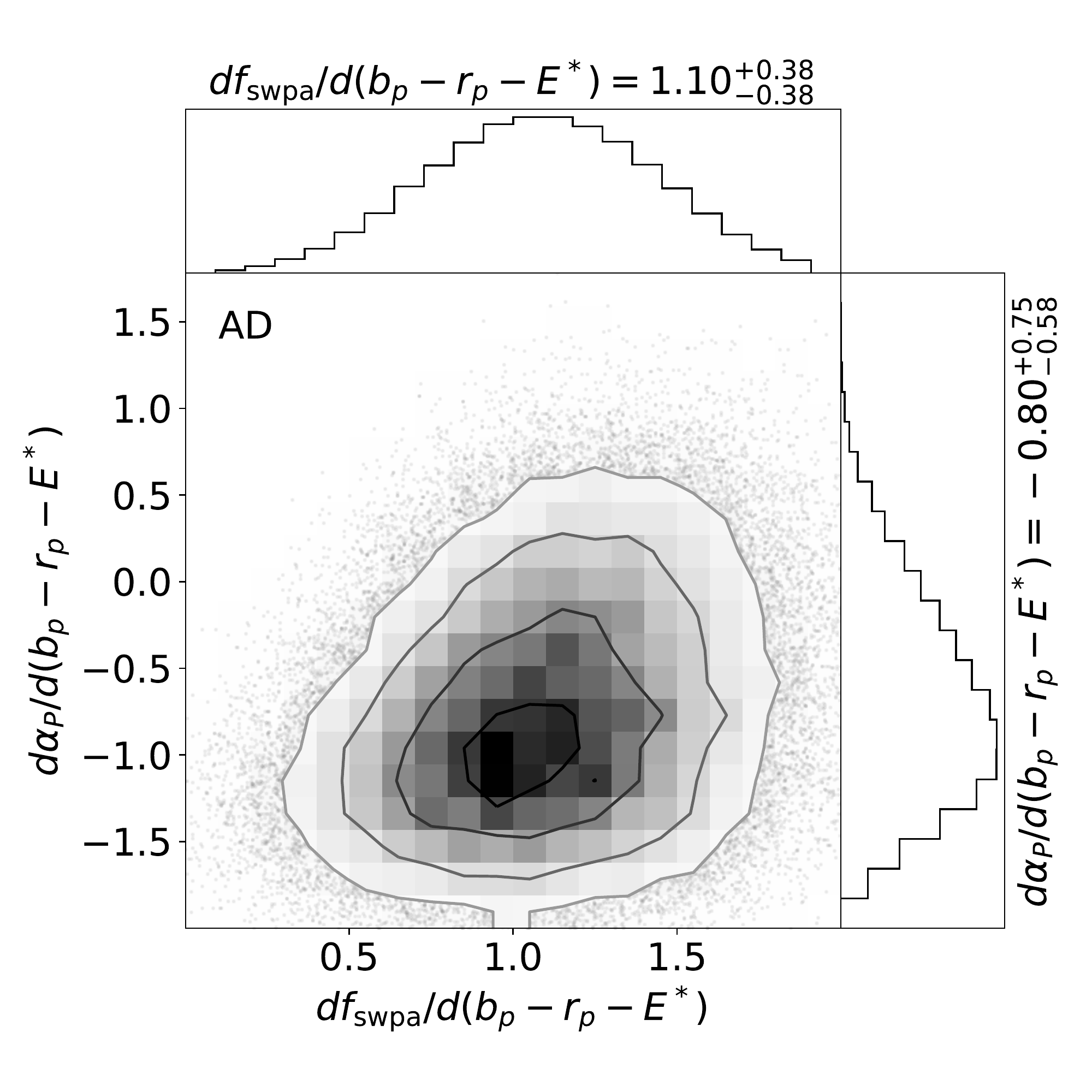}
\caption{Joint ABC posterior distributions of $df_{\rm swpa}/d(b_p-r_p-E^*)$ and $d\alpha_P/d(b_p-r_p-E^*)$ for a model in which we include both the linear $f_{\rm swpa}$ and linear $\alpha_P$ functions of stellar $b_p-r_p-E^*$ color, using KS (\textbf{top panel}) and AD (\textbf{bottom panel}) analyses. We used the same mean functions and distance thresholds for the GP emulator as for our other two linear models. While the slope for $f_{\rm swpa}$ is essentially unchanged compared to that of our linear $f_{\rm swpa}(b_p-r_p-E^*)$ model (and remains strongly positive, ruling out zero to more than $2\sigma$), the slope for $\alpha_P$ is significantly diminished using either KS or AD. The value of $d\alpha_P/d(b_p-r_p-E^*)$ appears slightly bimodal and consistent with zero using KS, and only slightly negative using AD. These results strengthen the linear $f_{\rm swpa}(b_p-r_p-E^*)$ model as the preferred model.}
\label{fig:joint_fswp_alphaP}
\end{figure}

For our model where $\alpha_P$ (instead of $f_{\rm swpa}$) is a linear function of $b_p-r_p-E^*$, we find a negative slope for the linear relation between $\alpha_P$ and $b_p-r_p-E^*$ color: $d\alpha_P/d(b_p-r_p-E^*) = -1.29_{-0.37}^{+0.45}$ using KS distances (similarly using AD). The period power--law index at the median color is $\alpha_{P,\rm med} = 0.67_{-0.44}^{+0.44}$. In Figure \ref{fig:linear_fswp_alphaP} (bottom panel), we also plot the best--fitting (KS) relations for $\alpha_P$ as a function of $b_p-r_p-E^*$. Similar to the middle panel, the bold red line shows an example (with $d\alpha_P/d(b_p-r_p-E^*) = -1.3$ and $\alpha_{P,\rm med} = 0.7$), while the thin red lines and shaded region show 100 models each passing our KS distance threshold and the 68.3\% region, respectively. For reference, $\alpha_P = -1$ corresponds to a flat distribution in log--period. Thus, we find that in this model, the occurrence of planets in our period range increases toward longer periods for nearly all FGK dwarfs, with a sharper rise for earlier type (bluer) stars and a shallower rise for later type stars. Since planets at shorter periods are more likely to transit and are easier to detect than planets at longer periods, this would imply that the observed rise in planet occurrence toward later spectral types can be explained by there being more short period planets for redder stars than for bluer stars. However, this relation is tentative at best, and we caution against using this model for reasons described below.

While the linear $\alpha_P(b_p-r_p-E^*)$ model provides overall best--fitting distances to the \Kepler{} planet catalog similar to the linear $f_{\rm swpa}(b_p-r_p-E^*)$ model (Appendix Figures \ref{fig:dists_KS} and \ref{fig:dists_AD}), a more detailed analysis shows that the linear relation between $\alpha_P$ and color is questionable. Compared to the $f_{\rm swpa}(b_p-r_p-E^*)$ model, the negative slope for $d\alpha_P/d(b_p-r_p-E^*)$ improves the fit to the period distribution of the bluer planet sample, but worsens the fit to the transit duration distribution of the same half. Most importantly, the value of $d\alpha_P/d(b_p-r_p-E^*)$ is not robustly determined when \textit{simultaneously} including both the linear $f_{\rm swpa}$ and linear $\alpha_P$ functions of $b_p-r_p-E^*$ in our model. In Figure \ref{fig:joint_fswp_alphaP}, we plot the joint posterior distributions of $df_{\rm swpa}/d(b_p-r_p-E^*)$ and $d\alpha_P/d(b_p-r_p-E^*)$ for such a model. We find that $d\alpha_P/d(b_p-r_p-E^*) = -0.26_{-0.92}^{+1.04}$ ($-0.80_{-0.58}^{+0.75}$) using KS (AD) analyses, results that are significantly closer to zero than in the model where only $\alpha_P$ is allowed to vary with color. The distribution using KS distances appears slightly bimodal, with one mode above and one below zero, although the nature is unknown and this is not seen in the AD results; in either case, $d\alpha_P/d(b_p-r_p-E^*)$ is consistent with no slope. In contrast, the slope for $f_{\rm swpa}$ with color is just as strong as in our linear $f_{\rm swpa}(b_p-r_p-E^*)$ model and still significantly positive: $df_{\rm swpa}/d(b_p-r_p-E^*) = 0.89_{-0.35}^{+0.36}$ ($1.10_{-0.38}^{+0.38}$) using KS (AD) analyses. Taken together, our results show that while varying $f_{\rm swpa}$ or $\alpha_P$ with color can have similar effects on the rate of observed planets as a function of stellar type (as we will further show in \S\ref{Observed_multiplicity}), a change in $f_{\rm swpa}$ is the more likely explanation. We conclude that the linear $f_{\rm swpa}(b_p-r_p-E^*)$ model is the preferred model, with the fraction of stars with planets clearly increasing toward later spectral types, as has been previously shown.

\begin{deluxetable*}{lcccccccccccc}
\centering
\tablecaption{Comparison of the observed multiplicity ($m$) distribution between the \Kepler{} data and our models.}
\tablewidth{0pt}
\tablehead{
 \colhead{$m$} & \multicolumn3c{\Kepler{} data} & \multicolumn3c{Constant $f_{\rm swpa}+\alpha_P$} & \multicolumn3c{Linear $f_{\rm swpa}(b_p - r_p - E^*)$} & \multicolumn3c{Linear $\alpha_P(b_p - r_p - E^*)$} \\
 & \colhead{All} & \colhead{Bluer} & \colhead{Redder} & \colhead{All} & \colhead{Bluer} & \colhead{Redder} & \colhead{All} & \colhead{Bluer} & \colhead{Redder} & \colhead{All} & \colhead{Bluer} & \colhead{Redder}
}
\decimalcolnumbers
\startdata
 1 & 1218 & 554 & 664 & $1239_{-132}^{+128}$ & $639_{-69}^{+70}$ & $599_{-65}^{+63}$ & $1252_{-109}^{+110}$ & $525_{-69}^{+68}$ & $726_{-78}^{+80}$ & $1238_{-122}^{+121}$ & $598_{-64}^{+62}$ & $635_{-65}^{+72}$ \\[5pt]
 2 & 261 & 120 & 141 & $270_{-30}^{+30}$ & $142_{-17}^{+18}$ & $128_{-17}^{+17}$ & $269_{-29}^{+29}$ & $116_{-18}^{+18}$ & $152_{-18}^{+21}$ & $266_{-27}^{+29}$ & $128_{-15}^{+16}$ & $138_{-17}^{+17}$ \\[5pt]
 3 & 101 & 38 & 63 & $91_{-14}^{+16}$ & $47_{-8}^{+10}$ & $44_{-8}^{+9}$ & $93_{-14}^{+14}$ & $39_{-8}^{+9}$ & $53_{-8}^{+10}$ & $92_{-13}^{+14}$ & $44_{-8}^{+8}$ & $48_{-8}^{+10}$ \\[5pt]
 4 & 30 & 12 & 18 & $29_{-6}^{+8}$ & $15_{-4}^{+5}$ & $14_{-4}^{+4}$ & $29_{-6}^{+9}$ & $13_{-4}^{+4}$ & $17_{-5}^{+5}$ & $29_{-6}^{+8}$ & $14_{-4}^{+5}$ & $15_{-4}^{+6}$ \\[5pt]
 5 & 7 & 4 & 3 & $8_{-3}^{+3}$ & $4_{-2}^{+3}$ & $3_{-1}^{+3}$ & $8_{-4}^{+3}$ & $3_{-1}^{+3}$ & $4_{-2}^{+3}$ & $8_{-3}^{+4}$ & $4_{-2}^{+2}$ & $4_{-2}^{+2}$ \\[5pt]
 6 & 3 & 1 & 2 & $1_{-1}^{+2}$ & $1_{-1}^{+1}$ & $1_{-1}^{+1}$ & $1_{-1}^{+2}$ & $1_{-1}^{+1}$ & $1_{-1}^{+1}$ & $1_{-1}^{+2}$ & $1_{-1}^{+1}$ & $1_{-1}^{+1}$ \\[5pt]
 7 & 0 & 0 & 0 & $0_{-0}^{+1}$ & $0_{-0}^{+0}$ & $0_{-0}^{+0}$ & $0_{-0}^{+1}$ & $0_{-0}^{+0}$ & $0_{-0}^{+0}$ & $0_{0}^{+1}$ & $0_{0}^{+0}$ & $0_{0}^{+0}$ \\[5pt]
 8 & 0 & 0 & 0 & $0_{-0}^{+0}$ & $0_{-0}^{+0}$ & $0_{-0}^{+0}$ & $0_{-0}^{+0}$ & $0_{-0}^{+0}$ & $0_{-0}^{+0}$ & $0_{0}^{+0}$ & $0_{0}^{+0}$ & $0_{0}^{+0}$ \\[5pt]
 \hline
 Totals & 2216 & 982 & 1234 & $2220_{-199}^{+206}$ & $1159_{-120}^{+117}$ & $1068_{-98}^{+101}$ & $2241_{-170}^{+166}$ & $951_{-112}^{+114}$ & $1284_{-116}^{+124}$ & $2213_{-174}^{+199}$ & $1066_{-99}^{+111}$ & $1148_{-98}^{+105}$ \\
\enddata
\tablecomments{The ``Bluer'' and ``Redder'' columns add up to the ``All'' columns. For each model, the 68.3\% credible intervals are computed from 1000 simulated catalogs passing the (KS) distance threshold (the results from our AD analyses, not shown, are similar but yield somewhat larger uncertainties). While all three models fit the overall (``All'') multiplicity distribution equally well (and produce nearly identical distributions), the linear models produce much better matches to the observed multiplicities of both the bluer and redder halves. We plot ratios of the model columns to the \Kepler{} columns in Figure \ref{fig:models_mult_ratios}.}
\label{tab:mult}
\end{deluxetable*}

\begin{figure}
 \includegraphics[scale=0.45,trim={0.6cm 0.2cm 0.6cm 0},clip]{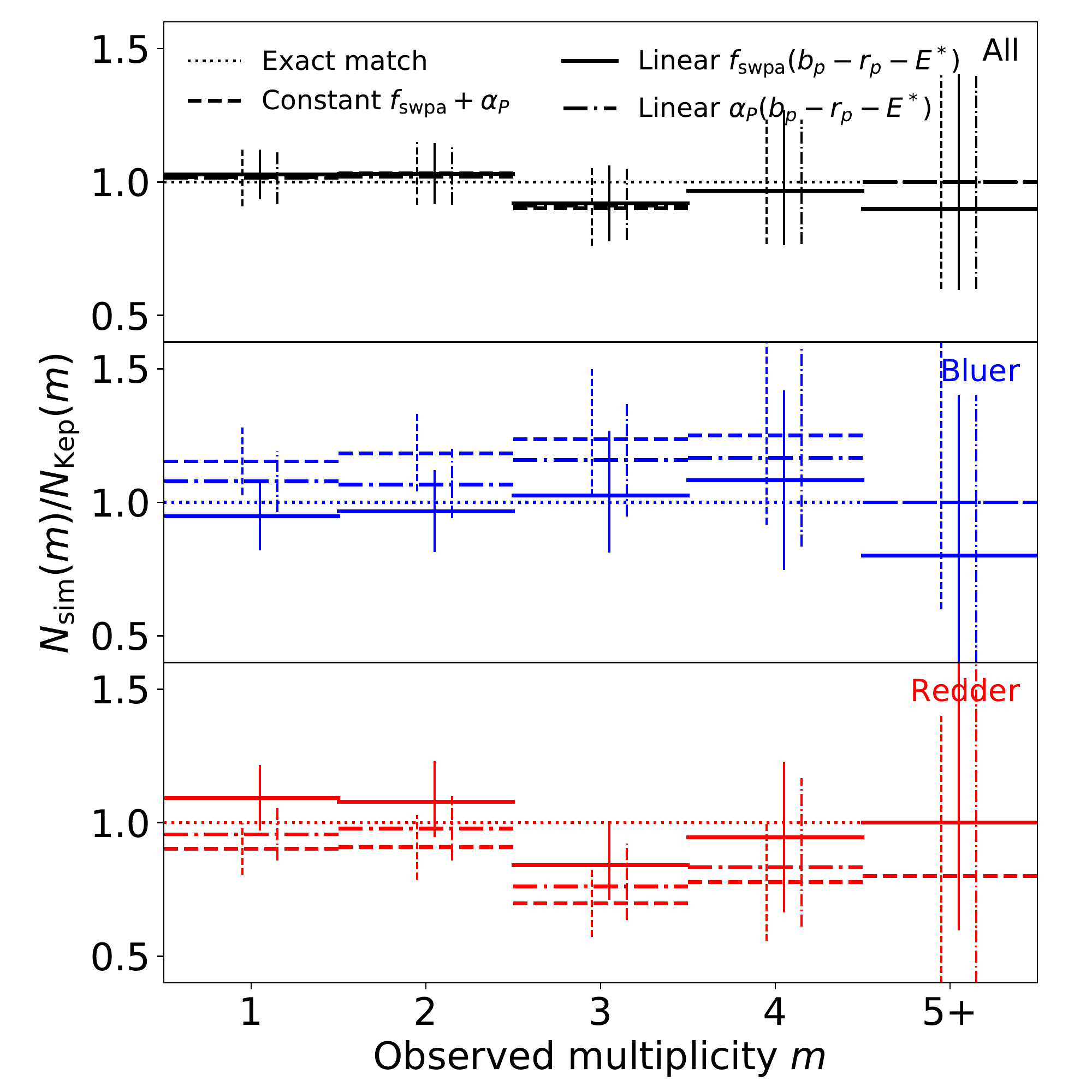}
\caption{Observed multiplicity distributions of our models, normalized by the \Kepler{} multiplicity distribution. The panels from top to bottom include the full sample (``All'') and the bluer and redder halves (respectively labeled and colored). In each panel, the dashed, solid, and dash--dotted lines represent the median multiplicities (normalized to the \Kepler{} counts) for our constant $f_{\rm swpa}+\alpha_P$, linear $f_{\rm swpa}(b_p-r_p-E^*)$, and linear $\alpha_P(b_p-r_p-E^*)$ models, respectively. Vertical error bars denote the 68.3\% credible regions. The multiplicity counts used to generate this figure are also listed in Table \ref{tab:mult}, and result from our KS analysis. We show a dotted horizontal line at $N_{\rm sim}(m)/N_{\rm Kep}(m) = 1$ as a reference for exact matches to the \Kepler{} data.
\newline While all three models fit the overall multiplicity distribution equally well (\textbf{top panel}), the constant $f_{\rm swpa}+\alpha_P$ model significantly overproduces observed systems around the bluer stars and underproduces systems around the redder stars in our sample. Our linear models are a better fit to the observed multiplicities for these two subsets, with the linear $f_{\rm swpa}(b_p-r_p-E^*)$ model providing the best fits.}
\label{fig:models_mult_ratios}
\end{figure}

\section{Discussion} \label{Discussion}

\subsection{The Observed Multiplicity Distribution} \label{Observed_multiplicity}

In Table \ref{tab:mult}, we list the observed multiplicity counts from the \Kepler{} planet catalog and our models (constant $f_{\rm swpa}+\alpha_P$, linear $f_{\rm swpa}(b_p-r_p-E^*)$, and linear $\alpha_P(b_p-r_p-E^*)$), in the total (``All'') sample as well as in the bluer and redder halves.
For the multiplicity counts observed from our models, we also compute and list uncertainties from generating 1000 simulated catalogs that pass our (KS) distance threshold. The results using our AD distance threshold are similar for the median values but provide somewhat larger uncertainties (which we have not listed here). We note that the 68.3\% credible regions we computed include two sources of uncertainty: (1) the Monte Carlo noise due to the finite number of targets used for each simulated catalog (which we set to be equal to the number of targets in our \Kepler{} stellar sample, 88,912 stars), and (2) the differences in the models due to the uncertainties in the model parameters, for which we only keep sets of parameters passing the distance threshold after one realization of a simulated catalog.

To facilitate a more direct comparison between the multiplicity distributions observed in the \Kepler{} catalog and our models, we also plot the ratios of our models to the \Kepler{} counts, $N_{\rm sim}(m)/N_{\rm Kep}(m)$, for each observed planet multiplicity order $m$ in Figure \ref{fig:models_mult_ratios}. We show panels for the total, bluer, and redder samples, computing the ratios of observed model counts to \Kepler{} counts for each sample. In each panel, horizontal bars denote the median ratio values (an exact match is given by a ratio of 1, guided by the dotted line), while vertical lines show the 68.3\% credible intervals (as listed in Table \ref{tab:mult}). The dashed, solid, and dash--dotted lines plot the results from our constant $f_{\rm swpa}+\alpha_P$, linear $f_{\rm swpa}(b_p-r_p-E^*)$, and linear $\alpha_P(b_p-r_p-E^*)$ models, respectively. Systems with $m \geq 5$ observed planets are binned together here because of their low counts (the same multiplicity binning is also done in our distance function when computing $D_{\rm mult}$ in equation \ref{eq_dist}).

The models explored in this paper fit the overall multiplicity distribution equally well. This is expected, as \citetalias{HFR2019} showed that a clustered model (with two populations of mutual inclinations) is necessary and fits the observed multiplicity distribution extremely well; the models in this paper are extensions of that model. For almost all multiplicity orders $m$, the median observed counts from our models very closely match the \Kepler{} count, and the latter falls within our 68.3\% credible intervals. The biggest difference is the number of triples ($m = 3$), as more triple systems are observed in the data (101, given our period--radius range and stellar sample) than what our models produce. However, the number is still within the 68.3\% credible intervals ($91_{-14}^{+16}$, $93_{-14}^{+14}$, and $92_{-13}^{+14}$ for our constant $f_{\rm swpa}+\alpha_P$, linear $f_{\rm swpa}(b_p-r_p-E^*)$, and linear $\alpha_P(b_p-r_p-E^*)$ models, respectively).

The need for a stellar--dependent $f_{\rm swpa}$ (or $\alpha_P$) is clear when considering the bluer and redder halves of our FGK dwarf sample. Our baseline (constant $f_{\rm swpa}+\alpha_P$) model consistently overproduces the rate of observed planetary systems around bluer stars and underproduces the rate around redder stars. This trend is significant at all observed multiplicity orders except the higher ones at $m = 4$ and $m \geq 5$, where the relatively low counts reduce the statistical power to distinguish between the two models. 
On the other hand, each of our linear models provides a significantly better match to the observed multiplicity distributions around both the bluer and redder halves compared to the constant model. The decreasing relation of $\alpha_P$ with color can appear to account for the observed rates of planets as a function of spectral type, in a similar manner as increasing the $f_{\rm swpa}$. However, the linear $f_{\rm swpa}(b_p-r_p-E^*)$ model (solid lines) appears to provide an overall slightly closer match than the linear $\alpha_P(b_p-r_p-E^*)$ model (dash--dotted lines). Interestingly, while the number of observed triples is a near exact match in the bluer sample, there is a higher occurrence of triples in the redder sample than what our model produces. Nevertheless, the excellent similarity between the multiplicity distributions of this model and the \Kepler{} data for both the bluer and redder halves is remarkable, given that our model assumes a very simple stellar dependence, the fraction of stars with planets as a simple linear function of the Gaia $b_p-r_p-E^*$ color, and that we have \textit{not} made any other parameters of the model depend on the stellar properties.

\begin{figure}
 \includegraphics[scale=0.42,trim={0 0.4cm 0 0},clip]{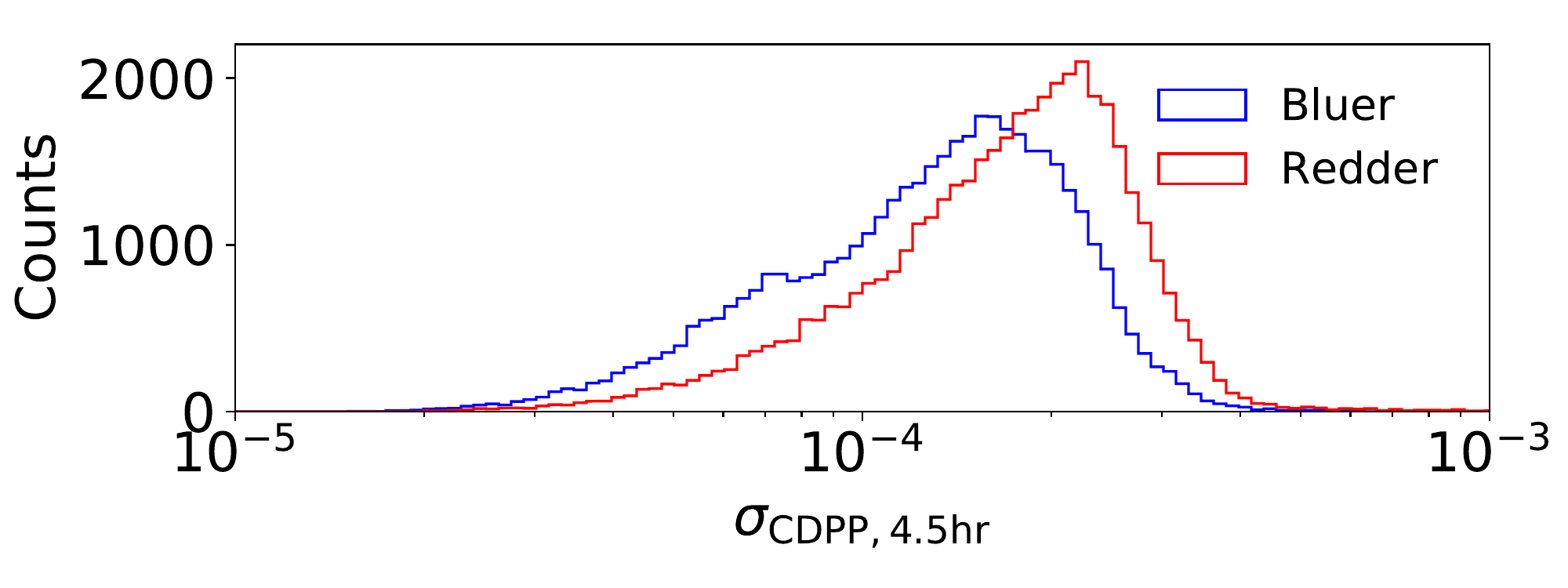}
 \includegraphics[scale=0.42,trim={0 0.4cm 0 0},clip]{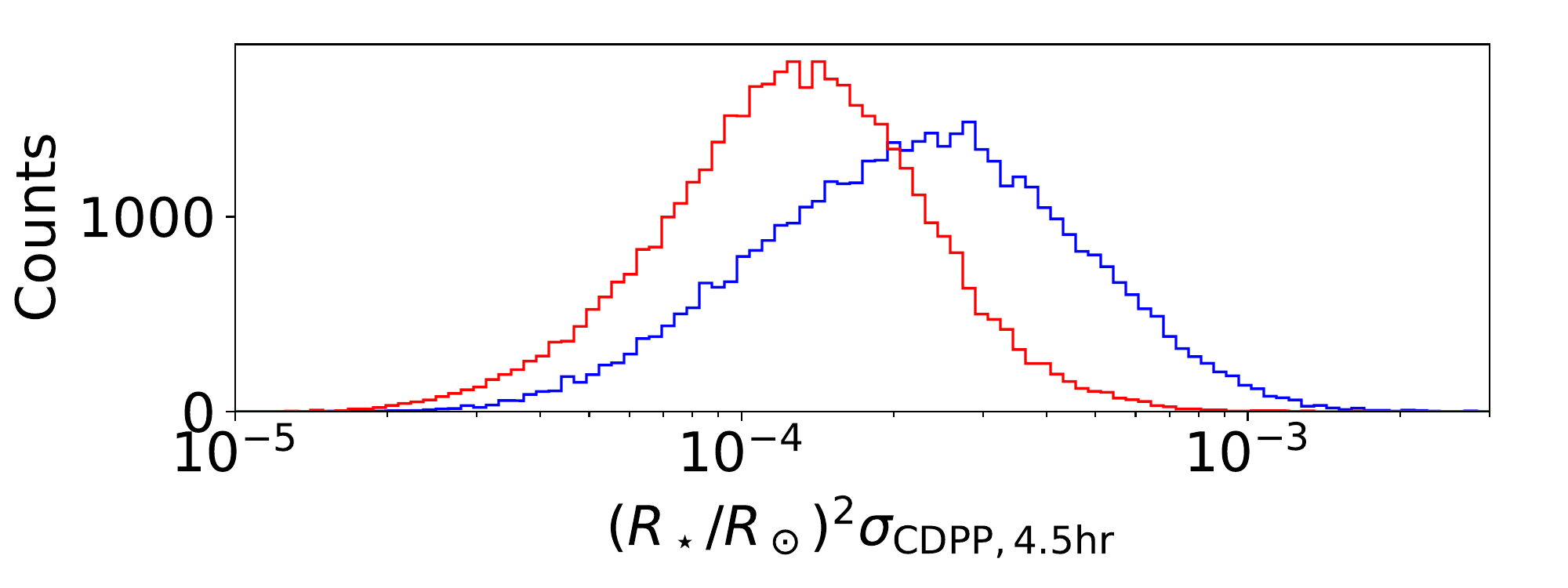}
 \includegraphics[scale=0.42,trim={0 0.4cm 0 0},clip]{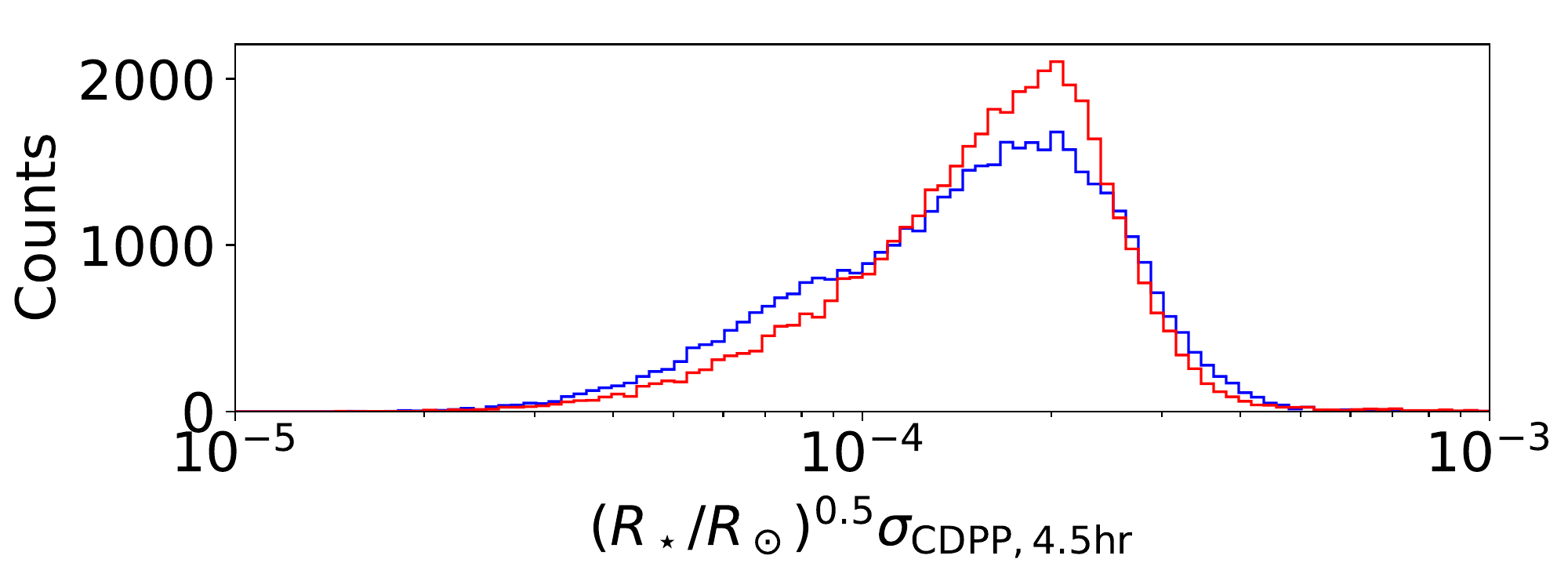}
\caption{\textbf{Top panel:} histograms of $\sigma_{\rm CDPP,4.5hr}$, a measure of photometric precision, for all 88,912 stars in our sample. Bluer stars are brighter and thus tend to have better (lower) photometric precision, making it easier to recover transits of a given depth.
\textbf{Middle panel:} histograms of $(R_\star/R_\odot)^2 \sigma_{\rm CDPP,4.5hr}$, related to the S/N of a single point in transit. Bluer stars are also larger in size, meaning that planets of a given size have much diminished transit depths ($\sim 1/R_\star^2$) leading to overall lower values of single--measurement S/N.
\textbf{Bottom panel:} histograms of $(R_\star/R_\odot)^{0.5} \sigma_{\rm CDPP,4.5hr}$. While larger stars cause shallower transit depths of a given size of planet, they also have greater geometric transit probability ($\sim R_\star$) and induce longer transit durations ($\sim R_\star^{0.5}$). Combining all four effects (CDPP, transit depth, transit probability, and transit duration), we have the most relevant quantity, $(R_\star/R_\odot)^{0.5} \sigma_{\rm CDPP,4.5hr}$ (a ``transit geometry--weighted S/N''). Finally, the planet radius distribution rises toward smaller sizes, meaning that the total planet yield is primarily influenced by the detection threshold of the ``best'' stars. Since there are more bluer stars than redder stars in our sample with small values of the transit geometry--weighted S/N (e.g., $R_\star^{0.5} \sigma_{\rm CDPP,4.5hr} \lesssim 10^{-4}$; 11,047 bluer vs. 8147 redder stars), we expect that \Kepler{} would detect more planets around stars from the bluer half than the redder half, if the distributions of planetary systems were the same for all stars (i.e. our constant $f_{\rm swpa}+\alpha_P$ model). In contrast, the actual catalog of \Kepler{} planet candidates includes more planets around redder FGK stars. This provides evidence supporting an increased occurrence rate of inner planetary systems around redder host stars.}
\label{fig:Rstar_cdpp_bprp}
\end{figure}

\subsection{Why Does Our Constant Model Produce More Detected Planets Around Hotter Stars?}

As we show in Table \ref{tab:mult}, splitting our \Kepler{} sample of stars into two equal--sized halves based on their Gaia $b_p-r_p-E^*$ colors includes more observed planets around redder stars than around bluer stars, at almost all multiplicity orders (except $m = 5$, although the total counts at high $m$ are small). On first glance, it is unclear if the higher rate of observed planets around redder stars is due to an observational bias in favor of smaller, cooler stars, or an inherent increase in the underlying planet occurrence rate, or some combination of both. However, in this paper we showed with our forward model that we get a modestly higher overall rate of observed planets around stars in the bluer half than in the redder half if we assume the same distribution of planetary systems (i.e. our baseline, constant $f_{\rm swpa}+\alpha_P$ model): $1159_{-120}^{+117}$ versus $1068_{-98}^{+101}$ planets in total for the bluer and redder samples, respectively. 
Indeed, the tension between these counts and that of the \Kepler{} data is why we require a dependence on the stellar color/effective temperature in order to produce the number of observed planets in both stellar samples. So why is it seemingly easier to detect planets of a given size and period around hotter stars?

On one hand, the larger sizes of bluer (hotter) stars provide one clear disadvantage for detecting transiting planets of a given size, since to first order the transit depth is given by the fraction of the stellar disk area blocked by the planet, $\delta \simeq (R_p/R_\star)^2$; that is, it is inversely proportional to the stellar radius squared. On the other hand, planets are more likely to transit larger stars simply due to the increased geometric transit probability ($\sim R_\star/a$). Larger stars may also induce longer transit durations, due to the distance a planet must travel in order to cross the stellar disk ($t_{\rm dur} \sim R_\star$ to first order).
Another factor contributing to the increased detection efficiency of planets around bluer stars in the \Kepler{} mission is their improved photometric precision over that of redder stars, due to their brightness.

To illustrate how each of these factors affect the relative detectability of planets around bluer versus redder stars in our FGK sample, we show a sequence of histograms of relevant quantities involving stellar radius and the root mean square Combined Differential Photometric Precision \citep{C2012}, $\sigma_{\rm CDPP}$ (4.5 hr duration for simplicity), for our bluer and redder stellar samples in Figure \ref{fig:Rstar_cdpp_bprp} (including all stars, not just known planet hosts, in our sample). The distribution of $\sigma_{\rm CDPP,4.5hr}$ for the bluer stars is shifted toward lower values than that of the redder stars, making it easier to detect transits of a given depth (top panel). Including the dependence on stellar radii of the transit depth, we have $R_\star^2 \sigma_{\rm CDPP,4.5hr}$, a ``single--measurement signal--to--noise ratio (S/N)''. 
This distribution (middle panel) shows that for a transiting planet with fixed size and transit duration, the decrease in transit depth is typically greater than the improvement in $\sigma_{\rm CDPP,4.5hr}$, thus favoring detections around redder stars. Factoring in the transit duration to the S/N ($\sim \sqrt{t_{\rm dur}}$) and the geometric transit probability ($\sim R_\star$) gives a ``transit geometry--weighted S/N'' that is proportional to $R_\star^{0.5} \sigma_{\rm CDPP,4.5hr}$ (bottom panel). While the distributions of the bluer and redder samples are quite similar, the lower tail of the distribution ($R_\star^{0.5} \sigma_{\rm CDPP,4.5hr} \lesssim 10^{-4}$) includes more bluer stars. 
This means that the rate of detections for the smallest detectable planets is enhanced for the bluer stars in our sample. Finally, since the occurrence of planets increases toward smaller sizes (e.g., we find $\alpha_{R1} \simeq -1.4$ for the power--law index of small planets $< 3 R_\oplus$ for the models presented in this paper), the smallest detectable planets contribute most to the overall rate of observed planets. This confirms our finding with the full model that if the fraction of stars with planets were independent of stellar color, that we would find somewhat more planets around bluer stars than redder stars (as in Table \ref{tab:mult}).

\subsection{Are Orbital Eccentricities Correlated with Stellar Type?} \label{secEcc}

In order to briefly explore correlations between stellar type and other parameters in our model, we also test a model in which the eccentricity scale $\sigma_e$ is allowed to vary as a function of $b_p-r_p-E^*$. For this analysis, we use the same procedure as described in \S\ref{Methods}, except we adopt the circular--normalized transit duration ($t_{\rm dur}/t_{\rm circ}$) in place of the transit duration ($t_{\rm dur}$) when computing the distance. The distribution of $t_{\rm dur}/t_{\rm circ}$ is more sensitive to the intrinsic eccentricity distribution, but requires the stellar properties (e.g. mean density) to be very well characterized \citep[e.g.,][]{M2011, PBC2014, vEA2015, X2016}. While we adopt stellar radii from \Gaia{} DR2, the stellar masses remain from \Kepler{} DR25; for this reason, we have avoided using $t_{\rm dur}/t_{\rm circ}$ in the analyses for all the other models. Nevertheless, we perform a cursory analysis to see if there are any clear trends between eccentricity and spectral type.

For simplicity, we also assume a linear relation parameterized by a slope $d\sigma_e/d(b_p-r_p-E^*)$ and $y$-intercept $\sigma_{e,\rm med}$ (at median color, $b_p-r_p-E^* \simeq 0.81$; analogous to the form of equation \ref{eq_fswp_bprp}).
We allow for $d\sigma_e/d(b_p-r_p-E^*)$ to vary in $[-0.1,0.1]$ and $\sigma_{e,\rm med}$ in $[0,0.1]$. We find no clear trend between $\sigma_e$ and $b_p-r_p-E^*$ in this analysis; while $\sigma_{e,\rm med}$ is constrained to similarly low values of $\sim 0.03$, the slope $d\sigma_e/d(b_p-r_p-E^*)$ takes on both positive and negative values. Interestingly, there appears to be a slight preference for positive or negative slopes compared to zero slope. We interpret this as suggesting that there is evidence for a higher-eccentricity population of exoplanets, but that this is not dependent on the host star color.

\begin{figure*}
\centering
\begin{tabular}{cc}
 \includegraphics[scale=0.43,trim={0 0.5cm 0 0.2cm},clip]{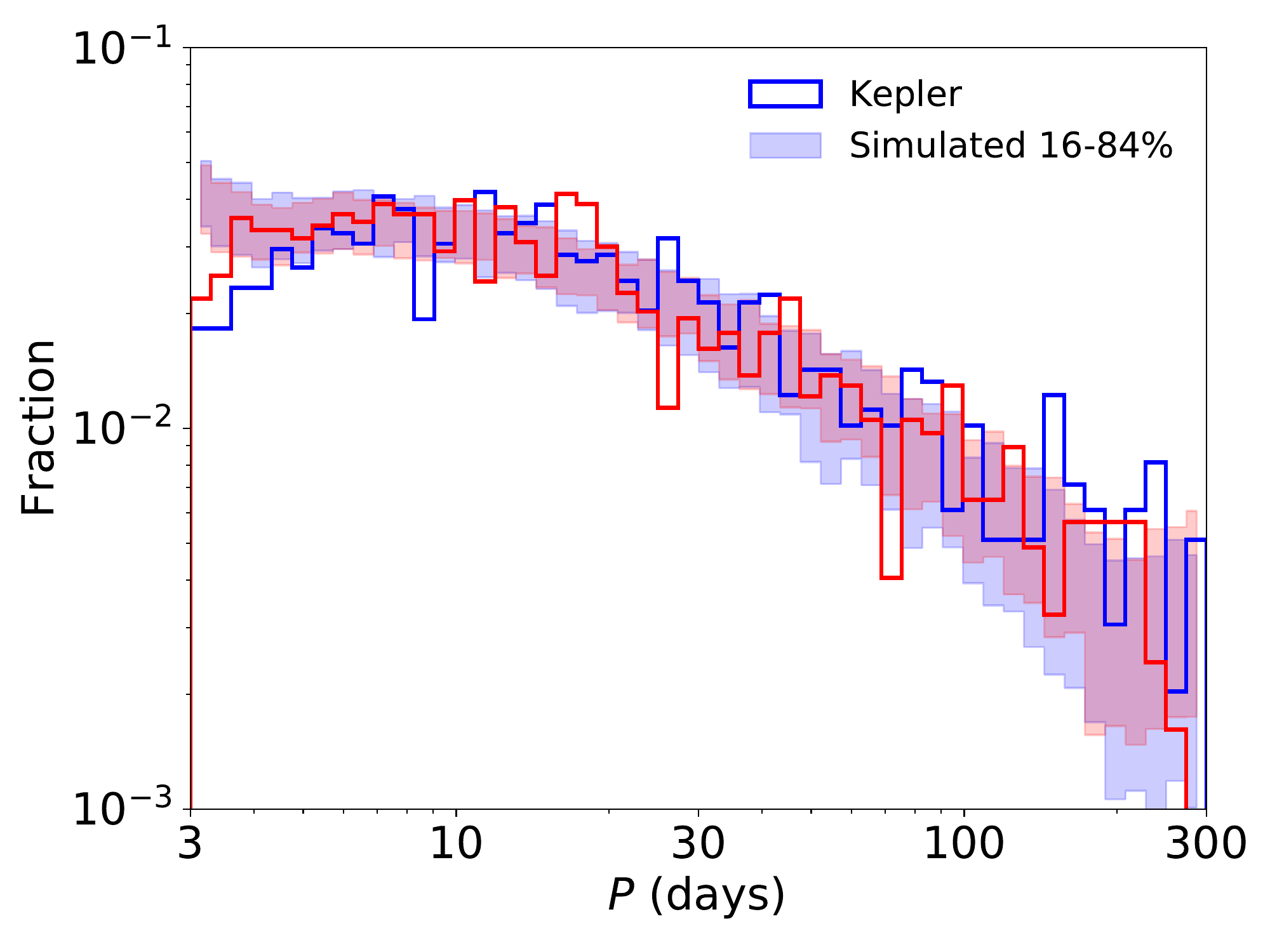} &
 \includegraphics[scale=0.43,trim={0 0.5cm 0 0.2cm},clip]{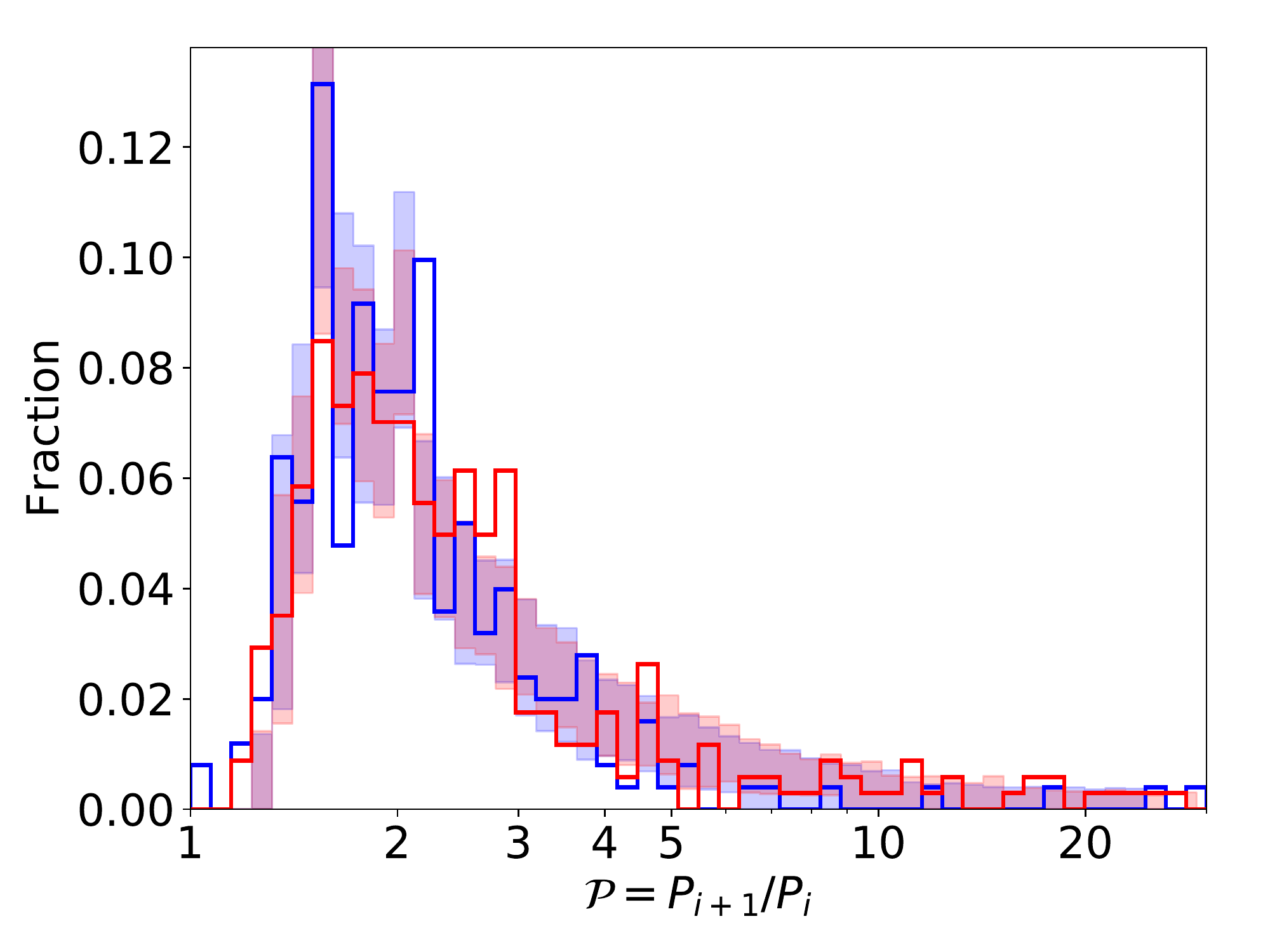} \\
 \includegraphics[scale=0.43,trim={0 0.5cm 0 0.2cm},clip]{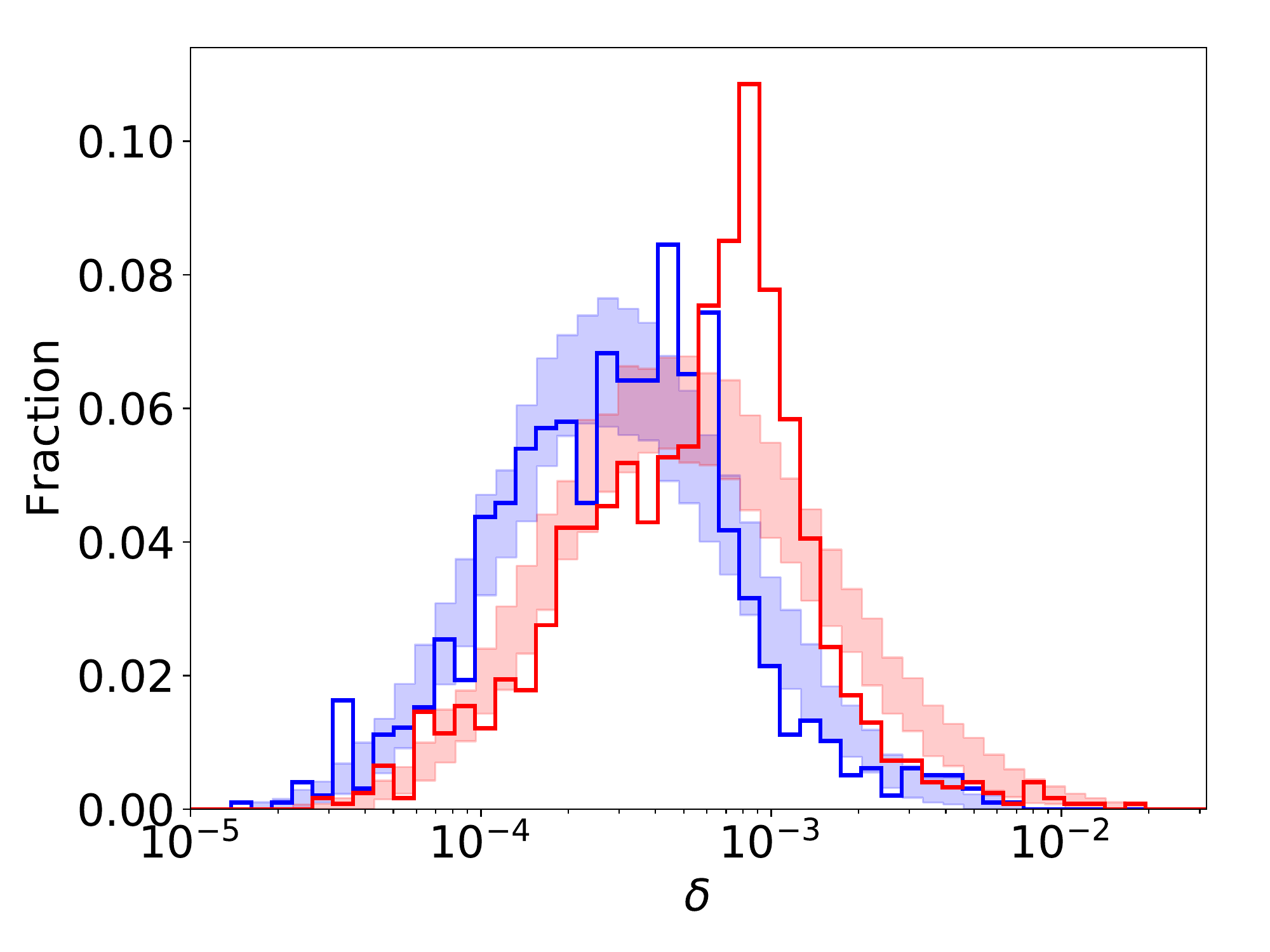} &
 \includegraphics[scale=0.43,trim={0 0.5cm 0 0.2cm},clip]{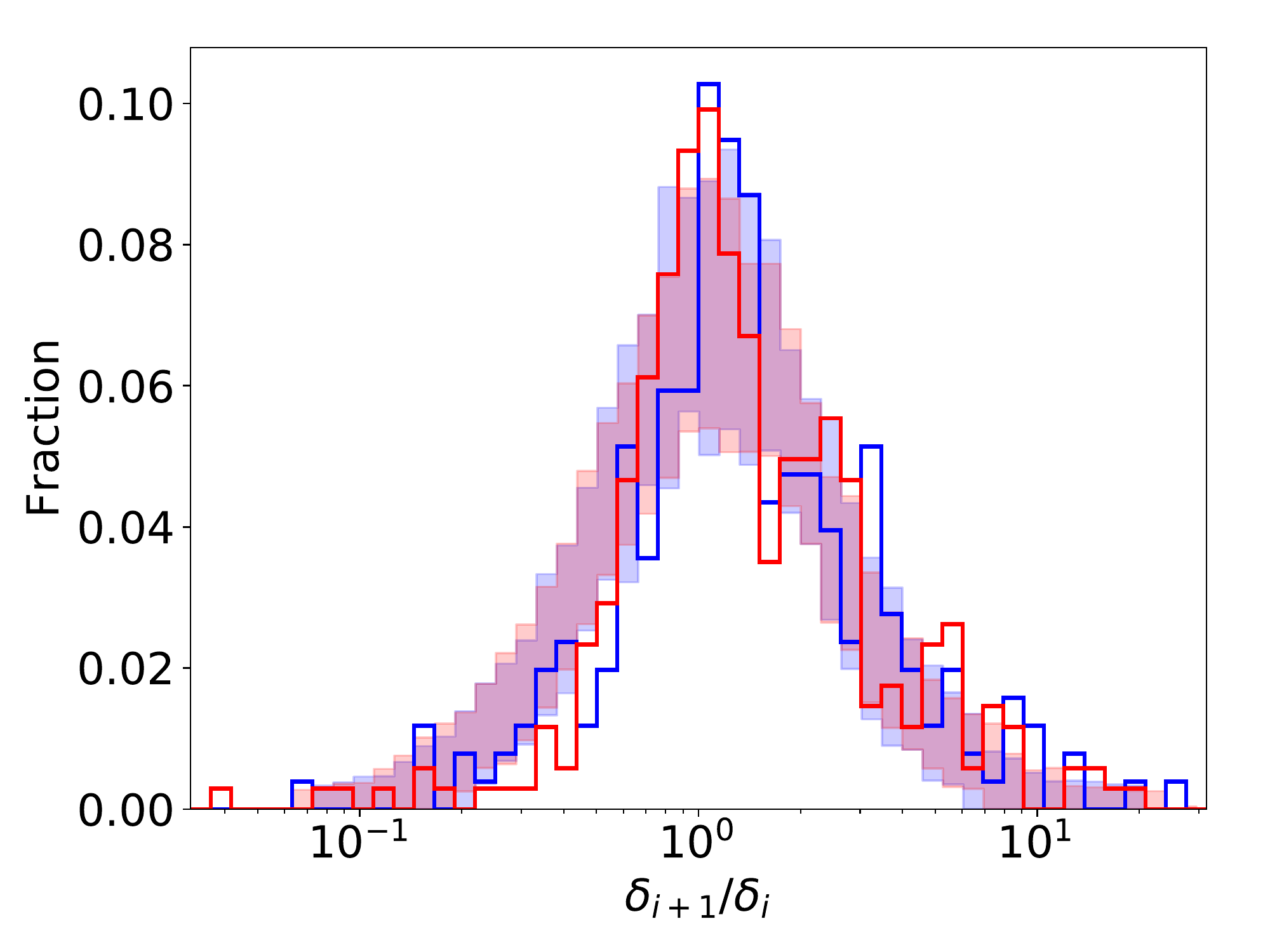} \\
 \includegraphics[scale=0.43,trim={0 0.5cm 0 0.2cm},clip]{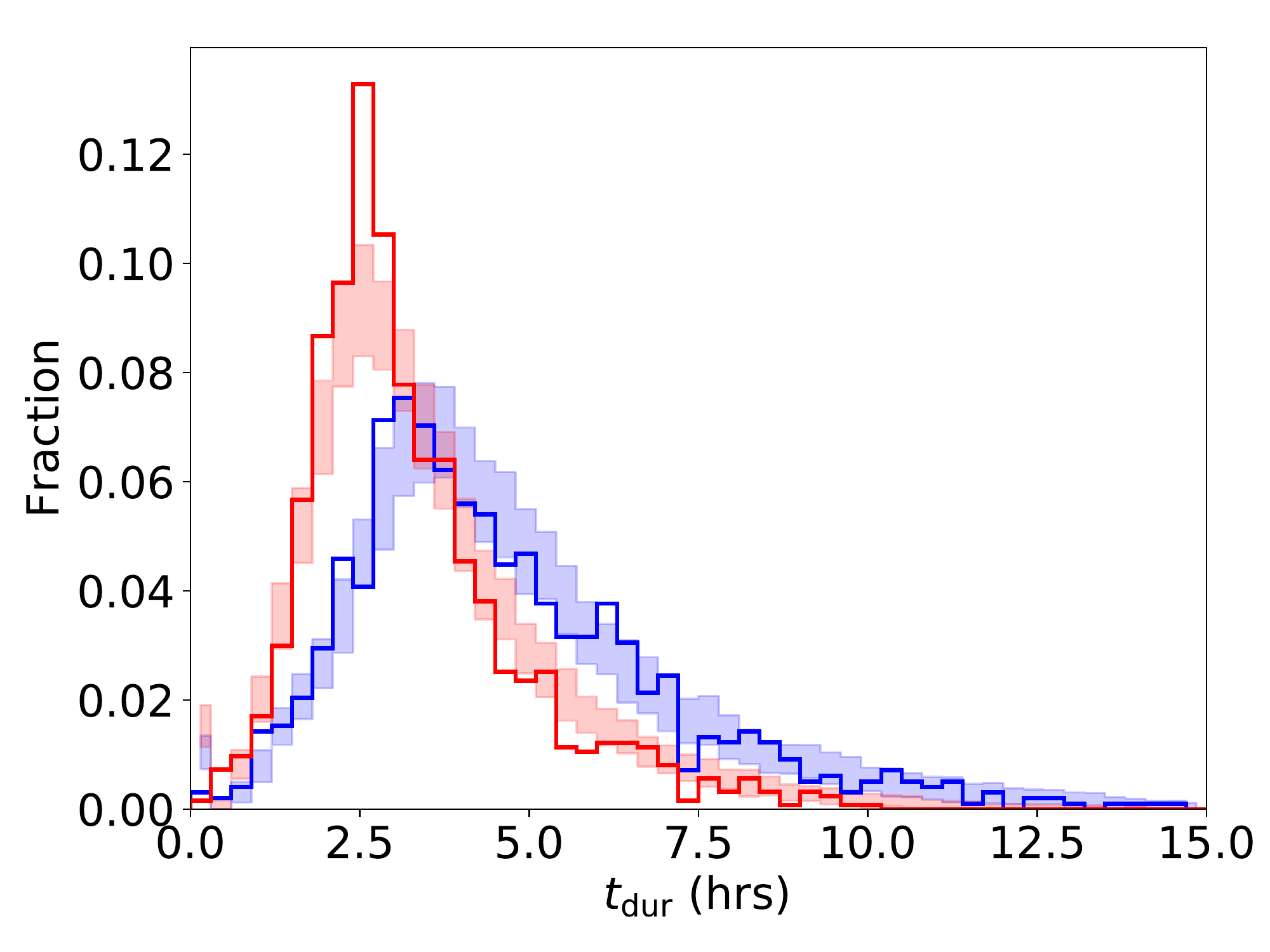} &
 \includegraphics[scale=0.43,trim={0 0.5cm 0 0.2cm},clip]{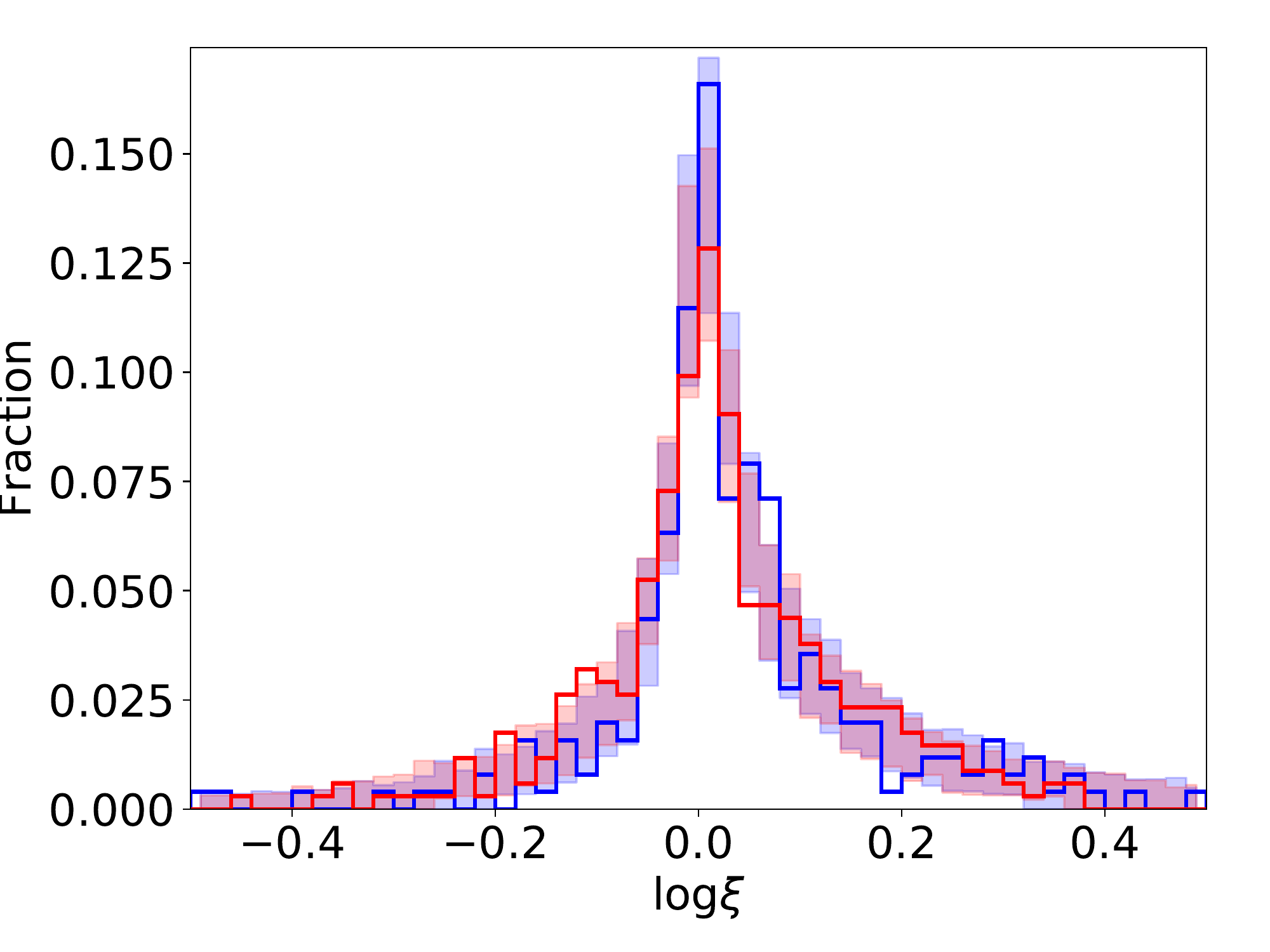} \\
\end{tabular}
\caption{Marginal distributions of the observable properties for our linear $f_{\rm swpa}(b_p - r_p - E^*)$ model as compared to the \Kepler{} data, split into bluer and redder halves as likewise colored. In each panel, the shaded regions denote the 16th to 84th percentiles of the model, while the solid line histograms denote the \Kepler{} data. \textbf{Left-hand panels:} from top to bottom, histograms of observed periods ($P$), transit depths ($\delta$), and transit durations ($t_{\rm dur}$). \textbf{Right-hand panels:} from top to bottom, histograms of observed period ratios ($\mathcal{P}$), transit depth ratios ($\delta_{i+1}/\delta_i$), and period--normalized transit duration ratios ($\xi$) for adjacent planet pairs in each system.}
\label{fig:model_fswp_bprp_split}
\end{figure*}

\subsection{Other Correlations in Planetary System Architectures with Stellar Type}

The models presented in \citetalias{HFR2019} and in this paper are driven by fits to the marginal distributions of a collection of key observables for the \Kepler{} DR25 catalog of exoplanet candidates, as listed in \S\ref{SummaryStats}. As discussed in \S\ref{Obs}, this study adopted a distance function that incorporates simultaneous fits to these summary statistics in the total, bluer, and redder samples in order to infer the best-fitting model parameters. While we did not explicitly explore how the planetary system properties vary with color for all of our other architectural model parameters, we can examine fits to the observed data for our linear $f_{\rm swpa}(b_p-r_p-E^*)$ (and linear $\alpha_P(b_p-r_p-E^*)$) models in order to discern possible differences in the architectures between the two samples. We note that we also attempted to fit the constant $f_{\rm swpa}+\alpha_P$ model \textit{to each of the bluer and redder samples}, independently, but found that we could not adequately constrain all of the model parameters, including the differences in $f_{\rm swpa}$. This is perhaps unsurprising, due to the reduced inference power of using half the data (or another way to think about it is that there are twice as many free parameters to describe the same data). Nevertheless, in this analysis we did not find any evidence for clear differences in the other parameters (i.e. the mean rates of clusters and planets per cluster, eccentricity and mutual inclination scales, and clustering scales in periods and radii) between the bluer and redder halves, suggesting that planetary system architectures are quite similar across all main sequence FGK spectral types.

We plot the marginal distributions of observed properties from our catalogs using our best--fitting linear $f_{\rm swpa}(b_p-r_p-E^*)$ model in Figure \ref{fig:model_fswp_bprp_split} (with the parameters listed in Table \ref{tab:param_fits}), where the model and the \Kepler{} data are split into the bluer and redder halves. A similar plot for the observed catalogs from our linear $\alpha_P(b_p-r_p-E^*)$ model is provided in Appendix Figure \ref{fig:model_alphaP_bprp_split}. The panels from left to right and top to bottom show the observed distributions of orbital periods ($P$), period ratios ($\mathcal{P}$), transit depths ($\delta$), transit depth ratios ($\delta_{i+1}/\delta_i$), transit durations ($t_{\rm dur}$), and period-normalized transit duration ratios ($\log\xi$). In each panel, the shaded regions denote the central 68.3\% credible interval (in each histogram bin) from 100 simulated catalogs passing the (KS) distance threshold, while the solid line histograms show the \Kepler{} population given our sample. Figures \ref{fig:dists_KS} and \ref{fig:dists_AD} in the Appendix show how $10^3$ simulated catalogs passing the distance thresholds for KS and AD, respectively, compare to the \Kepler{} catalog in terms of the individual (weighted) distance terms.

Overall, our linear $f_{\rm swpa}(b_p-r_p-E^*)$ model provides an excellent fit to the marginal distributions of the observed \Kepler{} planet candidates, for both the bluer and redder samples as well as the combined sample. The best distances are as low as $\mathcal{D}_{W,\rm KS} \simeq 10$ in both the bluer and redder samples and $\mathcal{D}_{W,\rm KS} \simeq 12$ for the total sample; for comparison, a ``perfect'' model would result in $\mathcal{D}_{W,\rm KS} \simeq 9$ for each of these samples given our number of distance terms. While many of the marginal distributions are fit nearly perfectly (i.e. have individual weighted distances around unity, given Monte Carlo noise), there are some differences between the \Kepler{} data and our model predictions, both subtle and significant (e.g. the transit depth and depth ratio distributions).
The bulk of the period distribution is well modeled, although there is a deviation at short periods suggesting a need for a more complicated model than the single power-law we have adopted (e.g., \citealt{H2012, Mu2015, Mu2018}). 
There are fewer planets at the shortest orbital periods we investigated ($\sim 3$ d) around the bluer stars than around the redder stars (note the log $y$-scale) and the fit to the period distribution is worse, suggesting that the inner edge of planetary systems may be stellar dependent as also found by \citet{PBC2014} and \citet{Mu2015}. Indeed, this is likely driving the results of our linear $\alpha_P(b_p-r_p-E^*)$ model; the period distribution is a slightly better fit for this model compared to the linear $f_{\rm swpa}(b_p-r_p-E^*)$ model, although only for the bluer sample. The fit to the period ratios is nearly identical for both linear models.

The transit depth distribution is very different between our two stellar samples as expected because of the different distributions of stellar radii and photometric precision, but the fit to the redder half is clearly worse than that of the bluer half, due to an apparent excess of transit depths at $\sim 10^{-3}$ in the data. On the other hand, the transit depth ratio distribution is very similar between the two stellar samples (in both the data and our models). As in \citetalias{HFR2019}, we note that while our simple clustering in planet sizes is necessary to explain the highly peaked nature of the distribution, it is not sufficient to explain the asymmetry caused by larger planets being more often the outer planet in observed adjacent pairs (\citealt{C2013, W2018a, GF2020}).
The intrinsic planet radius distribution is sculpted by the processes of photoevaporation \citep{OW2013, F2017, OW2017, vE2017, C2018} and/or heating from formation (e.g. core-powered mass-loss; \citealt{GSS2016, GSS2018, GS2018}), which our broken power-law with clustering for planet sizes cannot fully encapsulate.

The transit duration distributions differ between the bluer and redder halves as expected due to the dependencies on stellar radius and density. There are some tradeoffs in the fits to the $t_{\rm dur}$ distributions for each half between the models. While the linear $f_{\rm swpa}(b_p-r_p-E^*)$ model improves the fits to both samples compared to the constant $f_{\rm swpa}+\alpha_P$ model, it is intriguing that the linear $\alpha_P(b_p-r_p-E^*)$ model provides the best fit to the redder distribution and the worst fit to the bluer distribution (even compared to the constant model), which is a reversal to the period distribution fits. In any case, the linear $f_{\rm swpa}(b_p-r_p-E^*)$ model significantly improves the fit to the overall distribution over the constant $f_{\rm swpa}+\alpha_P$ or linear $\alpha_P(b_p-r_p-E^*)$ models. This is easily explained by the overall better matches to the frequency of observed planets of the bluer and redder halves in our linear $f_{\rm swpa}(b_p-r_p-E^*)$ model.
The constant $f_{\rm swpa}+\alpha_P$ model provides too many planets around the bluer stars, which have a different distribution than that of the planets around the redder stars, while the linear $f_{\rm swpa}(b_p-r_p-E^*)$ model provides just the right contributions. Finally, we see no difference in the fits to the period-normalized transit duration ratio distributions, which are modeled extremely well.

\section{Conclusions} \label{Conclusions}

We have extended our forward modeling methodology from \citet{HFR2019} (\citetalias{HFR2019}) to explore how the occurrence of planetary systems and their architectures may vary as a function of their host stars, using the \Gaia{} DR2 $b_p - r_p$ colors as a proxy for stellar effective temperature (i.e. spectral type). We use a clean sample of 88,912 \Kepler{} FGK dwarfs (a stellar catalog similar to the one defined in \citealt{H2019}, but with an additional reddening correction) and the \Kepler{} DR25 candidates with periods of $[3, 300]$ days and radii of $[0.5, 10] R_\oplus$ (the same range as explored in \citetalias{HFR2019}), and we define $f_{\rm swpa}$ to refer to the fraction of stars that host at least one planet in this range.

First, we adopt the clustered periods and sizes model from \citetalias{HFR2019} and re-parameterize the intrinsic multiplicity distribution by decoupling the fraction of stars with planets ($f_{\rm swpa}$) from the number of clusters and planets per cluster, to form our baseline (constant $f_{\rm swpa}+\alpha_P$) model. We show that this does not change our main results about the architectures of planetary systems, namely the underlying power-laws for the period and radius distributions, the extent of intra-cluster similarity in periods and planet radii, the role of two populations of multi-planet systems with different mutual inclination scales in order to explain the observed \Kepler{} dichotomy, and the overall $f_{\rm swpa}$ marginalized over all FGK stars.

Then, we generalize our baseline clustered model to test linear dependencies on stellar color for several model parameters, including the fraction of stars with planets ($f_{\rm swpa}$), the period power--law index ($\alpha_P$), and the eccentricity scale ($\sigma_e$). We use the \Gaia{} DR2 $b_p-r_p$ colors as a proxy for spectral type and correct for differential reddening using a simple model for $E^* \sim E(b_p-r_p)$. By splitting our stellar sample of FGK stars into two halves (bluer and redder) based on their $b_p-r_p-E^*$ colors, modifying our distance function to fit to the observed marginal distributions of both halves and the overall sample, and performing model inference using ABC, we find the following results:
\begin{itemize}[leftmargin=*]
 \item For our stellar and planet sample, dividing the stars into two equal--sized halves at the median color results in more \Kepler{} planets being observed around redder stars than around bluer stars. The relative counts are shaped by a combination of observational biases (arising from differing stellar properties, photometric precision, and so on) and intrinsic planet occurrence rates.
 \item The smallest sized planets are easier to detect around earlier type (hotter and bluer) stars in the \Kepler{} mission. While redder stars are smaller in size and thus induce larger transit depths for a given size of planet, bluer stars benefit from causing increased transit durations, geometric transit probability, and better photometric precision. Our forward model accounts for all of these effects. Assuming a constant rate of planetary systems across all stars (i.e., our constant $f_{\rm swpa}+\alpha_P$ model) produces more detected planets in the bluer sample and fewer in the redder sample, significantly at odds with the \Kepler{} counts.
 \item The two points above imply that the overall occurrence of planets increases toward later type dwarfs (cooler and redder stars), in agreement with the general trends reported in \citet{H2012} and \citet{Mu2015}. We find that this increase in planet occurrence is well described by a change in the fraction of stars with planets ($f_{\rm swpa}$), similar to the findings of \citet{YXZ2020}. Assuming a linear trend with $b_p-r_p-E^*$ color, we find a significant positive slope of $df_{\rm swpa}/d(b_p-r_p-E^*) = 0.84_{-0.35}^{+0.37}$ ($1.15_{-0.36}^{+0.35}$) using KS (AD) analyses. This implies that there is a substantial difference in the fraction of stars hosting planetary systems across FGK stars: $f_{\rm swpa} = 0.32_{-0.11}^{+0.12}$ for F2V dwarfs and $f_{\rm swpa} = 0.96_{-0.19}^{+0.04}$ for mid K dwarfs. The solar value is roughly half: $f_{\rm swpa} = 0.57_{-0.10}^{+0.14}$ for G2V. While our linear relation is likely an oversimplification at the extreme ends of our sample, extrapolating to later type stars suggests that planetary systems are ubiquitous around early M-dwarfs, as has been shown previously by \citet{BJ2016}. Note that this is in addition to the high rate of planets per M-dwarf star, as suggested by \citet{DC2013, Mu2015, Hu2019}.
 \item We verify that the increase in $f_{\rm swpa}$ toward later type stars is robust by exploring a step function in which the fraction of stars with planets is a constant $f_{\rm swpa,bluer}$ below and $f_{\rm swpa,redder}$ above the median color. In this step $f_{\rm swpa}$ model, we find a significant difference, with $f_{\rm swpa,redder}-f_{\rm swpa,bluer} = 0.20 \pm 0.09$ ($0.31 \pm 0.12$) using KS (AD) analyses. The higher $f_{\rm swpa}$ for redder stars is fully consistent with the results of our linear $f_{\rm swpa}(b_p-r_p-E^*)$ parameterization.
 \item We test an alternative explanation for the increased planet occurrence toward later types, by considering a change in the period power--law distribution, with a shallower rise in occurrence toward longer periods for planets around redder stars compared to bluer stars (characterized by a negative $d\alpha_P/d(b_p-r_p-E^*)$ where $\alpha_P$ is the power--law index). We find that this trend is only necessary when we hold $f_{\rm swpa}$ fixed with color, but disappears when simultaneously allowing for linear functions of $b_p-r_p-E^*$ for both $f_{\rm swpa}$ and $\alpha_P$. In either case, the strong positive slope of $df_{\rm swpa}/d(b_p-r_p-E^*)$ remains, strengthening the results of our linear $f_{\rm swpa}(b_p-r_p-E^*)$ model.
 \item While both linear $f_{\rm swpa}(b_p-r_p-E^*)$ and $\alpha_P(b_p-r_p-E^*)$ models improve the fits to the observed planet multiplicity distributions of our bluer and redder samples over the constant $f_{\rm swpa}+\alpha_P$ model, the linear $f_{\rm swpa}(b_p-r_p-E^*)$ model provides the best fit.
 \item The other architectural model parameters do not change significantly when including the stellar dependencies in our clustered model. Our constant $f_{\rm swpa}+\alpha_P$, linear $f_{\rm swpa}(b_p-r_p-E^*)$, and linear $\alpha_P(b_p-r_p-E^*)$ models result in similar rates of clusters and planets, radius distributions, eccentricity scales, and evidence for two populations of mutual inclinations.
 \item We find no clear correlation between orbital eccentricity $\sigma_e$ and spectral type, although there may be some evidence for a population of planets with more highly eccentric orbits than our single Rayleigh distribution with $\sigma_e \simeq 0.02$.
\end{itemize}

Our findings have consequences for informing future follow-up observations of exoplanet detections from the Transiting Exoplanet Survey Satellite (TESS) mission, which has completed its two-year primary mission \citep{Ri2015, S2015, S2018} and is poised for extended missions \citep{B2017, Hu2018}. With already over 1000 planet candidates collected, the TESS mission is expected to discover many more short-period ($\sim 10$ d) planets around nearby stars that are most amenable to RV follow-up \citep{BPQ2018, S2018}. Being a magnitude limited survey, the TESS mission will observe many more brighter targets (e.g. F stars, compared to later types) in its full field images. Our results show that while these nearby bright stars may be more tenable for transit recovery, the intrinsic rate of inner planetary systems is relatively low for these bluer stars and increases significantly toward later type stars. Thus, follow-up efforts should also target these fainter stars as multi-planet systems around such hosts are common \citep{BJ2016, B2019}. In any case, the primary and extended missions of TESS will likely boost our catalogs of planet candidates around a wide variety of stellar types, further enabling new studies on the architectures of planetary systems as a function of host star properties.

Additional planet companions in systems with short period transiting planets discovered by TESS can also provide stronger constraints on the mutual inclination distribution of multi-planet systems. This may allow future studies to differentiate between a dichotomous model, such as the one considered in this study, and other competing models \citep{Z2018, ZCH2019}.

The new catalogs generated from our models are available to the public, along with the core SysSim code (\url{https://github.com/ExoJulia/ExoplanetsSysSim.jl}), inputs collated from numerous data files (\url{https://github.com/ExoJulia/SysSimData}), and the code specific to the clustered models (\url{https://github.com/ExoJulia/SysSimExClusters}). We encourage other researchers to contribute model extensions via Github pull requests and/or additional public git repositories.

\acknowledgments

We thank the entire \Kepler{} team for years of work leading to a successful mission and data products critical to this study.  
We acknowledge many valuable contributions with members of the \Kepler{} Science Team's working groups on multiple body systems, transit timing variations, and completeness working groups.  
We thank Keir Ashby, Danley Hsu, and Robert Morehead for contributions to the broader SysSim project.  
We thank Derek Bingham, Earl Lawrence, Ilya Mandell, Dan Fabrycky, Gregory Gilbert, Jack Lissauer, and Gijs Mulders.

M.Y.H. acknowledges the support of the Natural Sciences and Engineering Research Council of Canada (NSERC), funding reference number PGSD3 - 516712 - 2018.
E.B.F. and D.R. acknowledge support from NASA Origins of Solar Systems grant No. NNX14AI76G and Exoplanet Research Program grant No. NNX15AE21G. 
E.B.F. acknowledges support from NASA Kepler Participating Scientist Program, grant Nos. NNX08AR04G, NNX12AF73G, and NNX14AN76G.
This work was supported by a grant from the Simons Foundation/SFARI (675601, E.B.F.).
E.B.F. acknowledges the support of the Ambrose Monell Foundation and the Institute for Advanced Study.
M.Y.H. and E.B.F. acknowledge support from the Penn State Eberly College of Science and Department of Astronomy \& Astrophysics, the Center for Exoplanets and Habitable Worlds, and the Center for Astrostatistics.  
E.B.F. acknowledges support and collaborative scholarly discussions during  residency at the Research Group on Big Data and Planets at the Israel Institute for Advanced Studies.  

The citations in this paper have made use of NASA's Astrophysics Data System Bibliographic Services.  
This research has made use of the NASA Exoplanet Archive, which is operated by the California Institute of Technology, under contract with the National Aeronautics and Space Administration under the Exoplanet Exploration Program.
This work made use of the stellar catalog from \citet{H2019} and thus indirectly the gaia-kepler.fun cross-match database created by Megan Bedell.
Several figures in this manuscript were generated using the \texttt{corner.py} package \citep{Fm2016}.
We acknowledge the Institute for Computational and Data Sciences (\url{http://icds.psu.edu/}) at The Pennsylvania State University, including the CyberLAMP cluster supported by NSF grant MRI-1626251, for providing advanced computing resources and services that have contributed to the research results reported in this paper.
This study benefited from the 2013 SAMSI workshop on Modern Statistical and Computational Methods for Analysis of \Kepler{} Data, the 2016/2017 Program on Statistical, Mathematical and Computational Methods for Astronomy, and their associated working groups.
This material was based upon work partially supported by the National Science Foundation under grant DMS-1127914 to the Statistical and Applied Mathematical Sciences Institute (SAMSI). Any opinions, findings, and conclusions or recommendations expressed in this material are those of the author(s) and do not necessarily reflect the views of the National Science Foundation.

\software{ExoplanetsSysSim \citep{F2018b},
          SysSimData \citep{F2019},
          Numpy \citep{Numpy2011},
          Matplotlib \citep{Matplotlib2007},
          Corner.py \citep{Fm2016}
          }


\bibliography{sample63}{}

\begin{thebibliography}{}

\bibitem[Anderson \& Darling(1952)]{AD1952} 
Anderson, T. W. \& Darling, D. A. 1952, The Annals of Mathematical Statistics, 23, 193

\bibitem[Andrae et al.(2018)]{A2018} 
Andrae, R., Fouesneau, M., Creevey, O., et al. 2018, A\&A, 616, A8

\bibitem[Bailer-Jones et al.(2018)]{BJ2018} 
Bailer-Jones, C. A. L., Rybizki, J., Fouesneau, M., et al. 2018, AJ, 156, 58

\bibitem[Ballard(2019)]{B2019} 
Ballard, S. 2019, AJ, 157, 113

\bibitem[Ballard \& Johnson(2016)]{BJ2016} 
Ballard, S. \& Johnson, J. A. 2016, ApJ, 816, 66

\bibitem[Barclay, Pepper, \& Quintana(2018)]{BPQ2018} 
Barclay, T., Pepper, J., \& Quintana, E. V. 2018, ApJS, 239, 2

\bibitem[Batalha et al.(2013)]{B2013} 
Batalha, N. M., Rowe, J. F., Bryson, S. T., et al. 2013, ApJS, 204, 24

\bibitem[Berger et al.(2020)]{B2020} 
Berger, T. A., Huber, D., van Saders, J. L., et al. 2020, AJ, 159, 280

\bibitem[Bonfils et al.(2013)]{Bon2013} 
Bonfils, X., Delfosse, X., Udry, S., et al. 2013, A\&A, 549, A109

\bibitem[Borucki et al.(2010)]{B2010} 
Borucki, W. J., Koch, D. G., Basri, G., et al. 2010, Science, 327, 977

\bibitem[Borucki et al.(2011a)]{B2011a} 
Borucki, W. J., Koch, D. G., Basri, G., et al. 2011a, ApJ, 728, 117

\bibitem[Borucki et al.(2011b)]{B2011b} 
Borucki, W. J., Koch, D. G., Basri, G., et al. 2011b, ApJ, 736, 19

\bibitem[Bouma et al.(2017)]{B2017} 
Bouma, L. G., Winn, J. N., Kosiarek, J., \& McCullough, P. R. 2017, arXiv:1705.08891

\bibitem[Bryson(2020)]{Bryson2020} 
Bryson, S. 2020 Res. Notes AAS, 4, 32

\bibitem[Burke \& Catanzarite(2017a)]{BC2017a} 
Burke, C. J., \& Catanzarite, J. 2017a, Planet Detection Metrics: Window and One-Sigma Depth Functions for Data Release 25, Tech. Rep. KSCI-19101-002

\bibitem[Burke \& Catanzarite(2017b)]{BC2017b} 
Burke, C. J., \& Catanzarite, J. 2017b, Planet Detection Metrics: Per-Target Flux-Level Transit Injection Tests of TPS for Data Release 25, Tech. Rep. KSCI-19109-002

\bibitem[Burke \& Catanzarite(2017c)]{BC2017c} 
Burke, C. J., \& Catanzarite, J. 2017c, Planet Detection Metrics: Per-Target Detection Contours for Data Release 25, Tech. Rep. KSCI-19111-002

\bibitem[Carrera et al.(2018)]{C2018} 
Carrera, D., Ford, E. B., Izidoro, A., et al. 2018, ApJ, 866, 104

\bibitem[Catanzarite \& Shao(2011)]{CS2011} 
Catanzarite, J. \& Shao, M. 2011, ApJ, 738, 151

\bibitem[Christiansen et al.(2012)]{C2012} 
Christiansen, J. L., Jenkins, J. M., Caldwell, D. A., et al. 2012, PASP, 124, 1279

\bibitem[Christiansen et al.(2020)]{C2020} 
Christiansen, J. L., Clarke, B. D., Burke, C. J., et al. 2020, AJ, 160, 159

\bibitem[Ciardi et al.(2013)]{C2013} 
Ciardi, D. R., Fabrycky, D. C., Ford, E. B., et al. 2013, ApJ, 763, 41

\bibitem[Coughlin(2017)]{Co2017} 
Coughlin, J. L. 2017, Planet Detection Metrics: Robovetter Completeness and Effectiveness for Data Release 25, Tech. Rep. KSCI-19114-001

\bibitem[Cressie \& Read(1984)]{CR1984} 
Cressie, N. \& Read, T. R. C. 1984, Journal of the Royal Statistical Society. Series B, 46, 440

\bibitem[Cumming et al.(2008)]{C2008} 
Cumming, A., Butler, R. P., Marcy, G. W., et al. 2008, PASP, 120, 531

\bibitem[Dressing \& Charbonneau(2013)]{DC2013} 
Dressing, C. D. \& Charbonneau, D. 2013, ApJ, 767, 95

\bibitem[Endl et al.(2006)]{E2006} 
Endl, M., Cochran, W. D., K\"urster, M., et al. 2006, ApJ, 649, 436

\bibitem[Fabrycky et al.(2014)]{F2014} 
Fabrycky, D. C., Lissauer, J. J., Ragozzine, D., et al. 2014, ApJ, 790, 146

\bibitem[Fang \& Margot(2012)]{FM2012} 
Fang, J., \& Margot, J.-L. 2012, ApJ, 761, 92

\bibitem[Ford et al.(2018b)]{F2018b} 
Ford, E. B., He, M. Y., Hsu, D. C., \& Ragozzine, D. 2018b, Planetary Systems Simulation \& Model of Kepler Mission for Characterizing the Occurrence Rates of Exoplanets and Planetary Architectures, v1.0, Zenodo, doi:10.5281/zenodo.1205172. \url{https://doi.org/10.5281/zenodo.1205172}

\bibitem[Ford(2019)]{F2019} 
Ford, E. B., 2019, ExoJulia/SysSimData: Initial Release of Data Files for the Exoplanet System Simulator, v1.0.0 Zenodo, doi:10.5281/zenodo.3255313. \url{https://doi.org/10.5281/zenodo.3255313}

\bibitem[Foreman-Mackey(2016)]{Fm2016} 
Foreman-Mackey, D. 2016, corner.py: Scatterplot matrices in Python, JOSS, 1, 24, doi:10.21105/joss.00024

\bibitem[Fressin et al.(2013)]{F2013} 
Fressin, F., Torres, G., Charbonneau, D., et al. 2013, ApJ, 766, 81

\bibitem[Fulton et al.(2017)]{F2017} 
Fulton, B. J., Petigura, E. A., Howard, A. W., et al. 2017, AJ, 154, 109

\bibitem[Gaia Collaboration et al.(2018)]{Gaia2018} 
Gaia Collaboration, Brown, A. G. A., Vallenari, A., et al. 2018, A\&A, 616, A1

\bibitem[Gaidos et al.(2013)]{G2013} 
Gaidos, E., Fischer, D. A., Mann, A. W., et al. 2013, ApJ, 771, 18

\bibitem[Gaidos et al.(2016)]{G2016} 
Gaidos, E., Mann, A. W., Kraus, A. L., et al. 2016, MNRAS, 457, 2877

\bibitem[Gilbert \& Fabrycky(2020)]{GF2020} 
Gilbert, G. \& Fabrycky, D. 2020, AJ, 159, 281

\bibitem[Ginzburg, Schlichting, \& Sari(2016)]{GSS2016} 
Ginzburg, S., Schlichting, H. E., \& Sari, R. 2016, ApJ, 825, 29

\bibitem[Ginzburg, Schlichting, \& Sari(2018)]{GSS2018} 
Ginzburg, S., Schlichting, H. E., \& Sari, R. 2016, MNRAS, 476, 759

\bibitem[Gupta \& Schlichting(2018)]{GS2018} 
Gupta, A. \& Schlichting, H. E. 2018, MNRAS, 487, 24

\bibitem[Hardegree-Ullman et al.(2019)]{Hu2019} 
Hardegree-Ullman, K. K., Cushing, M. C., Muirhead, P. S., \& Christiansen, J. L. 2019, AJ, 158, 75

\bibitem[He, Ford, \& Ragozzine(2019)]{HFR2019}
He, M. Y., Ford, E. B., \& Ragozzine, D. 2019, MNRAS, 490, 4575

\bibitem[Howard et al.(2012)]{H2012} 
Howard, A. W., Marcy, G. W., Bryson, S. T., et al. 2012, ApJS, 201, 15

\bibitem[Hsu et al.(2018)]{H2018} 
Hsu, D. C., Ford, E. B., Ragozzine, D., et al. 2018, AJ, 155, 205

\bibitem[Hsu et al.(2019)]{H2019} 
Hsu, D. C., Ford, E. B., Ragozzine, D., \& Ashby, K. 2019, AJ, 158, 109

\bibitem[Hsu, Ford, \& Terrien(2020)]{HFT2020} 
Hsu, D. C., Ford, E. B., \& Terrien, R. 2020, MNRAS, 498, 2249

\bibitem[Huang et al.(2018)]{Hu2018} 
Huang, C. X., Shporer, A., Dragomir, D., et al. 2018, Expected Yields of Planet discoveries from the TESS primary and extended missions, arXiv:1807.11129

\bibitem[Hunter(2007)]{Matplotlib2007}
Hunter, J. D. 2007, Computing in Science and Engineering, 9, 90

\bibitem[Johnson et al.(2018)]{J2017} 
Johnson, J. A., Petigura, E. A., Fulton, B. J., et al. 2017, AJ, 154, 108

\bibitem[Kolmogorov(1933)]{K1933} 
Kolmogorov, A. N. 1933, Giornale dell'Istituto Italiano degli Attuari, 4, 83

\bibitem[Lannier et al.(2016)]{L2016} 
Lannier, J., Delorme, P., Lagrange, A. M., et al. 2016, A\&A, 596, A83

\bibitem[Latham et al.(2011)]{La2011} 
Latham, D. W., Rowe, J. F., Quinn, S. N., et al. 2011, ApJL, 732, L24

\bibitem[Lissauer et al.(2011a)]{Li2011a} 
Lissauer, J. J., Fabrycky, D. C., Ford, E. B., et al. 2011a, Nature, 470, 53

\bibitem[Lissauer et al.(2011b)]{Li2011b} 
Lissauer, J. J., Ragozzine, D., Fabrycky, D. C., et al. 2011b, ApJS, 197, 8

\bibitem[Lissauer et al.(2014)]{Li2014} 
Lissauer, J. J., Marcy, G. W., Bryson, S. T., et al. 2014, ApJ, 784, 44

\bibitem[Moorhead et al.(2011)]{M2011} 
Moorhead, A. V., Ford, E. B., Morehead, R. C., et al. 2011, ApJS, 197, 1

\bibitem[Mulders, Pascucci, \& Apai(2015)]{Mu2015} 
Mulders, G., Pascucci, I., \& Apai, D. 2015, ApJ, 798, 112

\bibitem[Mulders et al.(2018)]{Mu2018} 
Mulders, G. D., Pascucci, I., Apai, D., et al. 2018, AJ, 156, 24

\bibitem[Ning, Wolfgang, \& Ghosh(2018)]{NWG2018} 
Ning, B., Wolfgang, A., \& Ghosh, S. 2018, ApJ, 869, 5

\bibitem[Owen \& Wu(2013)]{OW2013} 
Owen, J. E. \& Wu, Y. 2013, Kepler Planets: A Tale of Evaporation, ApJ, 775, 105

\bibitem[Owen \& Wu(2017)]{OW2017} 
Owen, J. E. \& Wu, Y. 2017, ApJ, 847, 29

\bibitem[Pecaut \& Mamajek(2013)]{PM2013} 
Pecaut, M. J. \& Mamajek, E. E. 2013, ApJS, 208, 9

\bibitem[Petigura, Marcy, \& Howard(2013b)]{PMH2013b} 
Petigura, E. A., Marcy, G. W., \& Howard, A. W. 2013b, ApJ, 770, 69

\bibitem[Pettitt(1976)]{P1976} 
Pettitt, A. N. 1976, Biometrika, 63, 161

\bibitem[Plavchan, Bilinski, \& Currie(2014)]{PBC2014} 
Plavchan, P., Bilinski, C., \& Currie, T. 2014, PASP, 126, 935

\bibitem[Ragozzine \& Holman(2010)]{RH2010} 
Ragozzine, D. \& Holman, M. J. 2010, arXiv:1006.3727

\bibitem[Rasmussen \& Williams(2006)]{RW2006} 
Rasmussen, C. E. \& Williams, C. K. I. 2006, Gaussian Processes for Machine Learning, MIT Press, ISBN 0-262-18253-X

\bibitem[Ricker et al.(2015)]{Ri2015} 
Ricker, G. R., Winn, J. N., Vanderspek, R., et al. 2015, JATIS, 1, 014003

\bibitem[Rowe et al.(2014)]{R2014} 
Rowe, J. F., Bryson, S. T., Marcy, G. W., et al. 2014, ApJ, 784, 45

\bibitem[Rowe et al.(2015)]{R2015} 
Rowe, J. F., Coughlin, J. L., Antoci, V., et al. 2015, ApJS, 217, 16

\bibitem[Sandford, Kipping, \& Collins(2019)]{SKC2019} 
Sandford, E., Kipping, D., \& Collins, M. 2019, MNRAS, 489, 3162

\bibitem[Smirnov(1948)]{S1948} 
Smirnov, N. 1948, The Annals of Mathematical Statistics, 19, 279

\bibitem[Stassun et al.(2018)]{S2018} 
Stassun, K. G., Oelkers, R. J., Pepper, J., et al. 2018, AJ, 156, 102

\bibitem[Steffen et al.(2010)]{S2010} 
Steffen, J. H., Batalha, N. M., Borucki, W., J., et al. 2010, ApJ, 725, 1226

\bibitem[Sullivan et al.(2015)]{S2015} 
Sullivan, P. W., Winn, J. N., Berta-Thompson, Z. K., et al. 2015, ApJ, 809, 77

\bibitem[Thompson et al.(2018)]{T2018} 
Thompson, S. E., Coughlin, J. L., Hoffman, K., et al. 2018, ApJS, 235, 38

\bibitem[van der Walt et al.(2011)]{Numpy2011}
van der Walt, S., Colbert, S. C., \& Varoquaux, G. 2011, CSE, 13, 22

\bibitem[Van Eylen \& Albrecht(2015)]{vEA2015} 
Van Eylen, V. \& Albrecht, S. 2015, ApJ, 808, 126

\bibitem[Van Eylen et al.(2017)]{vE2017} 
Van Eylen, V., Agentoft, C., Lundkvist, M. S., et al. 2018, MNRAS, 479, 4786

\bibitem[Weiss et al.(2018a)]{W2018a} 
Weiss, L. M., Marcy, G. W., Petigura, E. A., et al. 2018, AJ, 155, 48

\bibitem[Winn \& Fabrycky(2015)]{WF2015} 
Winn, J. N. \& Fabrycky, D. C. 2015, ARA\&A, 53, 407

\bibitem[Xie et al.(2016)]{X2016} 
Xie, J.-W., Subo, D., Zhu, Z., et al. 2016, PNAS, 113, 11431

\bibitem[Yang, Xie, \& Zhou(2020)]{YXZ2020} 
Yang, J.-Y., Xie, J.-W., \& Zhou, J.-L. 2020, AJ, 159, 164

\bibitem[Zhu et al.(2018)]{Z2018} 
Zhu, W., Petrovich, C., Wu, Y., et al. 2018, ApJ, 860, 101

\bibitem[Zink, Christiansen, \& Hansen(2019)]{ZCH2019} 
Zink, J. K., Christiansen, J. L., \& Hansen, B. M. S. 2019, MNRAS, 483, 4479

\bibitem[Zink \& Hansen(2019)]{ZH2019} 
Zink, J. K. \& Hansen, B. M. S. 2019, MNRAS, 487, 246

\end{thebibliography}
\bibliographystyle{aasjournal}



\appendix

\renewcommand{\thefigure}{A\arabic{figure}}
\renewcommand{\thetable}{A\arabic{table}}
\setcounter{figure}{0}
\setcounter{table}{0}

\begin{figure*}
\centering
\includegraphics[scale=0.25,trim={0 0 0 0},clip]{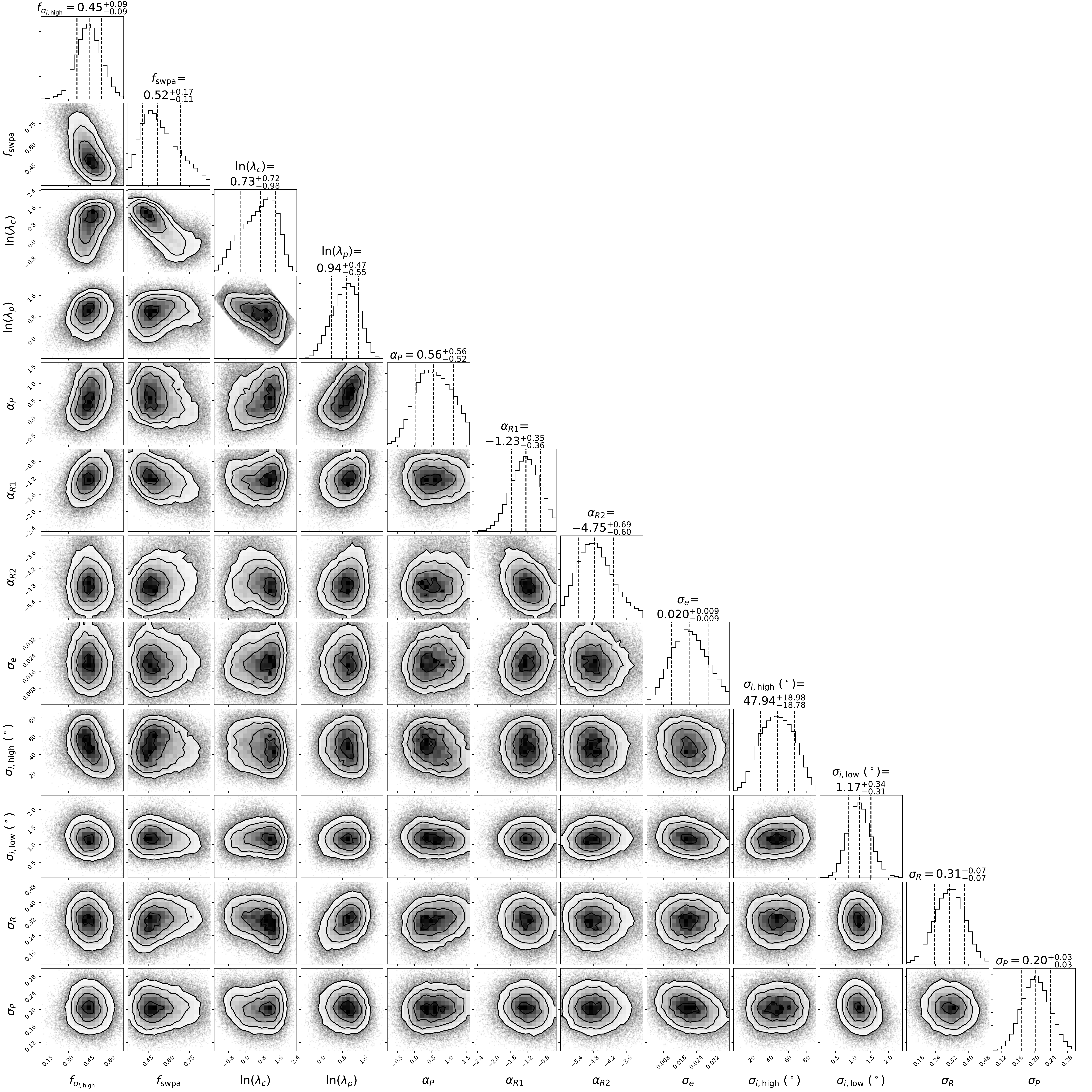}
\caption{ABC posterior distributions of the free model parameters of the constant $f_{\rm swpa}+\alpha_P$ model. A total of $5\times10^4$ points passing a distance threshold of $\mathcal{D}_{W,\rm KS} = 50$ as drawn from the GP emulator are shown. The prior mean function was set to a constant value of 75.}
\label{fig:const_fswp_corner_KS}
\end{figure*}

\begin{figure*}
\centering
\includegraphics[scale=0.25,trim={0 0 0 0},clip]{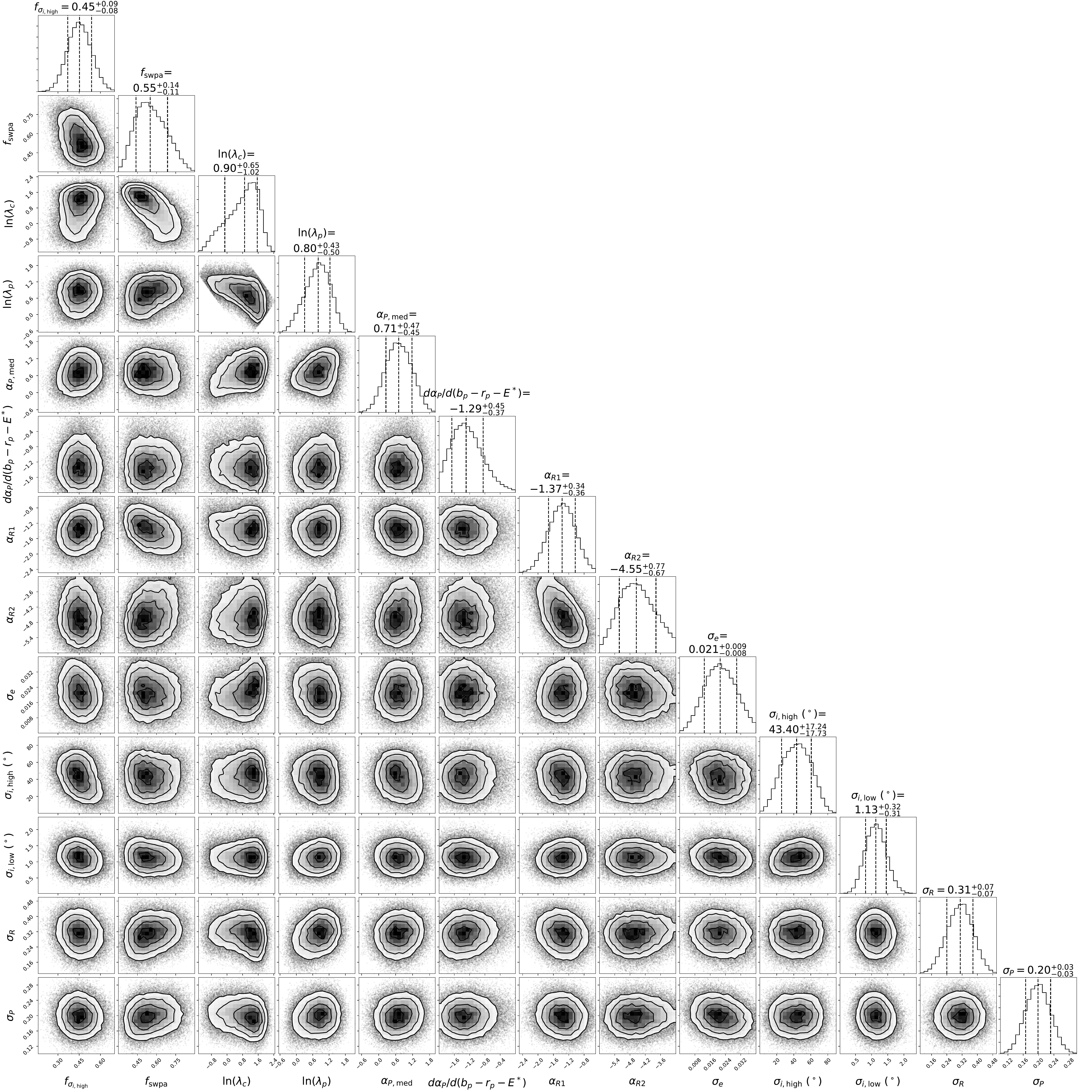}
\caption{ABC posterior distributions of the free model parameters of the linear $\alpha_P(b_p - r_p - E^*)$ model. A total of $5\times10^4$ points passing a distance threshold of $\mathcal{D}_{W,\rm KS} = 47$ as drawn from the GP emulator are shown. The prior mean function was set to a constant value of 75.}
\label{fig:linear_alphaP_corner_KS}
\end{figure*}

\begin{figure*}
\centering
\begin{tabular}{cc}
 \includegraphics[scale=0.43,trim={0 0.5cm 0 0.2cm},clip]{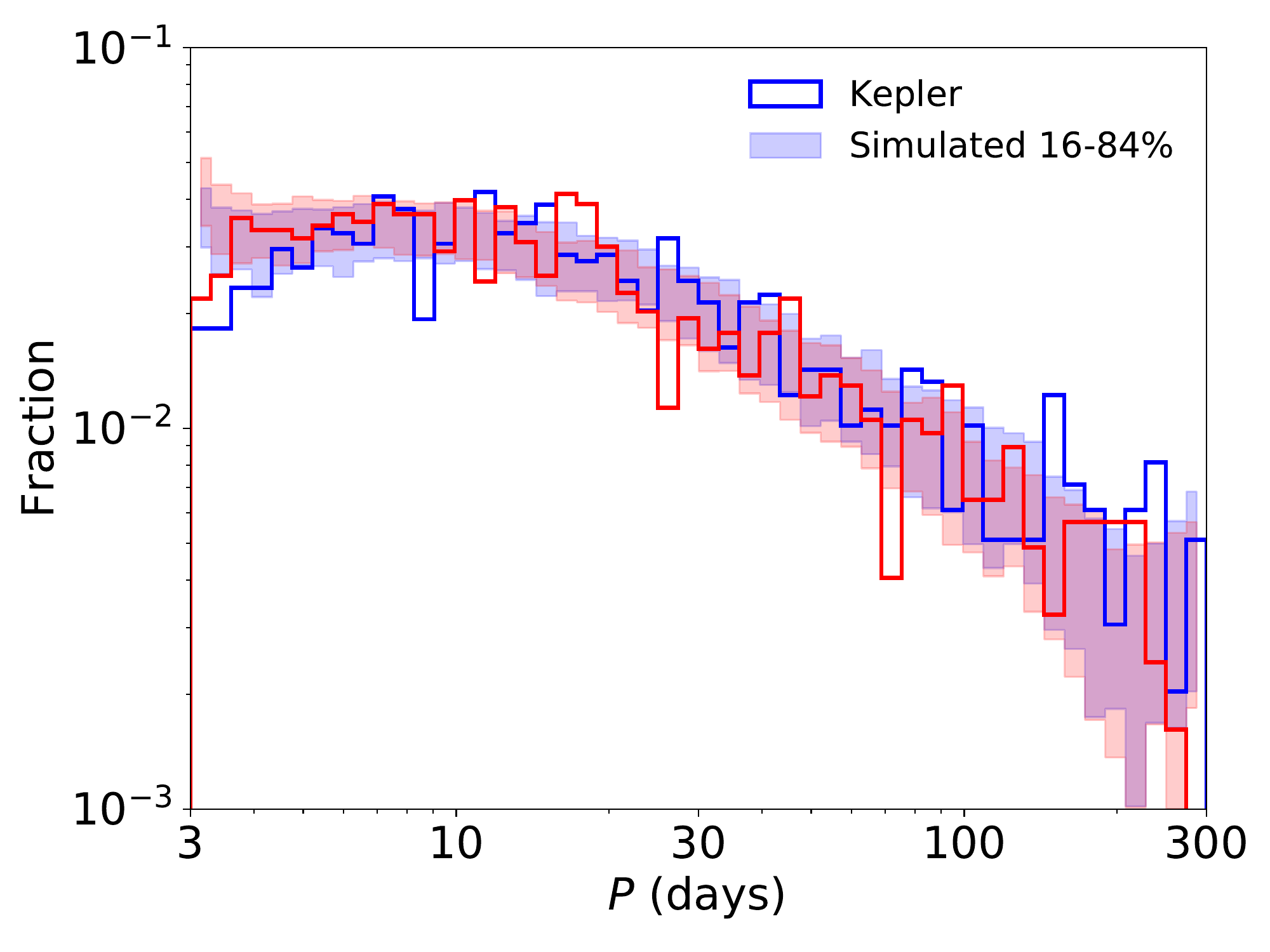} &
 \includegraphics[scale=0.43,trim={0 0.5cm 0 0.2cm},clip]{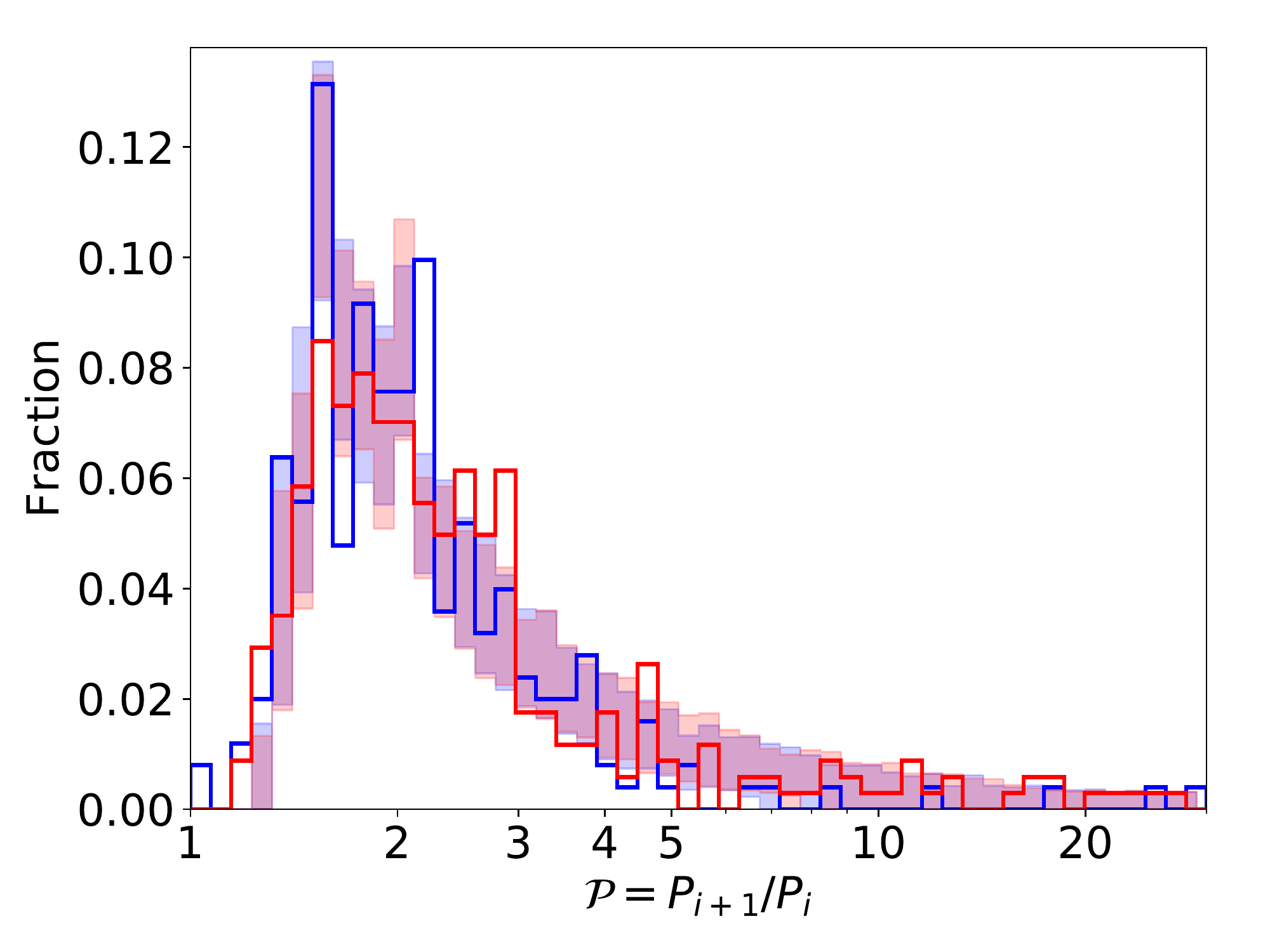} \\
 \includegraphics[scale=0.43,trim={0 0.5cm 0 0.2cm},clip]{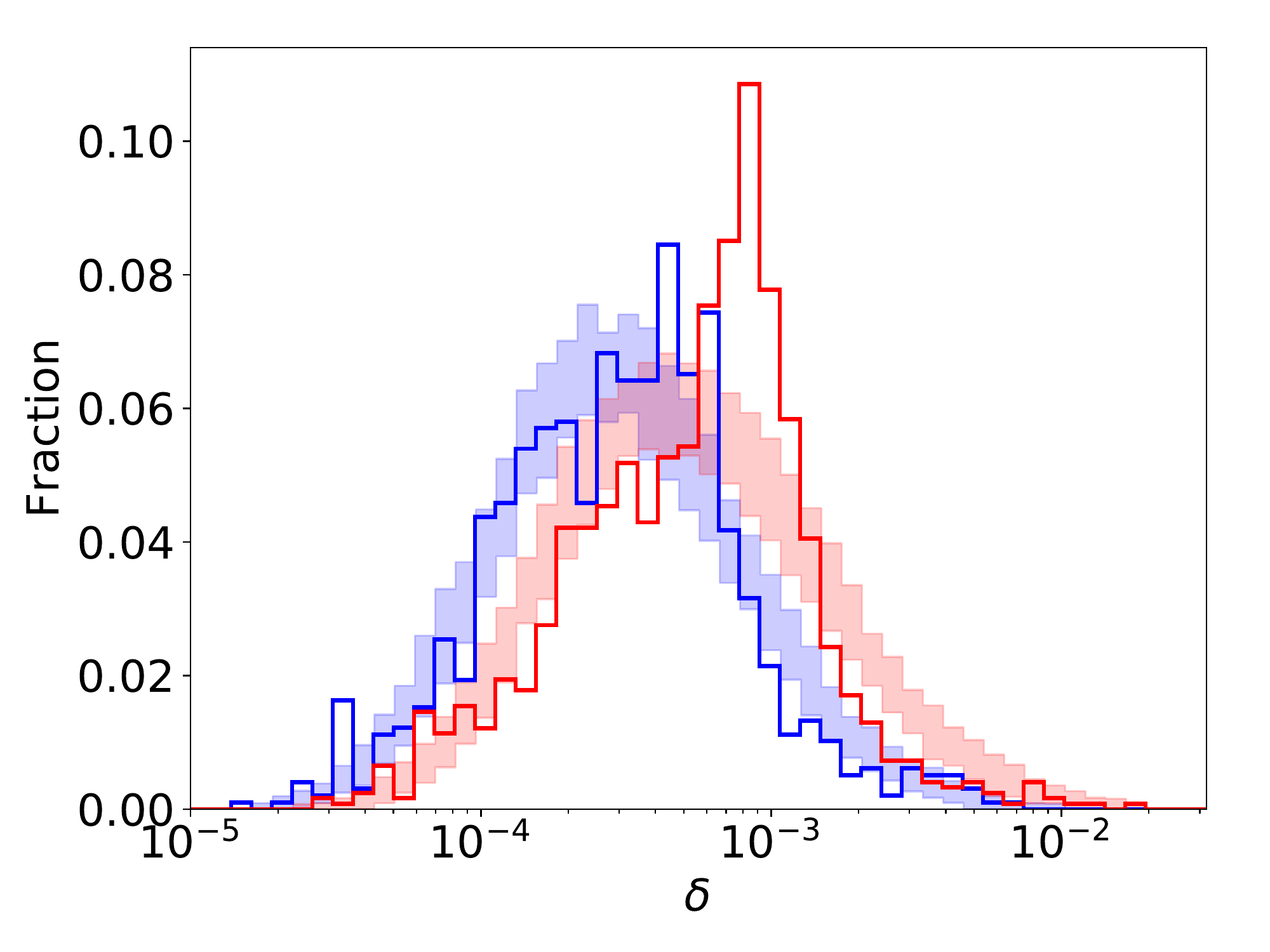} &
 \includegraphics[scale=0.43,trim={0 0.5cm 0 0.2cm},clip]{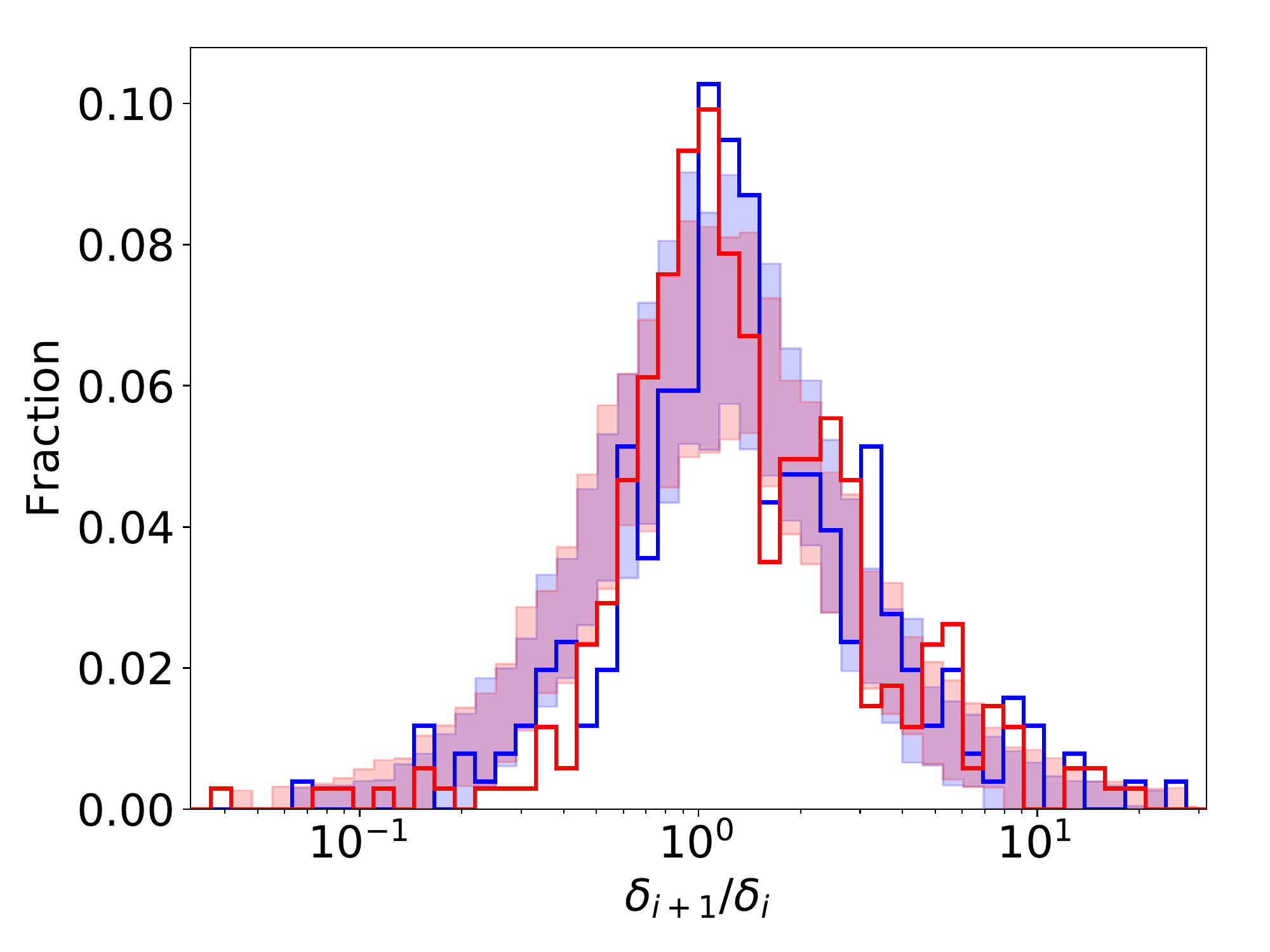} \\
 \includegraphics[scale=0.43,trim={0 0.5cm 0 0.2cm},clip]{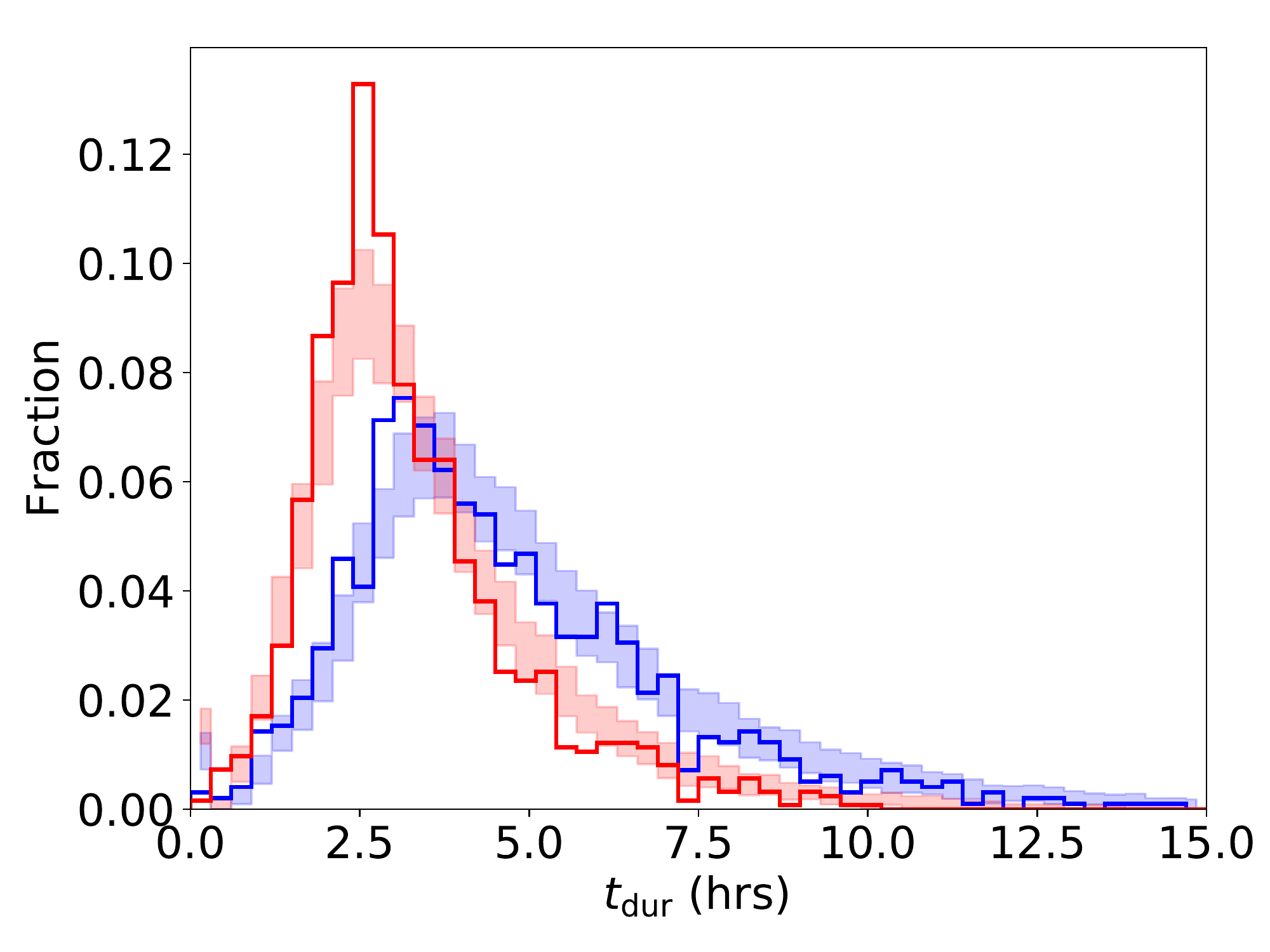} &
 \includegraphics[scale=0.43,trim={0 0.5cm 0 0.2cm},clip]{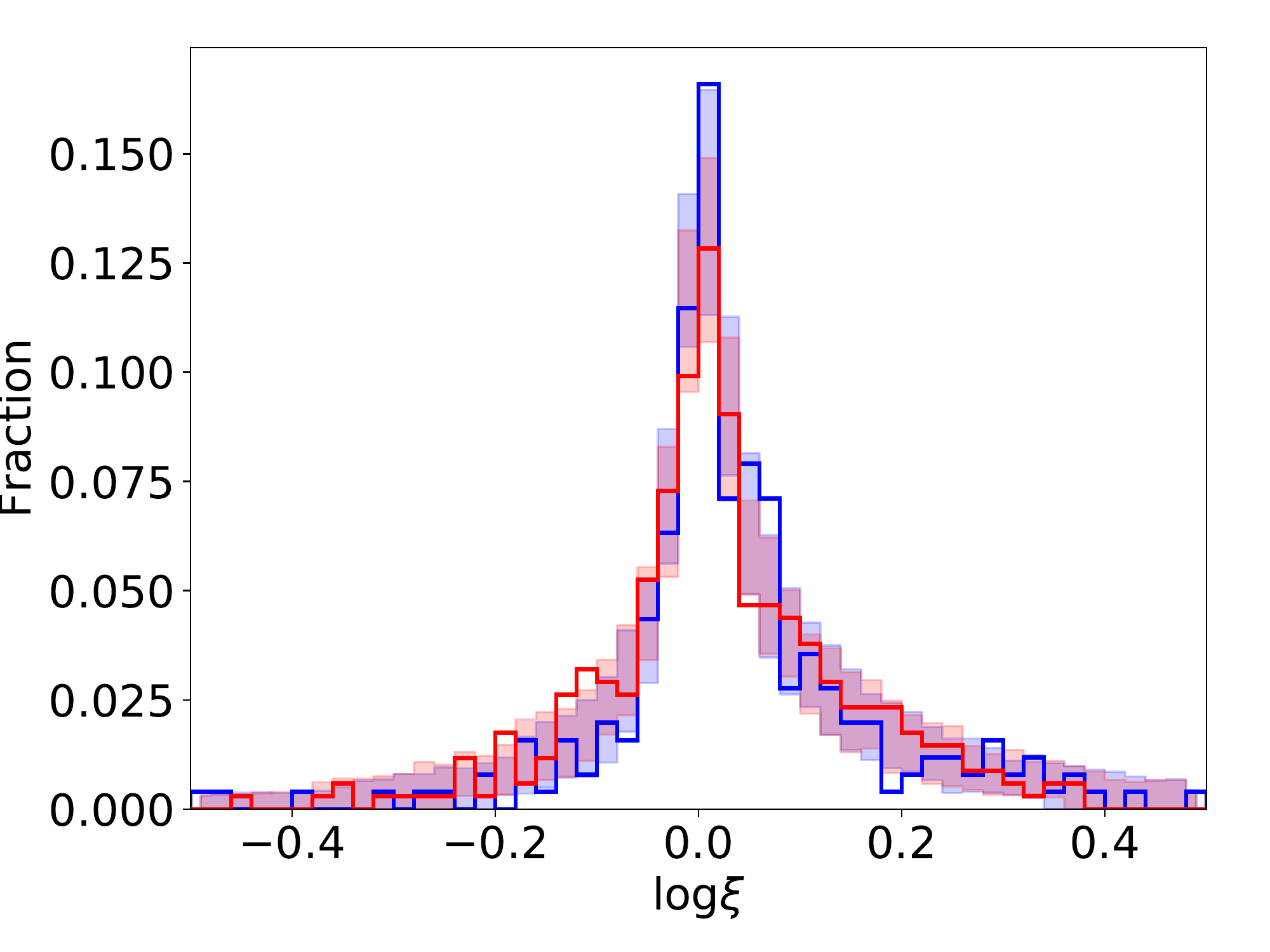} \\
\end{tabular}
\caption{Same as Figure \ref{fig:model_fswp_bprp_split}, but for our linear $\alpha_P(b_p - r_p - E^*)$ model instead of the linear $f_{\rm swpa}(b_p - r_p - E^*)$ model. As before, the shaded regions denote the 16th to 84th percentiles of the model, while the solid line histograms denote the \Kepler{} data.}
\label{fig:model_alphaP_bprp_split}
\end{figure*}

\begin{figure*}
\centering
\begin{tabular}{cc}
 \includegraphics[scale=0.28,trim={1.5cm 0.2cm 0 0.2cm},clip]{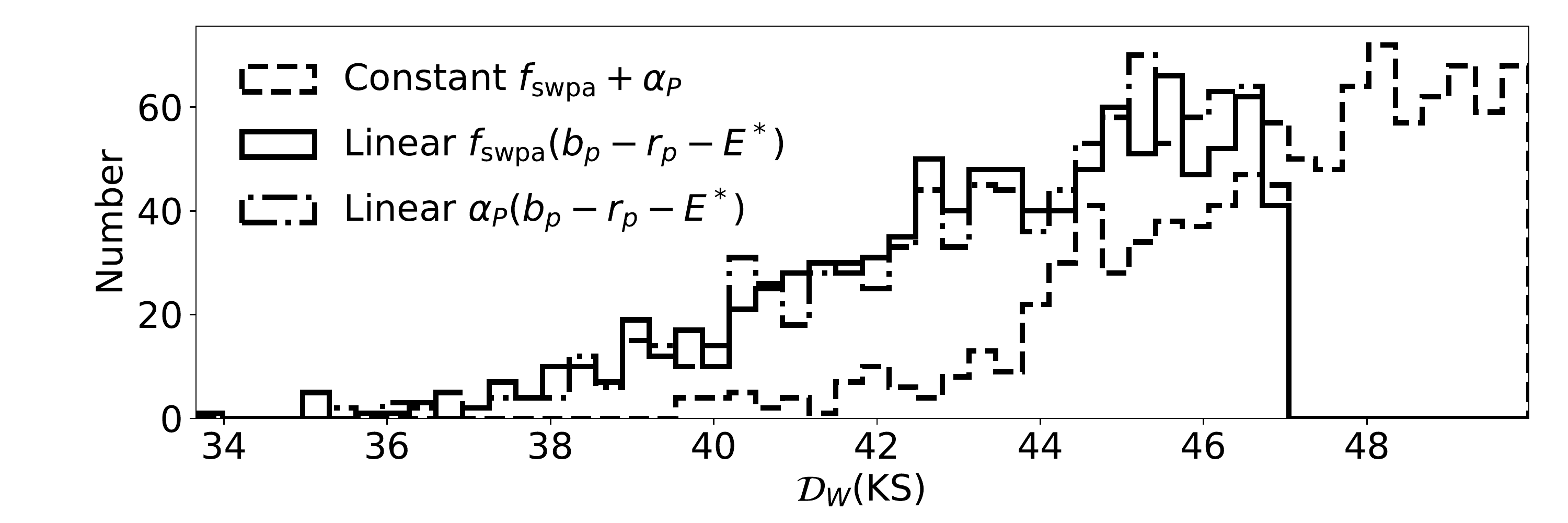} &
 \includegraphics[scale=0.28,trim={1.5cm 0.2cm 0 0.2cm},clip]{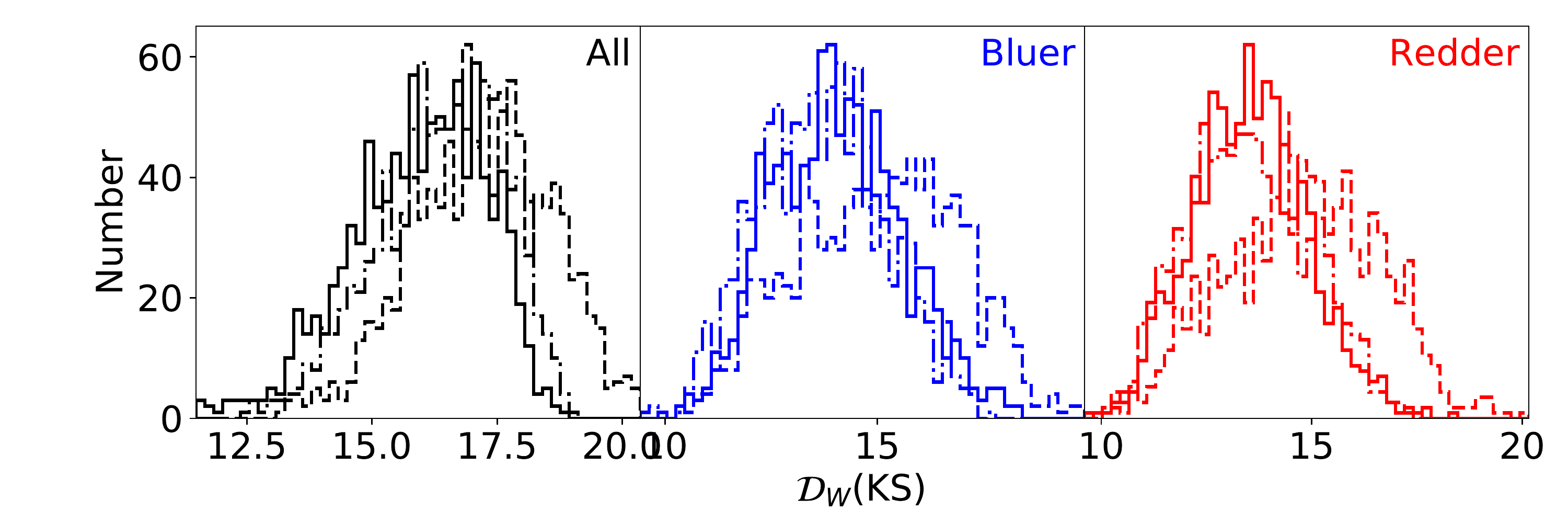} \\
\end{tabular}
\begin{tabular}{ccc}
 \includegraphics[scale=0.28,trim={0.2cm 0.2cm 0.4cm 0.2cm},clip]{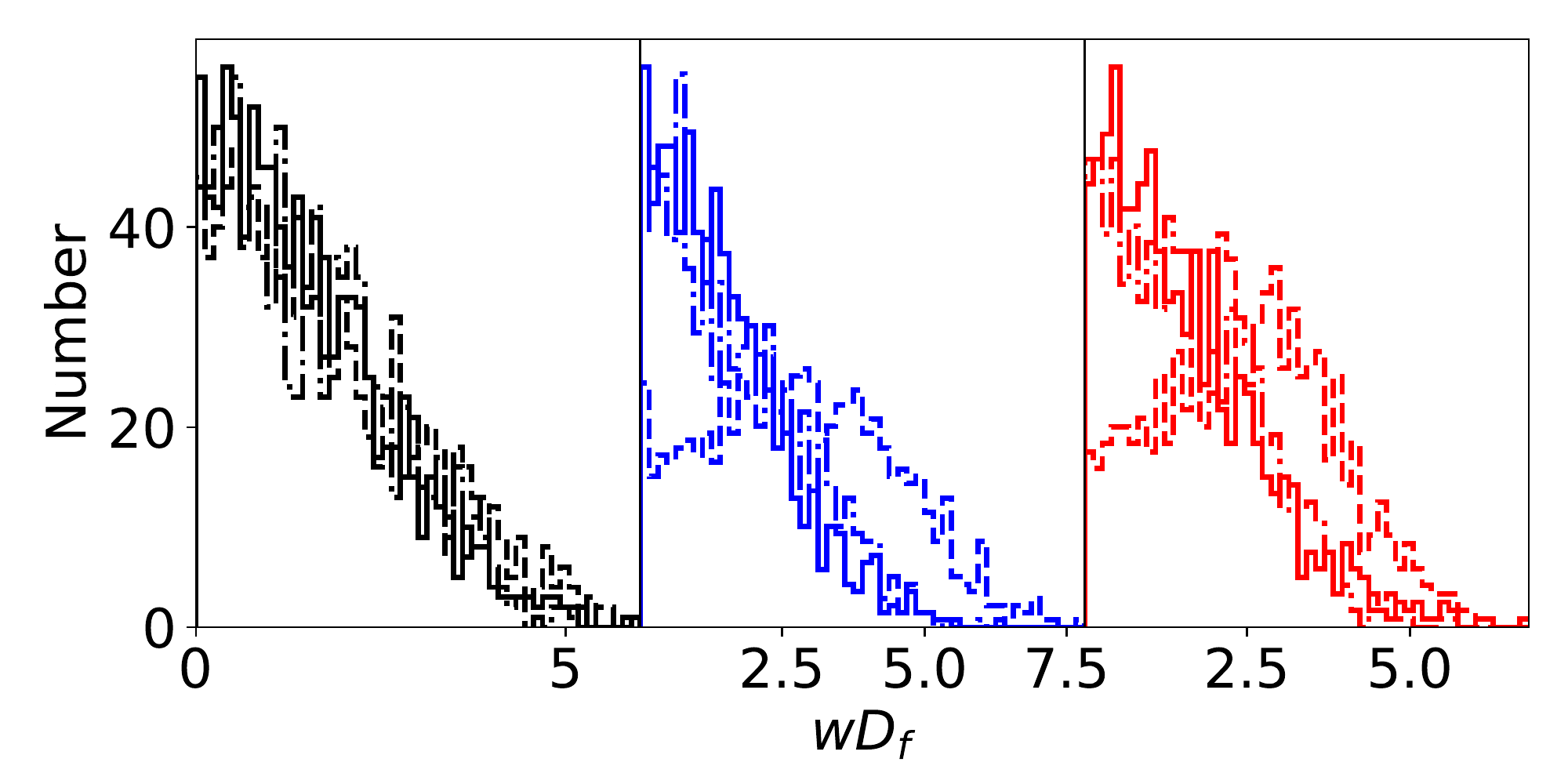} &
 \includegraphics[scale=0.28,trim={0.2cm 0.2cm 0.4cm 0.2cm},clip]{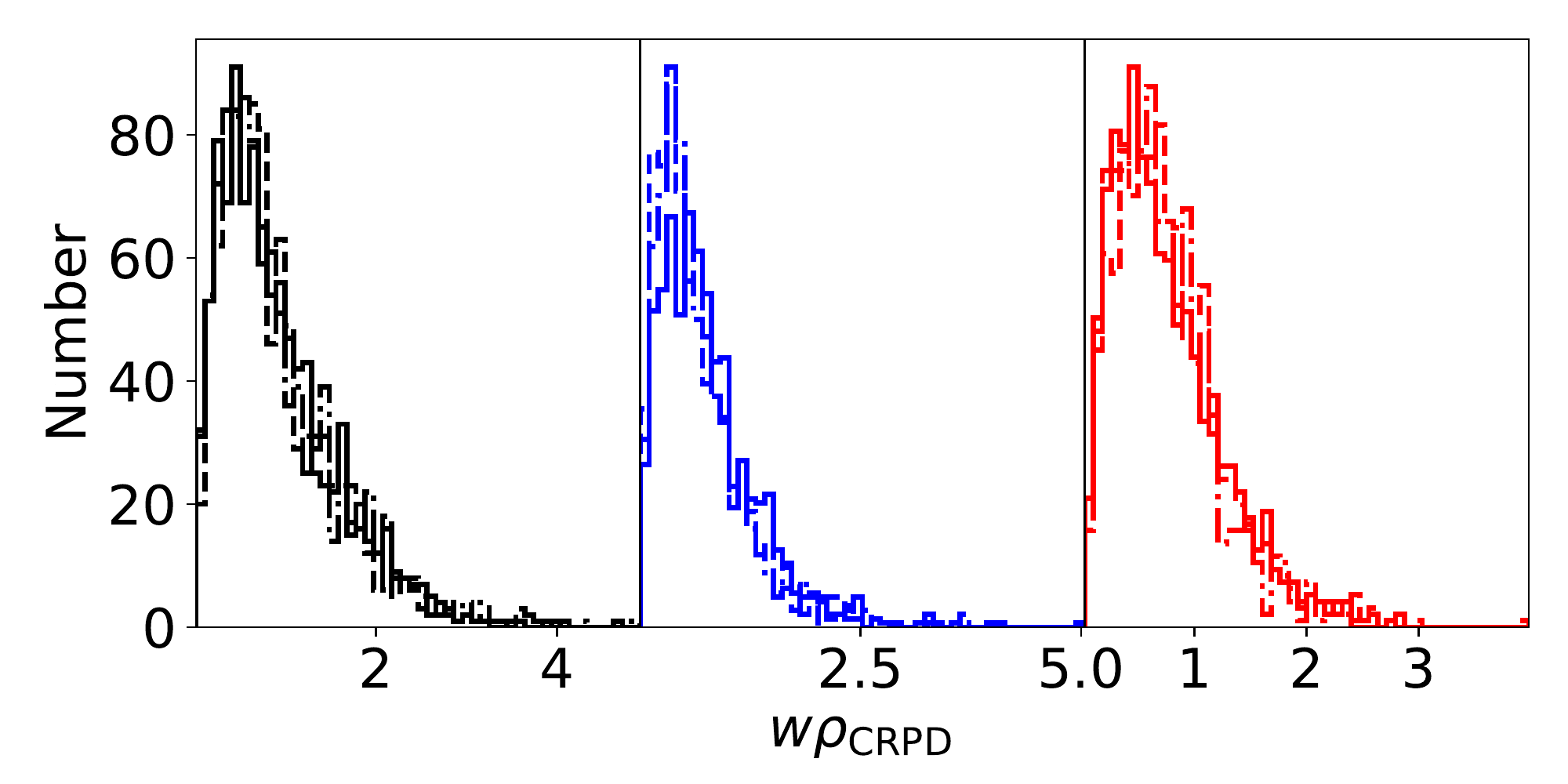} &
 \includegraphics[scale=0.28,trim={0.2cm 0.2cm 0.4cm 0.2cm},clip]{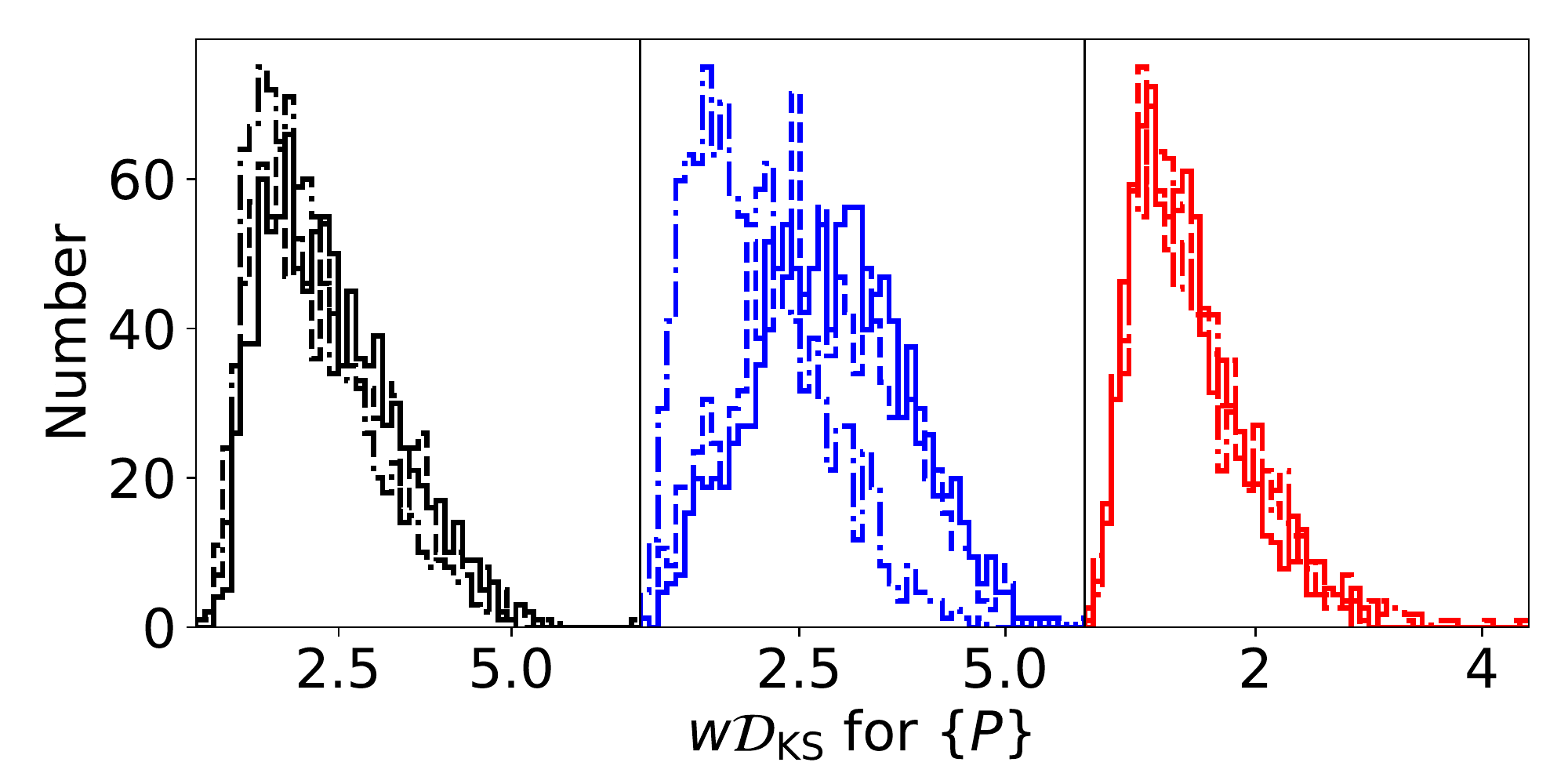} \\
 \includegraphics[scale=0.28,trim={0.2cm 0.2cm 0.4cm 0.2cm},clip]{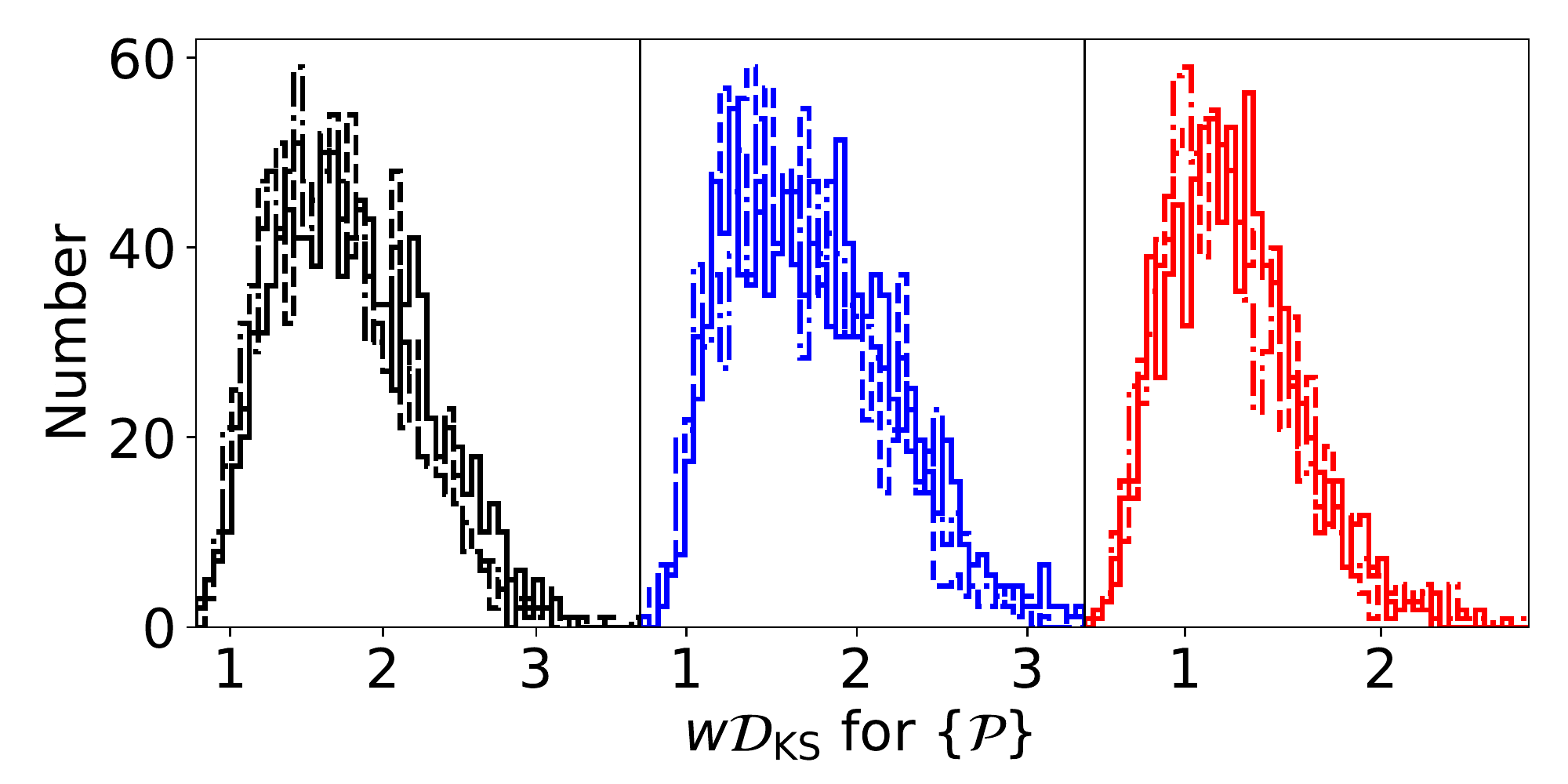} &
 \includegraphics[scale=0.28,trim={0.2cm 0.2cm 0.4cm 0.2cm},clip]{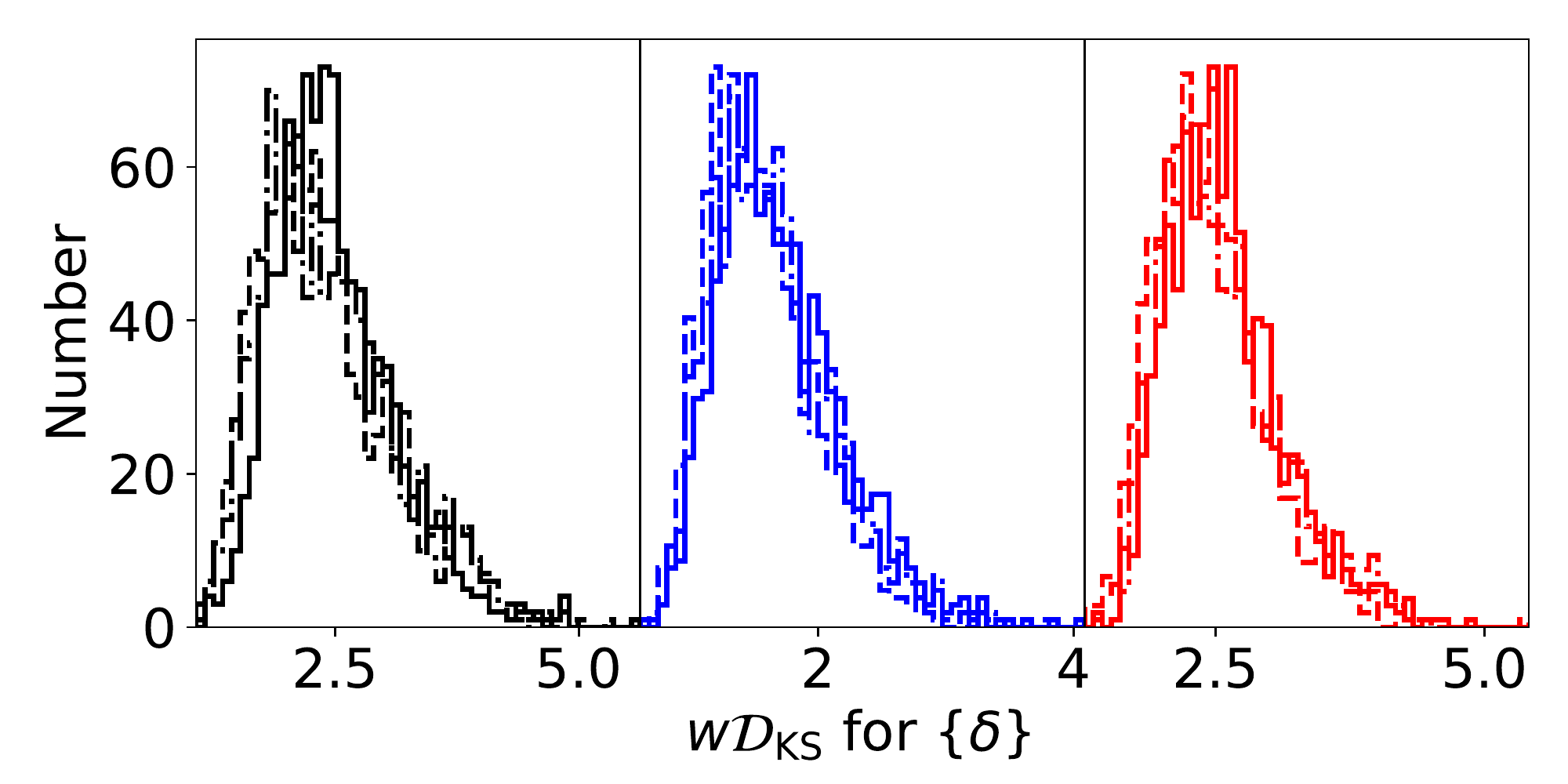} &
 \includegraphics[scale=0.28,trim={0.2cm 0.2cm 0.4cm 0.2cm},clip]{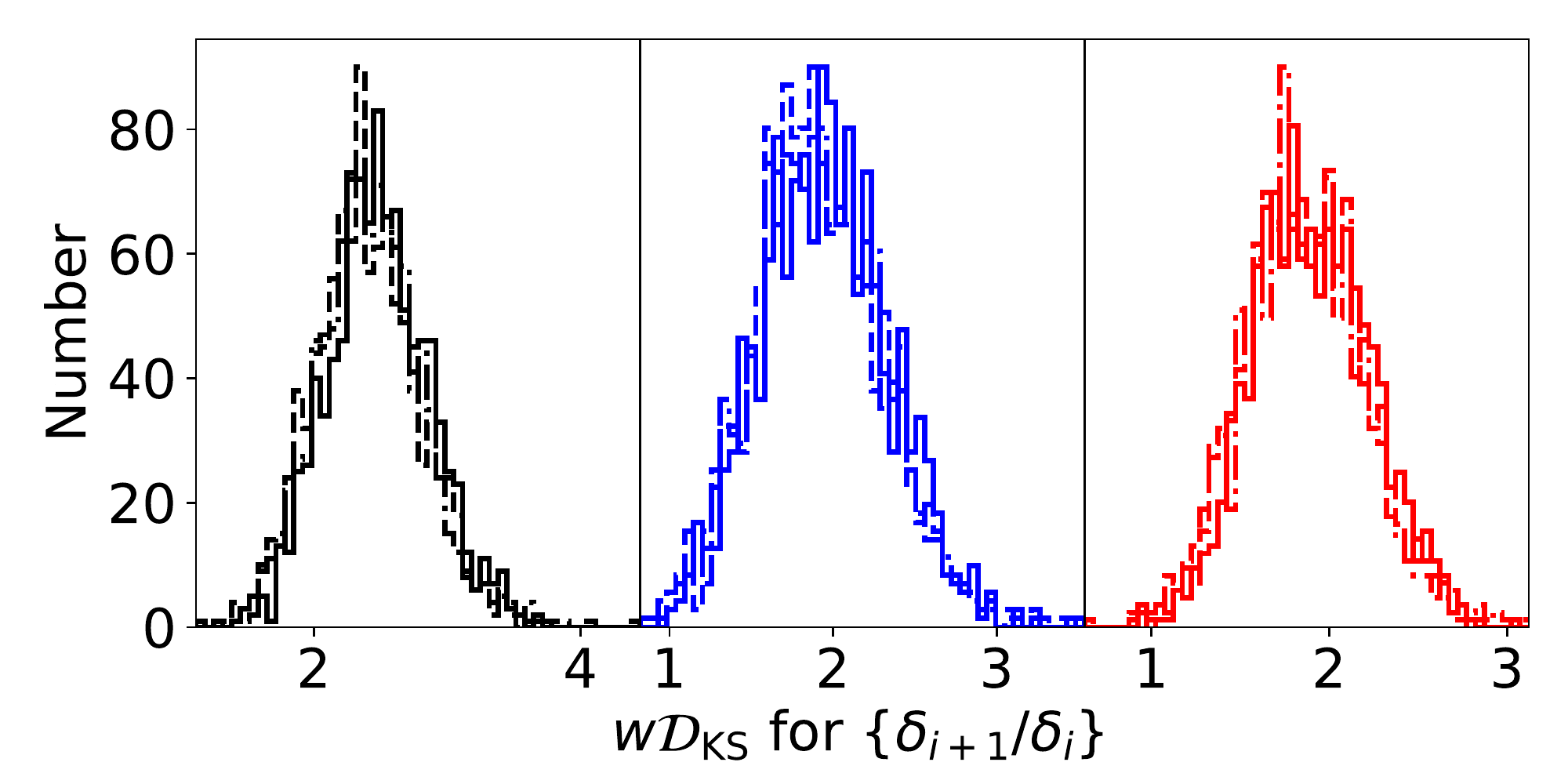} \\
 \includegraphics[scale=0.28,trim={0.2cm 0.2cm 0.4cm 0.2cm},clip]{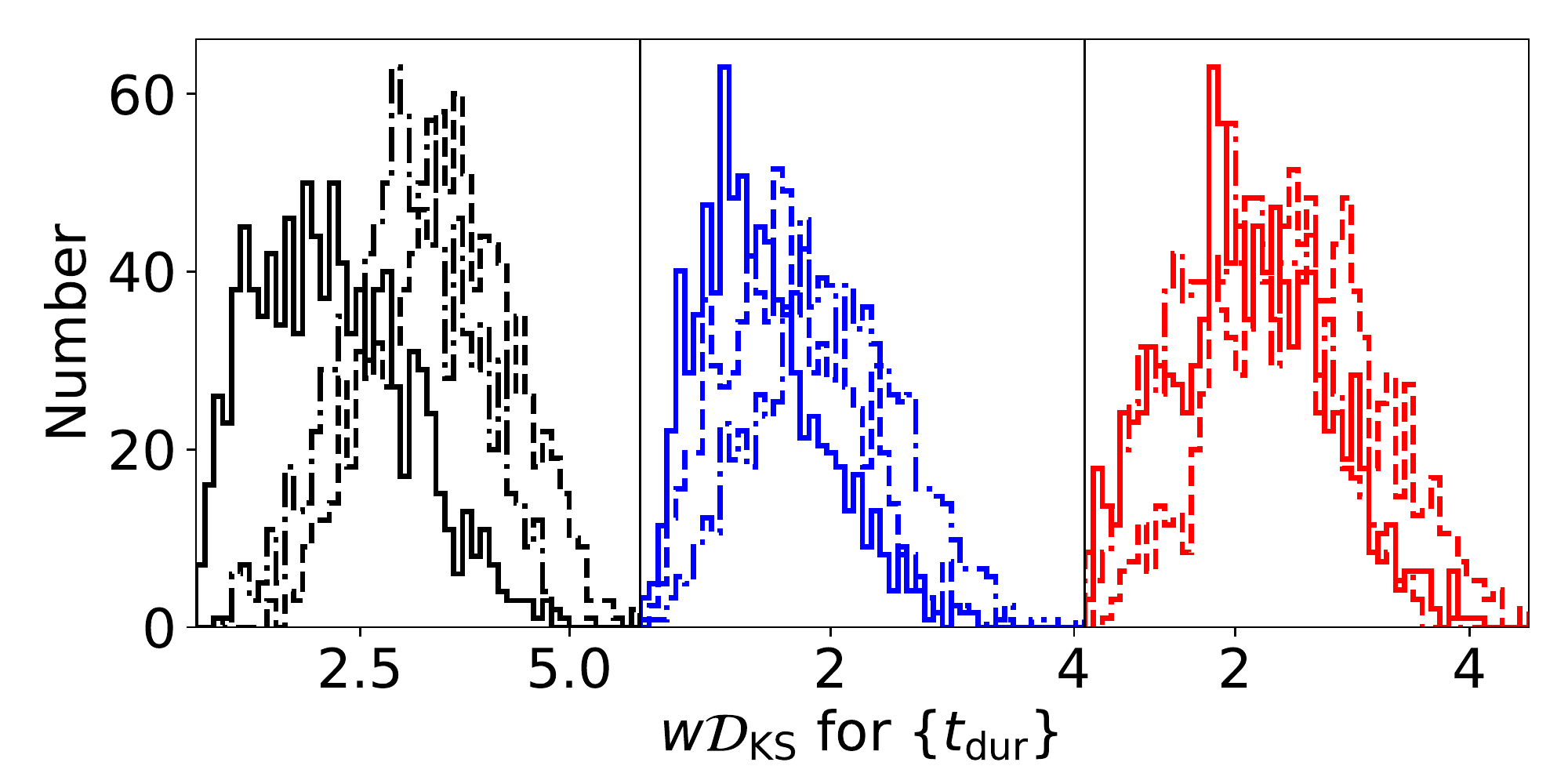} &
 \includegraphics[scale=0.28,trim={0.2cm 0.2cm 0.4cm 0.2cm},clip]{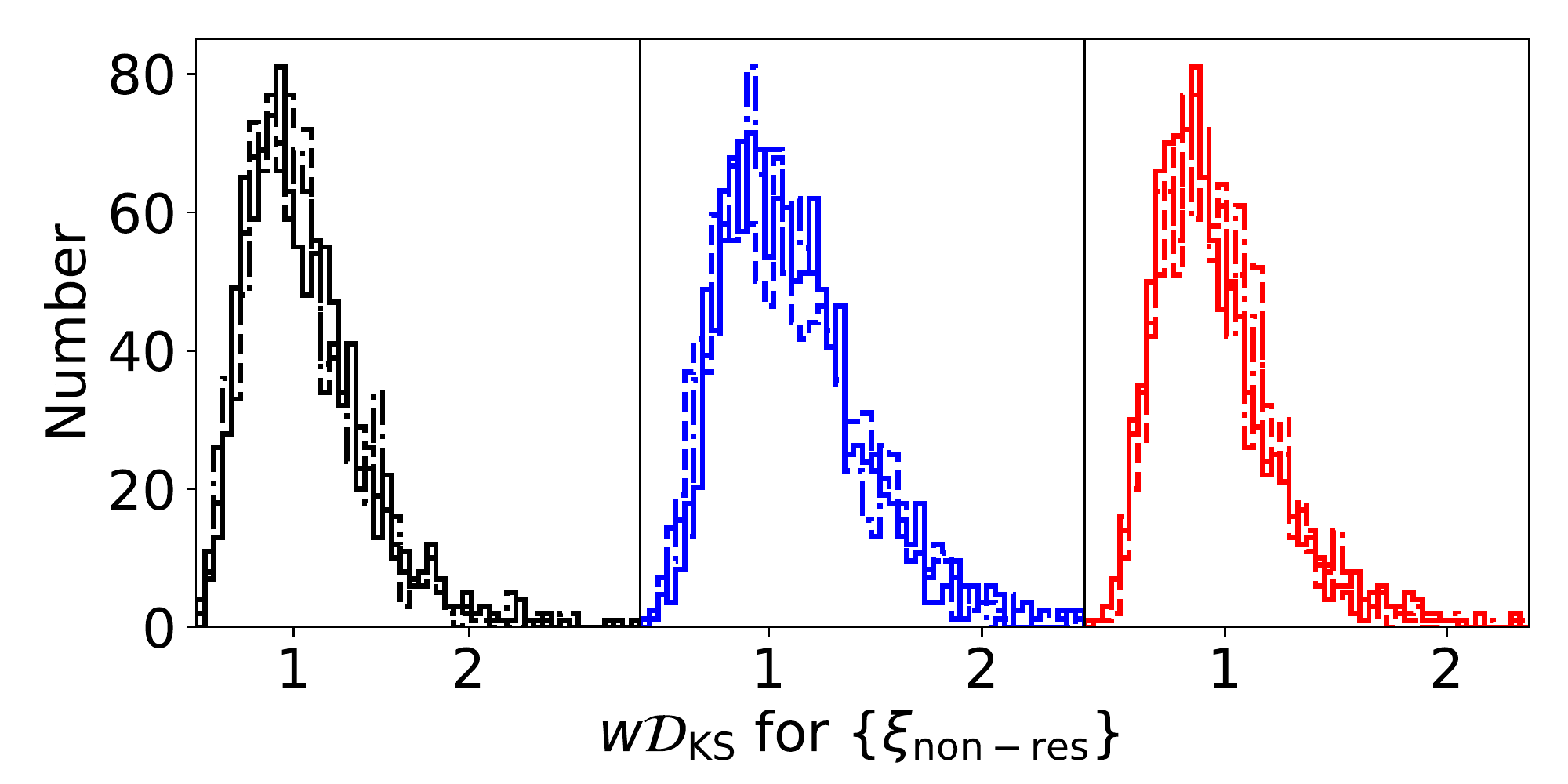} &
 \includegraphics[scale=0.28,trim={0.2cm 0.2cm 0.4cm 0.2cm},clip]{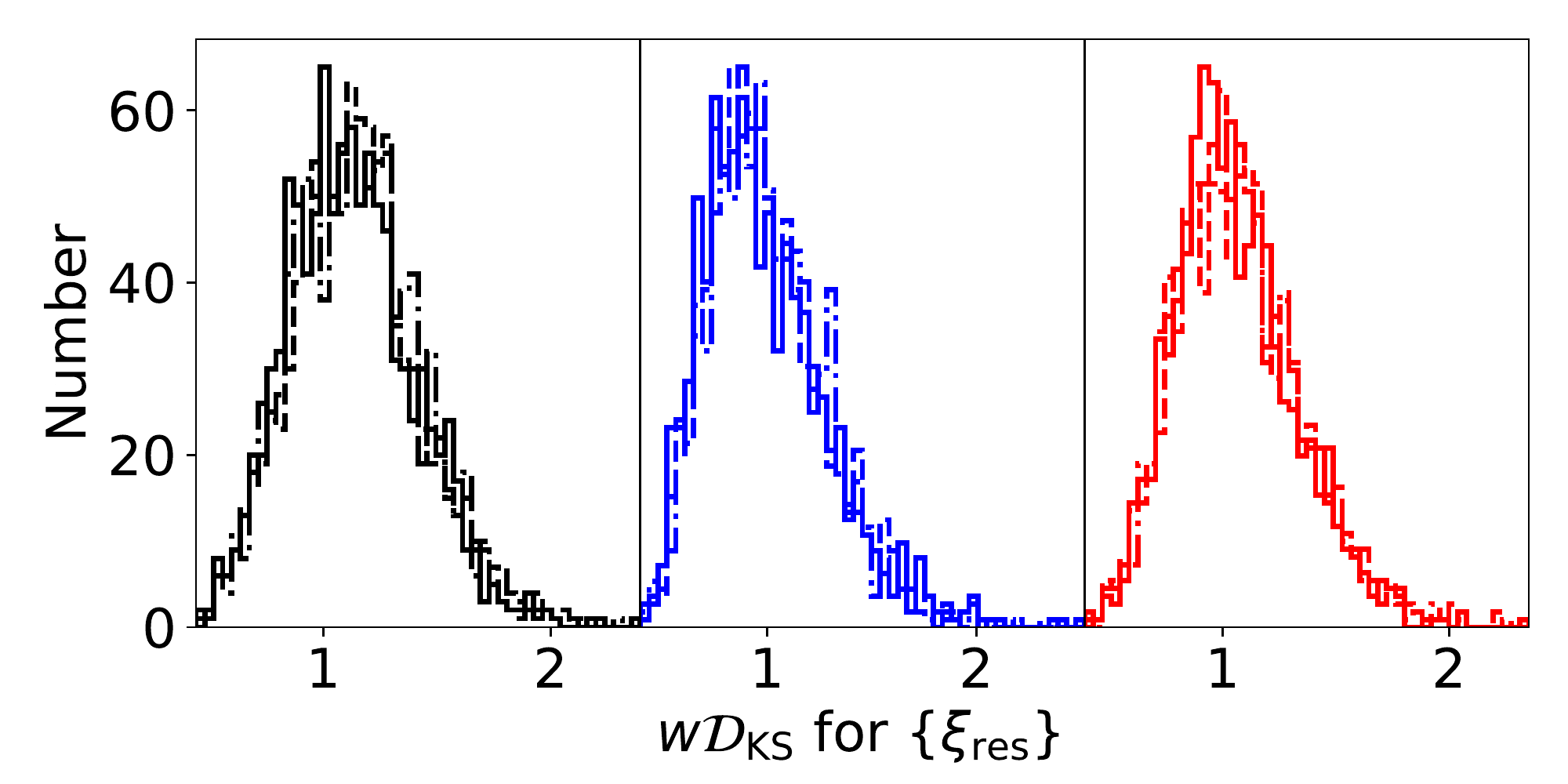} \\
\end{tabular}
\caption{Histograms of the weighted total distances (top row) and individual distance terms (second row and below) for our models as compared to the \Kepler{} data, including 1000 simulated catalogs that pass our distance thresholds of $\mathcal{D}_{W,\rm KS} = 50$, 47, and 47 for the constant $f_{\rm swpa}$ (dashed histograms), linear $f_{\rm swpa}(b_p-r_p-E^*)$ (solid histograms), and linear $\alpha_P(b_p-r_p-E^*)$ (dash-dotted histograms) models, respectively. In the top row, the right-hand panel shows the weighted sum of the individual distance terms for each subset (all, bluer, and redder stars in our sample, colored black, blue, and red, respectively), while the left-hand panel shows the sum of these three components. The panels in the second row and below show the (weighted) individual distance terms for each subset. Note that the $x$-axes for each subplot are not necessarily the same.}
\label{fig:dists_KS}
\end{figure*}

\begin{figure*}
\centering
\begin{tabular}{cc}
 \includegraphics[scale=0.28,trim={1.5cm 0.2cm 0 0.2cm},clip]{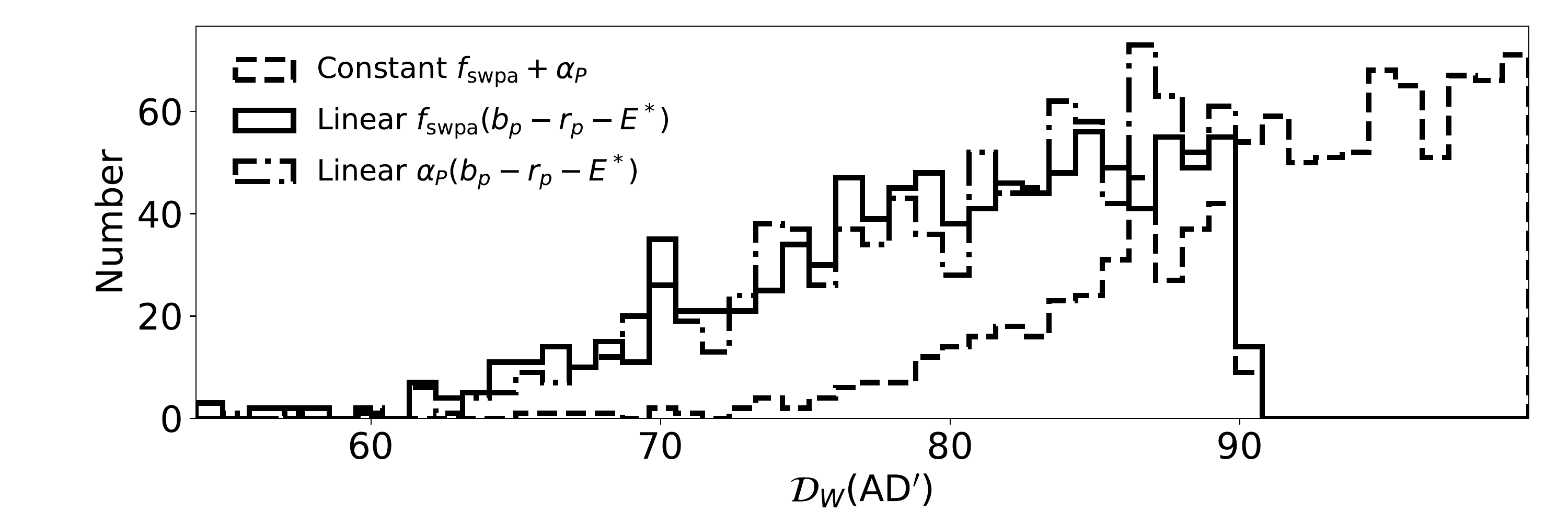} &
 \includegraphics[scale=0.28,trim={1.5cm 0.2cm 0 0.2cm},clip]{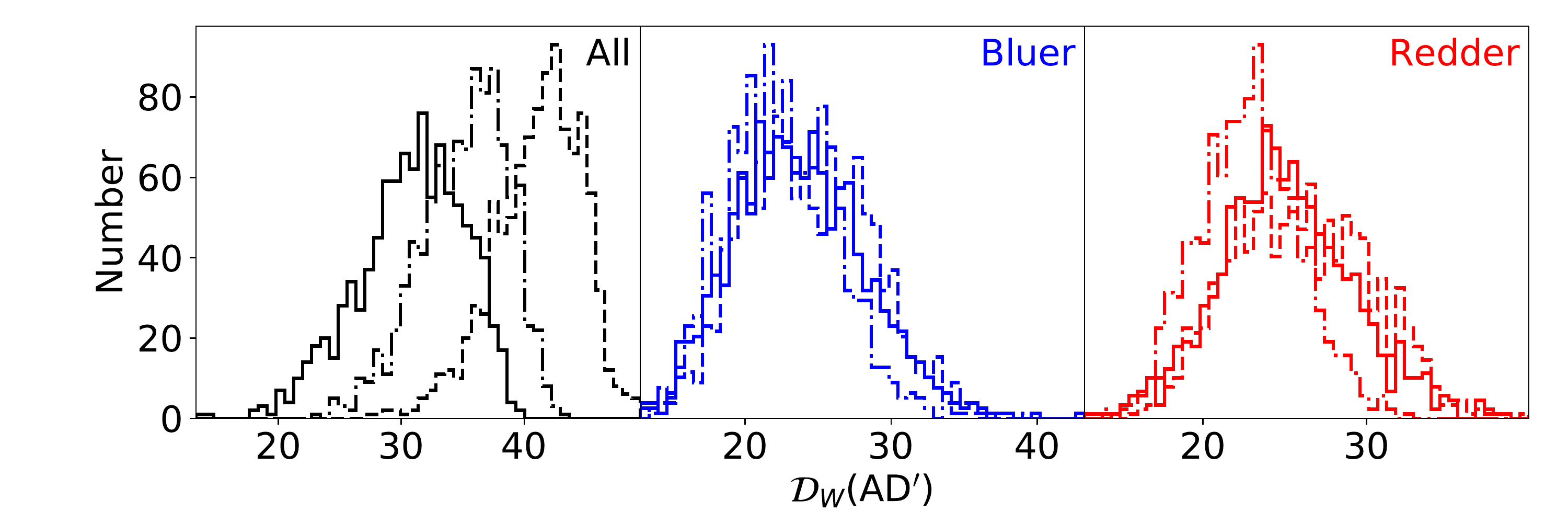} \\
\end{tabular}
\begin{tabular}{ccc}
 \includegraphics[scale=0.28,trim={0.2cm 0.2cm 0.4cm 0.2cm},clip]{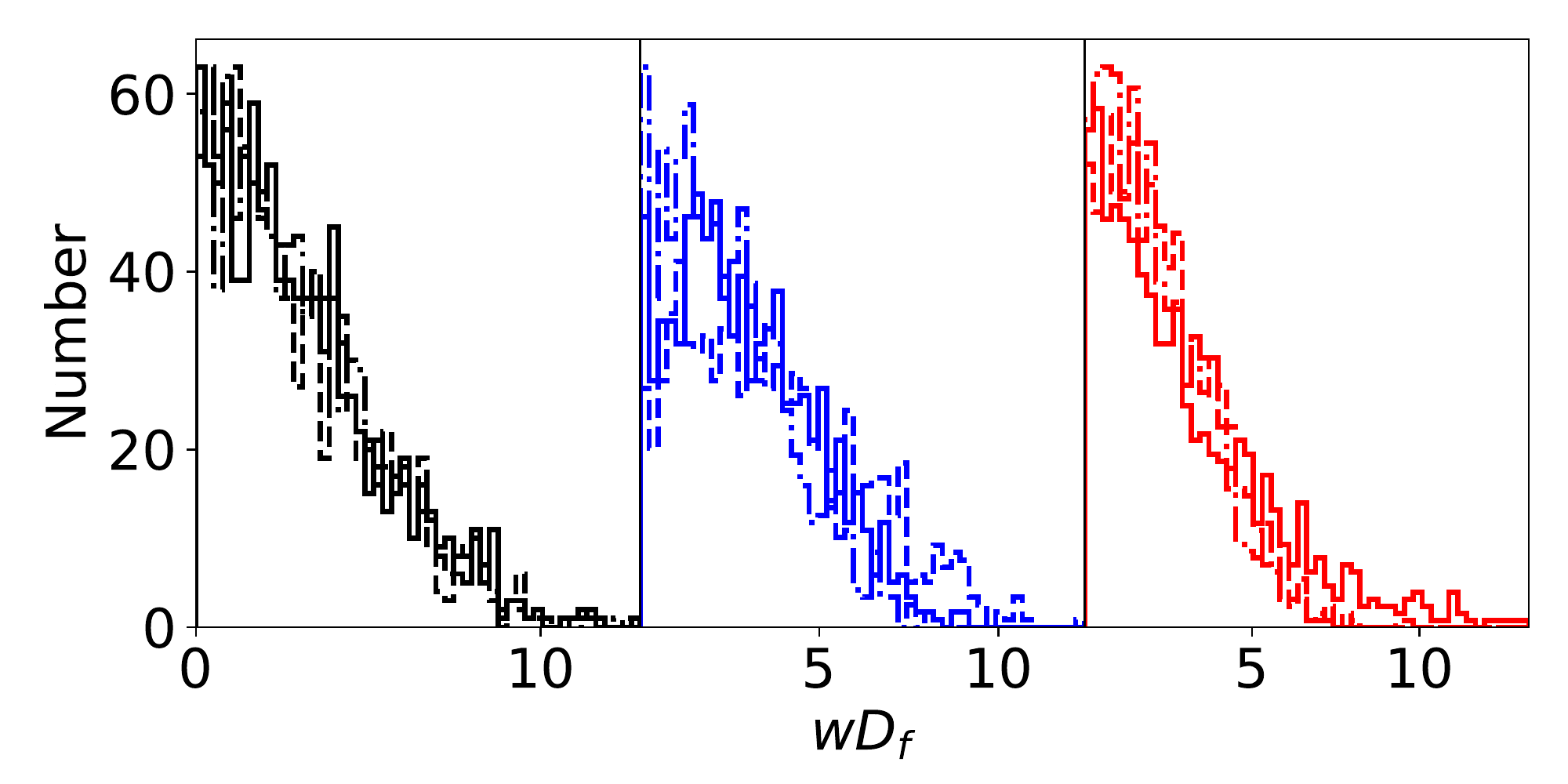} &
 \includegraphics[scale=0.28,trim={0.2cm 0.2cm 0.4cm 0.2cm},clip]{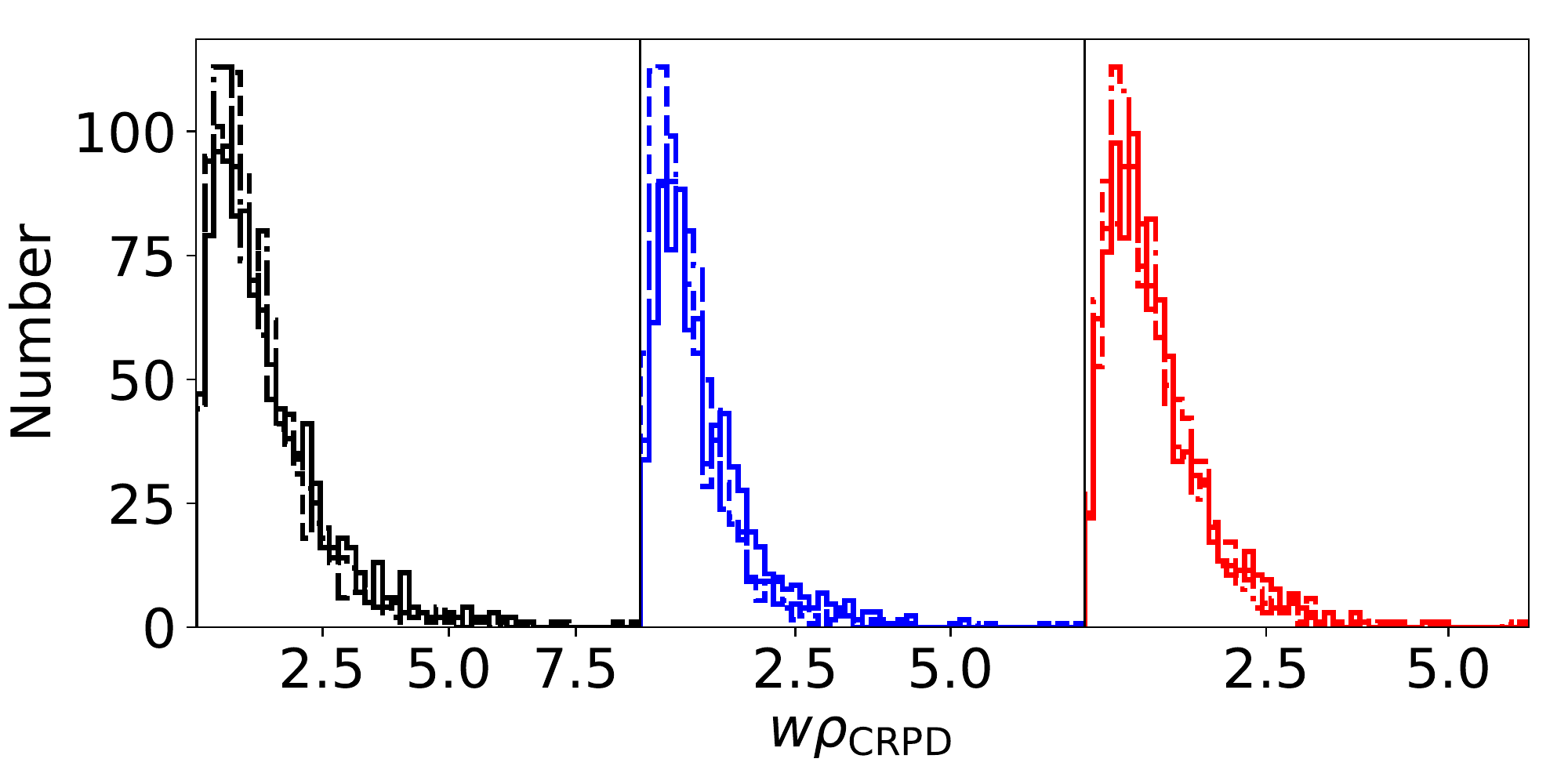} &
 \includegraphics[scale=0.28,trim={0.2cm 0.2cm 0.4cm 0.2cm},clip]{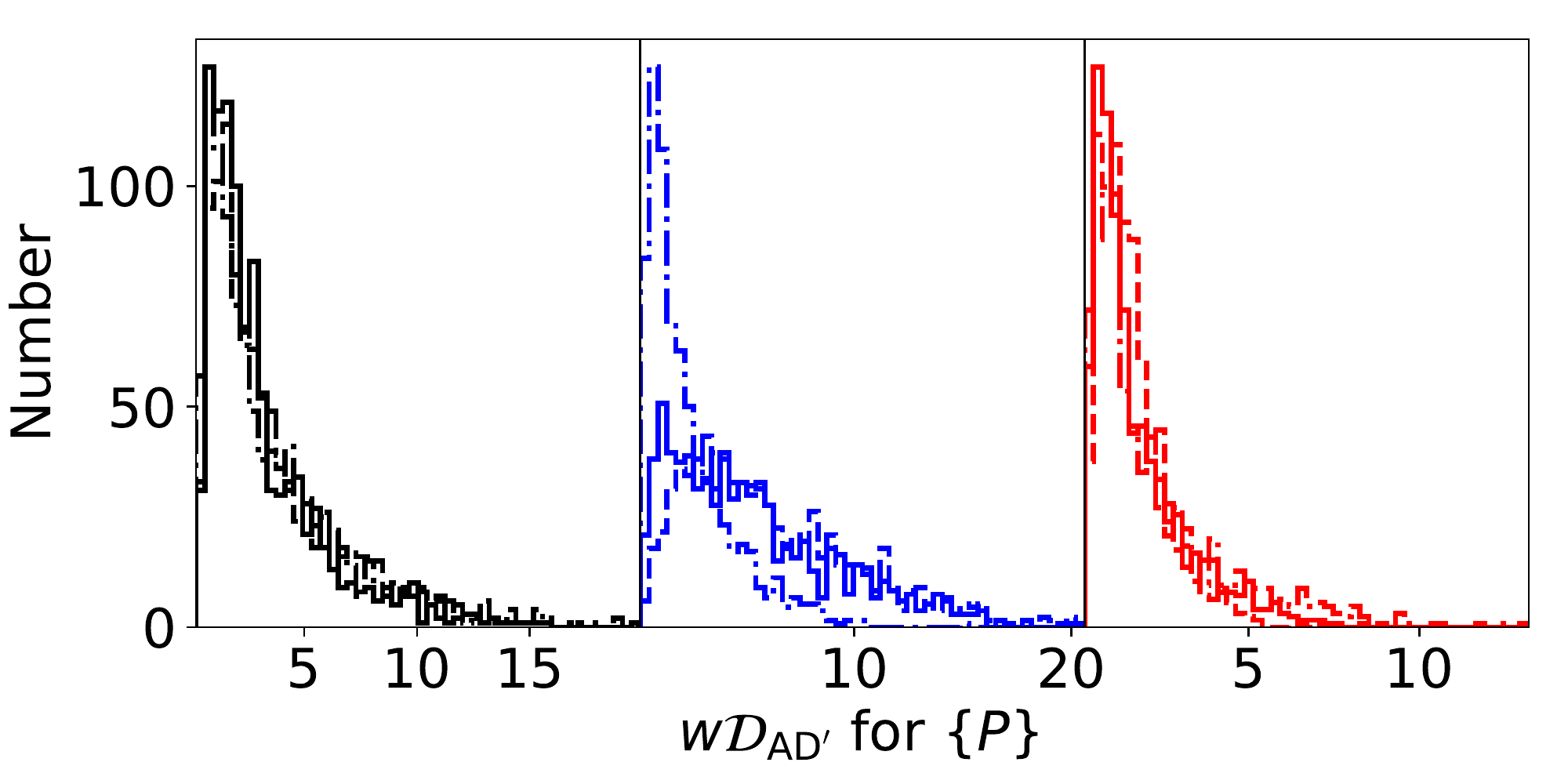} \\
 \includegraphics[scale=0.28,trim={0.2cm 0.2cm 0.4cm 0.2cm},clip]{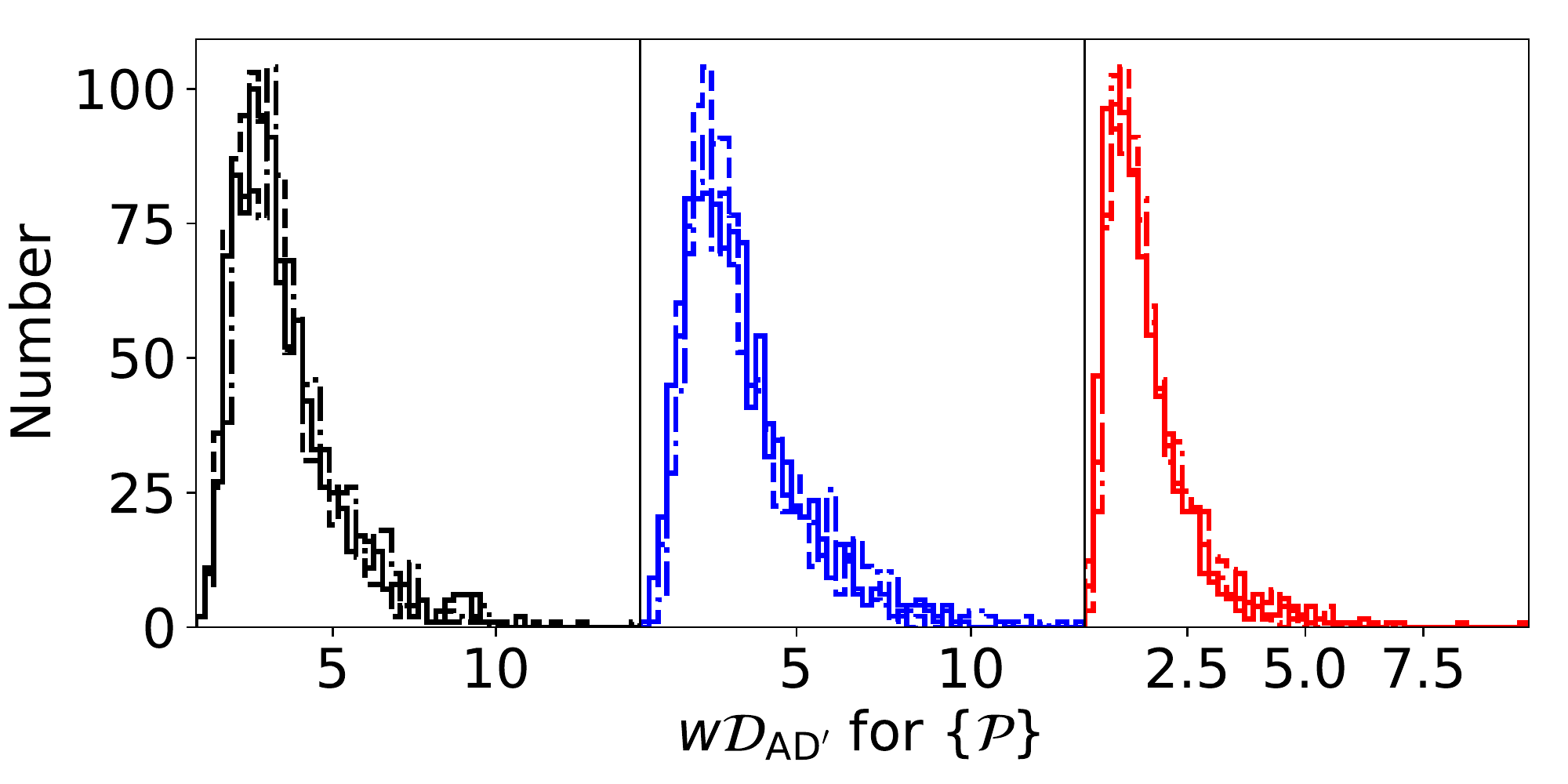} &
 \includegraphics[scale=0.28,trim={0.2cm 0.2cm 0.4cm 0.2cm},clip]{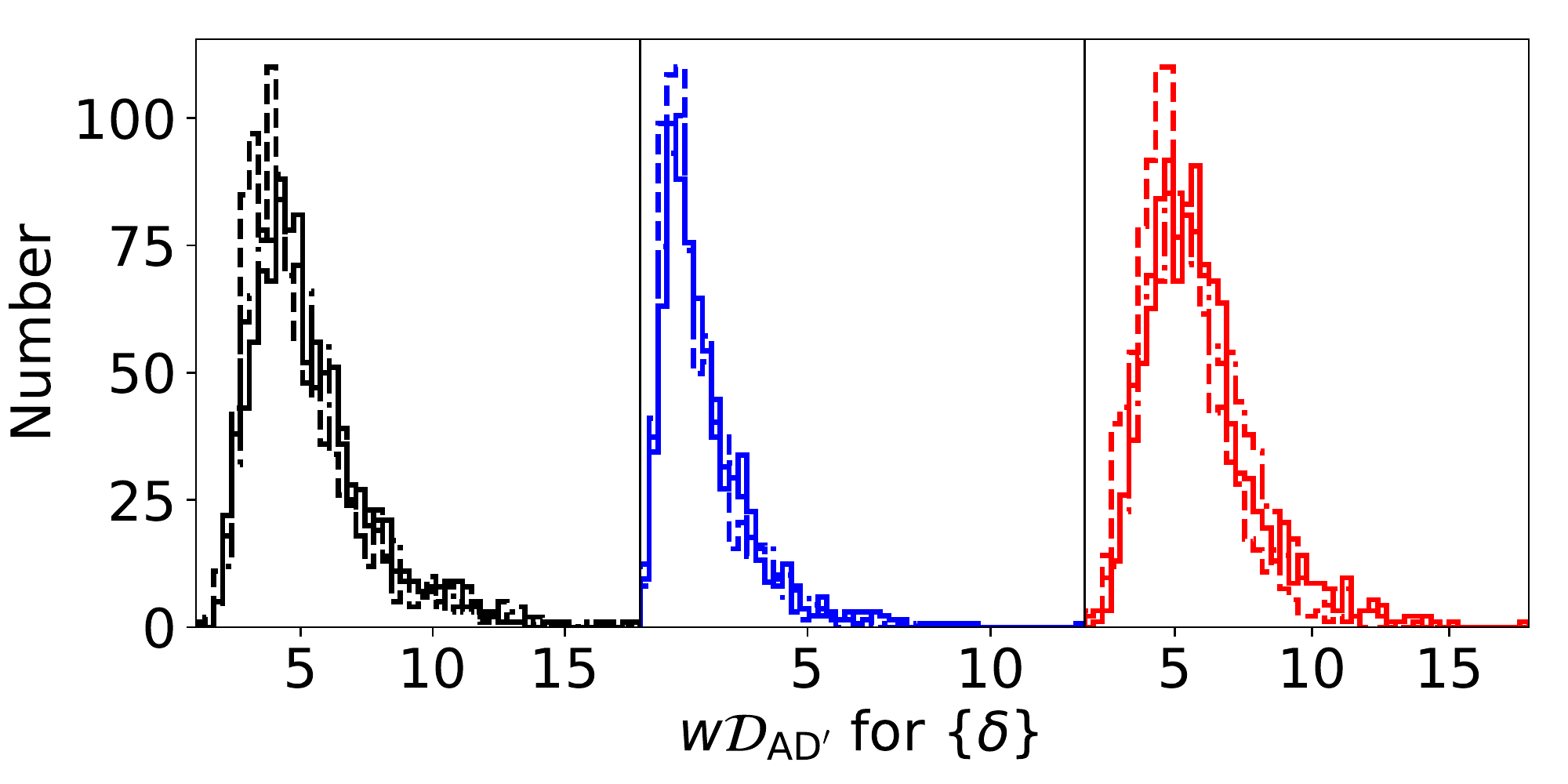} &
 \includegraphics[scale=0.28,trim={0.2cm 0.2cm 0.4cm 0.2cm},clip]{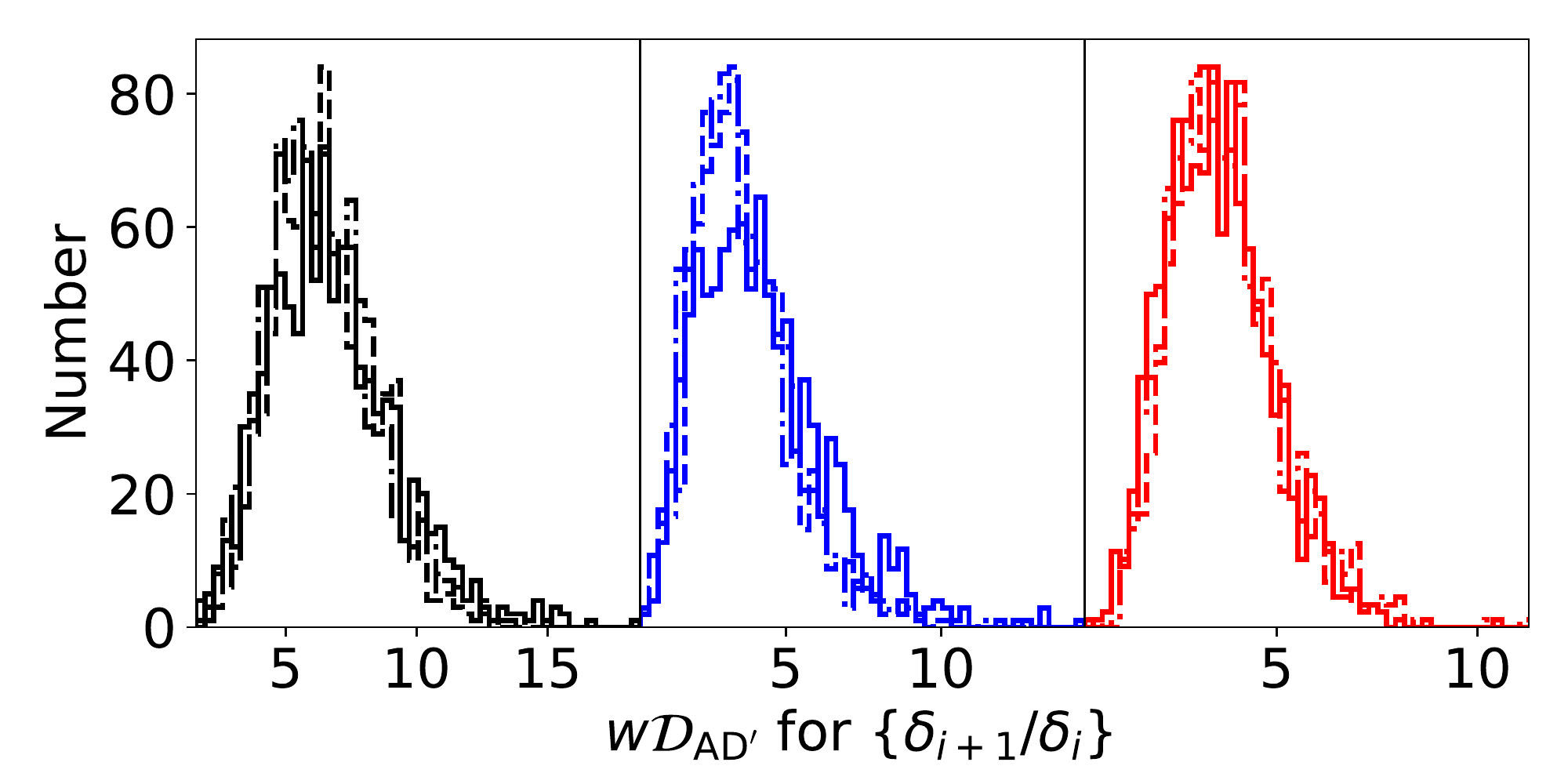} \\
 \includegraphics[scale=0.28,trim={0.2cm 0.2cm 0.4cm 0.2cm},clip]{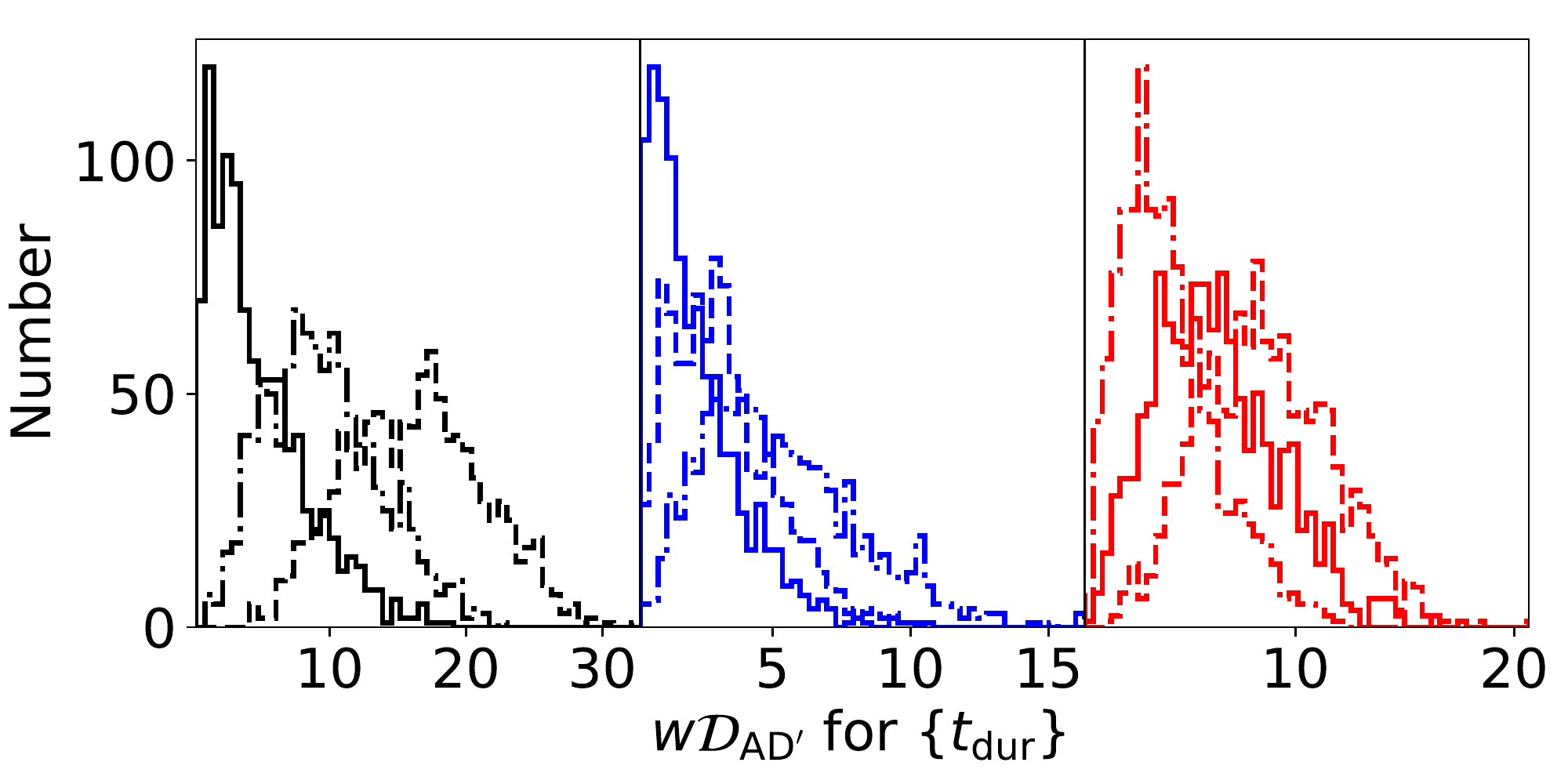} &
 \includegraphics[scale=0.28,trim={0.2cm 0.2cm 0.4cm 0.2cm},clip]{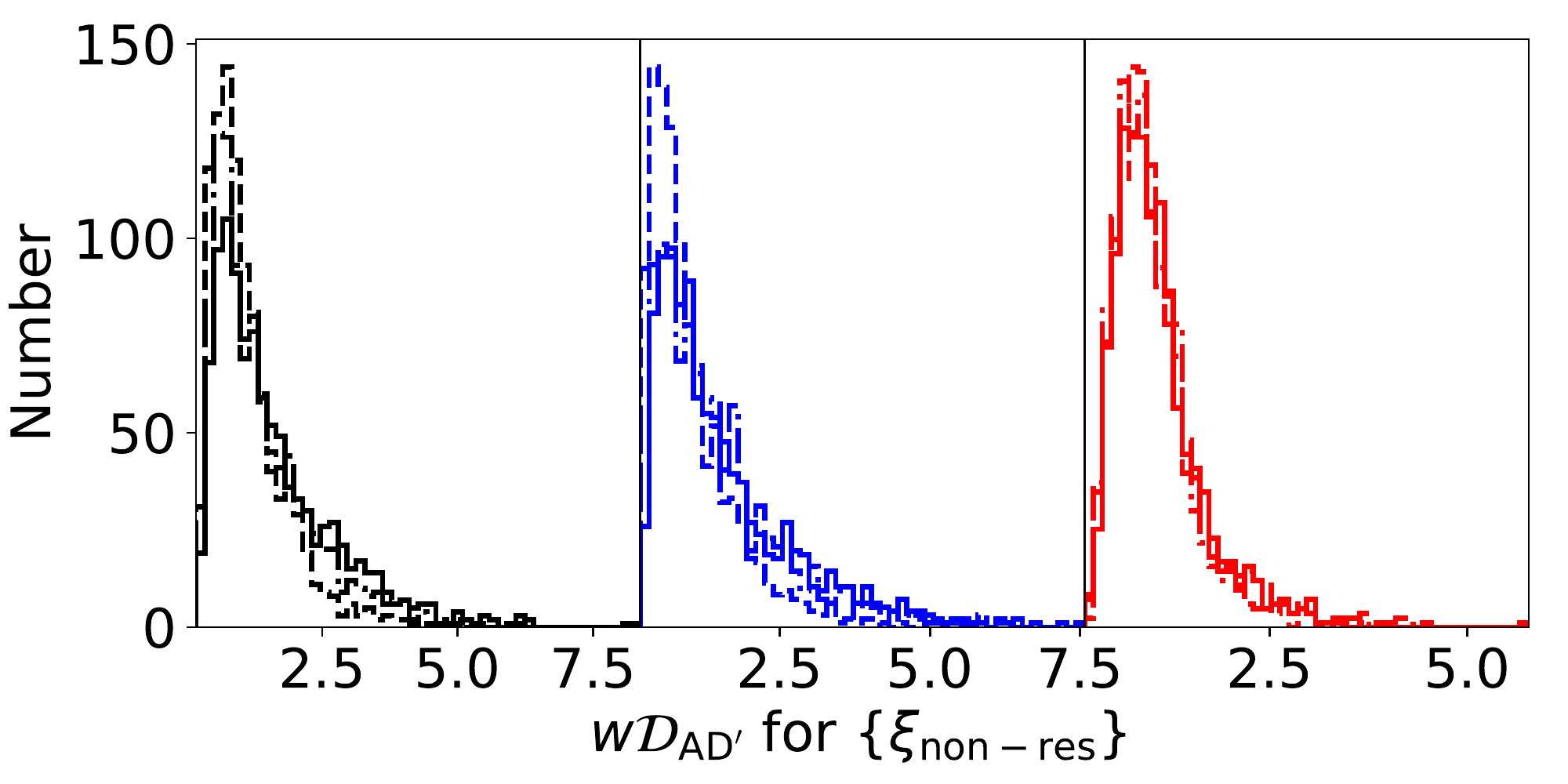} &
 \includegraphics[scale=0.28,trim={0.2cm 0.2cm 0.4cm 0.2cm},clip]{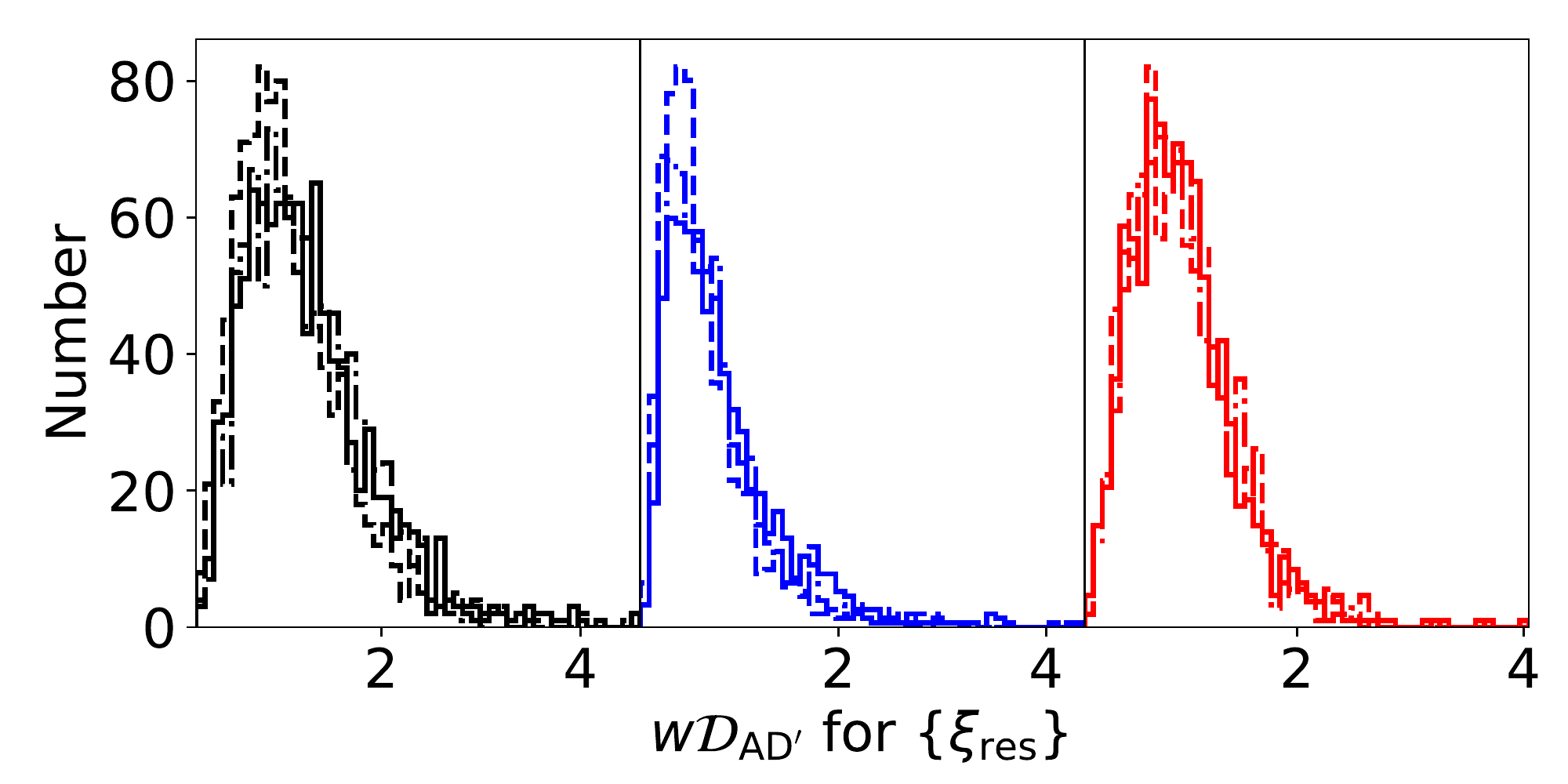} \\
\end{tabular}
\caption{Histograms of the weighted total distances (top row) and individual distance terms (second row and below) for our models as compared to the \Kepler{} data, including 1000 simulated catalogs that pass our distance thresholds of $\mathcal{D}_{W,\rm AD^\prime} = 100$, 90, and 90 for the constant $f_{\rm swpa}$ (dashed histograms), linear $f_{\rm swpa}(b_p - r_p - E^*)$ (solid histograms), and linear $\alpha_P(b_p-r_p-E^*)$ (dash-dotted histograms) models, respectively. The panels, lines, and colors are the same as those in Figure \ref{fig:dists_KS}.}
\label{fig:dists_AD}
\end{figure*}

Figures \ref{fig:const_fswp_corner_KS} and \ref{fig:linear_alphaP_corner_KS} show the ABC posterior distributions for the model parameters of our constant $f_{\rm swpa}+\alpha_P$ and linear $\alpha_P(b_p-r_p-E^*)$ models, respectively.
In Figure \ref{fig:model_alphaP_bprp_split}, we show the observed marginal distributions of our linear $\alpha_P(b_p-r_p-E^*)$ model, for comparison with the linear $f_{\rm swpa}(b_p-r_p-E^*)$ model in Figure \ref{fig:model_fswp_bprp_split}.
While the observed distributions appear very similar between the two models, there are subtle differences in how they fit the marginal distributions of the \Kepler{} planet candidates. These differences are more clearly seen in Figures \ref{fig:dists_KS} and \ref{fig:dists_AD}, where we show how a large number of simulated catalogs from each of the three primary models considered in this study compare to the \Kepler{} catalog in terms of the individual (weighted; KS and AD) distance terms for each observable distribution.

\end{document}